\documentclass[12pt]{article}
\usepackage[dvips]{graphicx}
\usepackage{amssymb}
\usepackage{epsfig}
\usepackage{here}
\evensidemargin=\oddsidemargin
\renewcommand{\baselinestretch}{0.93}
\renewenvironment{thebibliography}[1]
    {\begin{list}{\arabic{enumi}.}
    {\usecounter{enumi}\setlength{\parsep}{0pt}
\setlength{\leftmargin 1.25cm}{\rightmargin 0pt}
     \setlength{\itemsep}{0pt} \settowidth
    {\labelwidth}{#1.}\sloppy}}{\end{list}}

\def\NPB#1#2#3{Nucl. Phys. B\,{\bf #1}, #3 (19#2)}

\def\PLB#1#2#3{Phys. Lett. B\,{\bf #1}, #3 (19#2)}

\def\PRD#1#2#3{Phys. Rev. D\,{\bf #1}, #3 (19#2)}
\def\PRDMM#1#2#3{Phys. Rev. D\,{\bf #1}, #3 (20#2)}
\def\PRL#1#2#3{Phys. Rev. Lett. {\bf#1}, #3 (19#2)}

\newcommand{\SU}[2]{{SU(#1)}_{#2}}

\def\VEV#1{\left\langle #1\right\rangle}

\def\slash#1{#1\!\!\!/\!\,\,}
\def\beq{\begin{equation}}
\def\eeq{\end{equation}}
\def\bea{\begin{eqnarray}}
\def\eea{\end{eqnarray}}
\def\half{\frac{1}{2}}

\def\bq{\begin{quote}}
\def\eq{\end{quote}}

\def\half{\frac{1}{2}}     
\def \lta {\mathrel{\vcenter
     {\hbox{$<$}\nointerlineskip\hbox{$\sim$}}}}
 
\relax

\def\D0{D\O~}

\def\be{\begin{equation}}
\def\ee{\end{equation}} 
\def\beq{\begin{equation}}
\def\eeq{\end{equation}}
\def\eq{\beq\eeq}
\def\beqn{\begin{eqnarray}}
\def\eeqn{\end{eqnarray}}
\def\bea{\begin{eqnarray}}
\def\eea{\end{eqnarray}}
\relax
\def\ba{\begin{array}}
\def\ea{\end{array}}

\def\be{\begin{equation}}
\def\ee{\end{equation}}
\def\bea{\begin{eqnarray}}
\def\eea{\end{eqnarray}}

\def\to{\rightarrow}

\def\dis{\displaystyle}
\def\f{\frac}
\def\ov{\overline}
\def\[{\left[}
\def\]{\right]}
\def\({\left(}
\def\){\right)}

\def\ra{\rightarrow}
\def\ov{\overline}

\def\ta2{\frac{\tau^a}{2}}

\def\sb{{\cal B}}

\def\sq2{\sqrt{2}}

\def\la2{\frac{\lambda^{\alpha}}{2}}

\def\tanb{\tan\hspace*{-0.8mm}\beta}
\def\cb{c_{\beta}}
\def\sb{s_{\beta}}

\def\({\left(}
\def\){\right)}
\def\[{\left[}
\def\]{\right]}

\def\Zbb{{Zb\bar{b}}}

\def\la{\lambda}

\def\x{{\chi}}
\def\w{{\omega}}
\def\O{{\cal O}}

\def\muxx{\mu_{\chi\chi}}
\def\muww{\mu_{\omega\omega}}
\def\cut{\Lambda}
\def\MX{M_\chi}
\def\MW{M_\omega}
\def\MB{{\overline{M}}}

\def\kBI{\f{\kappa_c}{\kappa}}
\def\kB{\f{\kappa}{\kappa_c}}
\def\mtt{m_{tt}}
\def\mtx{m_{t\chi}}
\def\mbw{m_{b\omega}}
\def\muwb{\mu_{\omega b}}
\def\muxt{\mu_{\chi t}}
\def\k{\kappa}

\def\ep{\epsilon}     
\def\Sigma{{\mathbf \Phi}}
\def\sq2{\sqrt{2}}

\def\noi{\noindent}
\def\End{\end{document}}
\def\thisday{August, 2001}

\begin{document}

\thispagestyle{empty}
 
\hspace{-0.38cm}  \thisday  \hfill {\large hep-ph/0108041} \\ [1mm]
\rightline{UTEXAS-HEP-01-013}\\
\rightline{FERMILAB-Pub-01/164-T}
\rightline{ANL-HEP-PR-01-047}\\

\vspace{1.8cm}
 
\renewcommand{\thefootnote}{\fnsymbol{footnote}}
\setcounter{footnote}{1}
 
%%%%%%%%%%%%%%%%%%%%%%%%%%%%%%%%%%%%%%%%%%%%%%%%%%%%%%%%%%%%%%%%%%%%%%%%
%%%                           Titlepage
%%%
%%%%%%%%%%%%%%%%%%%%%%%%%%%%%%%%%%%%%%%%%%%%%%%%%%%%%%%%%%%%%%%%%%%%%%%%
 
\begin{center}
{\Large {\bf Top Quark Seesaw, Vacuum Structure}}
\\[2.5mm]
{\Large {\bf and Electroweak Precision Constraints}}

\vspace{1.5cm}
{\large 
{\sc Hong-Jian He}$\,^{1,2}$,~~ 
{\sc Christopher T. Hill}$\,^{2}$,~~
{\sc Tim M.P. Tait}$\,^3$  
}\\[4mm]
$^1$~{\it The University of Texas at Austin, Austin, Texas 78712, USA}\\[2mm]
$^2$~{\it Fermi National Accelerator Laboratory, 
          Batavia, Illinois 60510, USA}\\[2mm]
%$^3$~{\it Enrico Fermi Institute, University of Chicago,
%          Chicago, Illinois 60637, USA}\\[2mm]
$^3$~{\it Argonne National Laboratory, Argonne, Illinois 60439, USA}

\vspace*{3.8cm}
\begin{abstract}
\vspace*{2mm}
\noindent
We present a complete study of the vacuum 
structure of Top Quark Seesaw models of the Electroweak 
Symmetry Breaking, including bottom quark mass generation. 
Such models emerge naturally from extra dimensions. 
We perform a systematic gap equation analysis and develop
an improved broken phase formulation for including exact 
seesaw mixings.
The composite Higgs boson spectrum is studied in the 
large-$N_c$ fermion-bubble approximation and an improved 
renormalization group approach. 
The theoretically allowed parameter space is restrictive,
leading to well-defined predictions. 
We further analyze the electroweak precision constraints. 
Generically, a heavy composite Higgs boson 
with a mass of $\sim\!1$\,TeV is predicted,
yet fully compatible with the precision data.
\\[3mm]
\noindent PACS number(s): 12.60.Nz, 11.15.Ex, 12.15.Ff \\[0.2cm]
\end{abstract}
\end{center}

\baselineskip20pt   % stretch linespacing in main text
%%%%%%%%%%%%%%%%%%%%%%%%%%%%%%%%%%%%%%%%%%%%%%%%%%%%%%%%%%%%%%%%%%%%%%%%
%%% Text
%%%%%%%%%%%%%%%%%%%%%%%%%%%%%%%%%%%%%%%%%%%%%%%%%%%%%%%%%%%%%%%%%%%%%%%%
\newpage
\renewcommand{\baselinestretch}{0.95}
\setcounter{footnote}{0}
\renewcommand{\thefootnote}{\arabic{footnote}}
\setcounter{page}{2}

\section{\hspace*{-2mm}{\large Introduction}}
%%%%%%%%%%%%%%%%%%%%%%%%%%%%%%%%%%%%%%%%%%%%%%%%%%%%%%%%%%%%%%

Unraveling the mystery of  electroweak symmetry breaking (EWSB) is
the most compelling challenge facing particle physics today. 
It is of central importance
because it devolves into the question of the fundamental
{\em  organizing principle}
for the dynamics at or above the electroweak scale.

Supersymmetry provides an excellent
candidate for this organizing principle. 
It is an extra-dimensional theory in
which the extra dimensions are fermionic, or Grassmannian.
Supersymmetry can lead
naturally, upon ``integrating out'' the extra fermionic dimensions
(i.e., descending from a superspace action to a space-time action),
to perturbative extensions of the Standard Model (SM), 
such as the Minimal Supersymmetric SM (MSSM).  
In such a scheme
the Higgs sector contains at least two weak doublets,
and the lightest Higgs boson is expected to be in a range determined
by the {\em perturbative} electroweak constraints, $\lta 140$ GeV.
From a ``bottom-up'' perspective a lesson
from the supersymmetry is that an  {\em organizing principle} for
physics beyond the Standard Model can 
be derived from hidden extra-dimensions
which are then integrated out.
Upon specifying the algebraic properties of the extra-dimensions
one is led to a particular symmetry structure 
and a class of dynamics for the EWSB.

On the other hand, the organizing principle for physics beyond the
Standard Model may descend from hidden extra
dimensions other than fermionic, and
thus different from the supersymmetry.  
It could, for instance, be a theory of compactified
bosonic extra dimensions with gauge fields in the bulk.  
By using
the transverse lattice technique\,\cite{wang0,wang1,trans,acg}, 
one can ``integrate out'' the bosonic extra dimensions,
preserving gauge invariance and arrive at an effective Lagrangian including
Kaluza-Klein (KK) modes 
(in a sense the KK modes are analogues of superpartners).
This leads naturally to a strong dynamical origin
of the EWSB\,\cite{wang2,cohen}.  Topcolor\,\cite{TCATC,Eichten}
and in particular, the Top Seesaw Model\,\cite{seesaw}, emerge naturally 
from extra dimensions in this way\,\cite{wang2}, following
the original suggestion in \cite{dob1}.  
Top Seesaw models are particularly favored
from our perspective because they have a natural dynamics 
with minimal fine-tuning and are consistent
with the electroweak precision constraints.

The organizing principle of bosonic extra dimensions
leading to strong dynamical electroweak symmetry breaking
can be described in the sequence of Figures 1-4, in analogy with
\cite{wang2}.
In Fig.\,1, we show a lattice approximation to the
fifth dimension of a $1+4$ theory in which the gauge fields, in
particular from QCD, and SM fermions propagate
in the bulk. The lattice description
reveals the $SU(3)\times SU(3)\times \cdots$, one
gauge group per lattice brane, 
the Topcolor structure\,\cite{wang0,wang1,wang2}. 
A Dirac fermion has both
left- and right-handed chiral modes on each lattice brane 
and hopping links to nearest neighbor branes. 

\vspace*{-2cm}
\begin{figure}[H]
\label{fig1}
\begin{center}
\hspace*{-9mm}
\includegraphics[width=10.5cm, height=16.5cm, angle=-90]{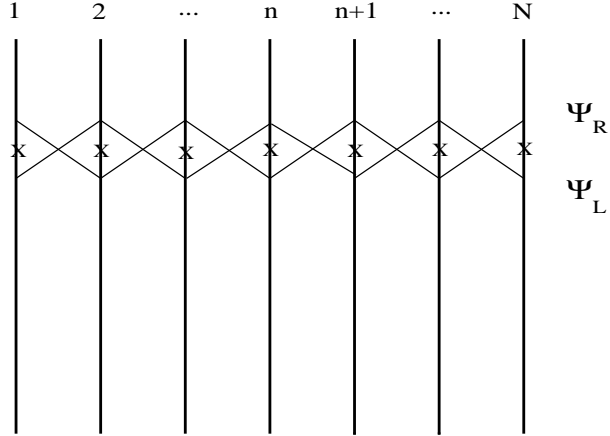} %\\[-20mm]  
%\vspace*{7cm}
%\special{psfile=stplothe.ps
%          angle=0 hscale=80 vscale=80 hoffset=70 voffset=-0}
%\vspace*{-4mm}
\end{center}
\vspace*{-2cm}
\caption[]{Dirac fermion corresponding to constant $\phi$
has both chiral modes on all branes. The $\times$ symbols denote the
$\phi$ couplings on each brane, and the links are the 
latticized fermion kinetic terms which become Wilson
links when gauge fields are present.}
\end{figure}

It is well known that chiral fermions can
be localized in the fifth dimension by background 
fields\,\cite{jackiw,schmaltz}.   
A free fermion has the action,
\beq
\int d^5x \; \bar{\Psi}(i\slash{\partial}
-\partial_5\gamma^5 - \phi(x^5) )\Psi
\eeq
where $\phi$ is a background-field giving mass.
(Here we neglect the gauge interactions.)
From the lattice viewpoint,
we must decompose $\partial_5$ into ``fast''
components (high momentum) and ``slow'' components (low momentum).
The fast components correspond to distance scales much shorter 
than the lattice spacing, and the dynamics
in the lattice description corresponding
to the slow scale must match onto a Lagranigian
which implements the fast scale behavior. 
If the background  field
is approximately constant 
then we impose $\partial_{5\;{\rm fast}}\Psi =0$,
i.e., we discard high momentum field components of $\Psi$
in the lattice approximation,
and both chiral components are kept on each lattice brane.
We thus have the Dirac fermion depicted in Fig.\,1.

If, on the other hand, $\phi(x^5)$
swings through zero rapidly in the vicinity of brane $n$, then we impose
$(\gamma^5 \partial_{5\;{\rm fast}} + \phi(x^5))\Psi =0$ in
the vicinity of this brane, and one chiral component of
$\Psi$ (corresponding to the non-normalizeable solution) is thus 
projected to zero on the brane.
A single chiral component is thus kept on the brane $n$, 
as shown in Fig.\,2. The chiral
zero mode is essentially a localized dislocation in the lattice.

\vspace*{-2cm}
\begin{figure}[H]
\label{fig2}
\begin{center}
\hspace*{-9mm}
\includegraphics[width=10.5cm, height=16.5cm, angle=-90]{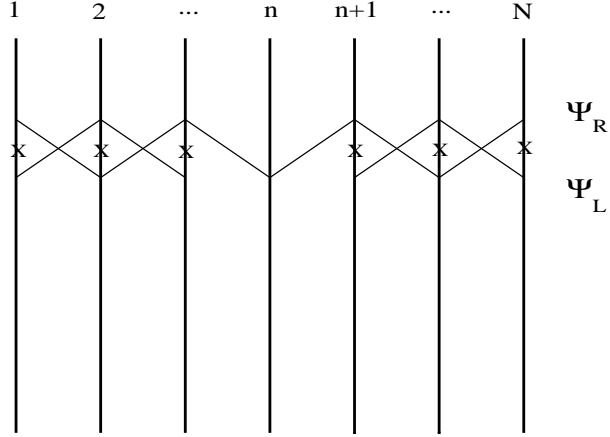} 
%\\[-20mm]  
%\vspace*{7cm}
%\special{psfile=stplothe.ps
%          angle=0 hscale=80 vscale=80 hoffset=70 voffset=-0}
%\vspace*{-4mm}
\end{center}
\vspace*{-2cm}
\caption[]{A chiral fermion occurs on brane $n$ where 
$\phi(x^5)$ swings rapidly through zero. The chiral fermion has
kinetic term (Wilson links) connecting to adjoining branes. 
}
\end{figure}

We can furthermore demand the coupling strength of 
$SU(3)_n$ on the $n$-th brane to
be arbitrary, hence it can be super-critical. 
This can be triggered by renormalization
effects due to the $\phi$ field as well, e.g., a background field
coupling as in $\phi(x^5)(G^a_{\mu\nu})^2$,
will renormalize the coupling on the brane $n$ \cite{wang2}.
It is, therefore,  not coincidental to
expect this to happen; indeed a variety of
effects are expected near the
dislocation, e.g., the chiral fermions themselves
can feed-back onto the gauge fields to produce such renormalization
effects. 
The result is a chiral condensate on the brane $n$ forming between
chiral fermions. Identifying $\Psi=(t,b)_L$  and $t_R$ as
the chiral zero-modes on the brane $n$ and,
in the limit that we take the compact extra dimension very small,
the nearest neighbor links decouple at low energies. 
As shown in Fig.\,3, under this limit
we recover a Topcolor model with pure top quark 
condensation\,\cite{nambu,miransky,BHL,condrev}. \\

\vspace*{-2cm}
\begin{figure}[H]
\label{fig3}
\begin{center}
\hspace*{-9mm}
\includegraphics[width=10.5cm, height=16.5cm, angle=-90]{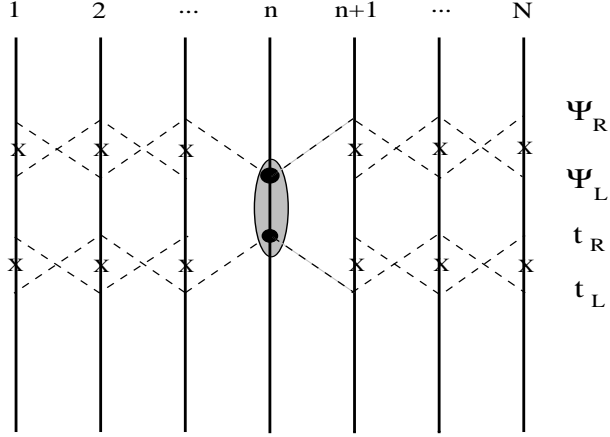} 
%\\[-20mm]  
%\vspace*{7cm}
%\special{psfile=stplothe.ps
%          angle=0 hscale=80 vscale=80 hoffset=70 voffset=-0}
%\vspace*{-4mm}
\end{center}
\vspace*{-2cm}
\caption[]{Pure top quark condensation by Topcolor is obtained in the limit
of critical coupling on brane $n$ and
decoupling to the nearest neighbors.  Decoupling corresponds to taking the
compactification mass scale large; the links are then denoted by dashed lines.
}
\end{figure}
\vspace*{-2cm}
\begin{figure}[H]
\label{fig4}
\begin{center}
\hspace*{-9mm}
\includegraphics[width=10.5cm, height=16.5cm, angle=-90]{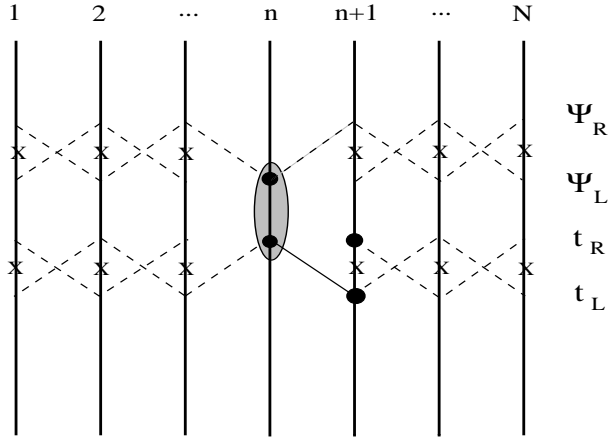} 
%\\[-20mm]  
%\vspace*{7cm}
%\special{psfile=stplothe.ps
%          angle=0 hscale=80 vscale=80 hoffset=70 voffset=-0}
%\vspace*{-4mm}
\end{center}
\vspace*{-2cm}
\caption[]{Top Seesaw Model arises when the effects of nearest
neighbor vector-like fermions are retained, i.e., when these
heavier states are only partially decoupled. Keeping more
links maintains the seesaw. Usually we denote $t_{Rn}\sim \chi_R$,
 $t_{Ln+1}\sim \chi_L$, $t_{Rn+1}\sim t_R$ \cite{seesaw}.   
}
\end{figure}

In Fig.\,4, we consider the case that some of the 
links to nearest neighbors are not completely decoupled.
Again, this can arise from renormalizations due to background fields, or
due to warping\,\cite{wang2}. Thus the mixing with heavy vector-like
fermions occurs in addition to the chiral dynamics
on the brane $n$. In this limit, we naturally obtain an effective
Top Quark Seesaw Model\,\cite{seesaw}.

In the present paper we will undertake a complete
and systematic analysis of the effective 4-dimensional
Top Seesaw vacuum structure and the precision electroweak constraints.
This also extends the earlier works 
in Refs.\,\cite{seesaw,georgi,chiv} which studied
the precision bounds on the seesaw scheme.
The Higgs boson in this scheme is
composite and heavy, with a mass $\sim\!1$\,TeV, 
and the theory would seemingly be
ruled out by the precision constraints on the
oblique parameters $S$-$T$ \cite{STU}.
We have, however, necessary compensating
positive $T$ contributions coming from the additional seesaw quarks
($\x$), and the size of these effects can be well predicted 
by systematically solving the gap equations. 
Remarkably, a heavy Higgs boson is derived and
naturally consistent with precision constraints
in the Top Seesaw model.

In the recent classification of various models 
by Peskin and Wells\,\cite{pw},  such compensating effects
have been characterized as ``conspiratorial''.  Certainly many
models introduce such compensating effects in an {\em ad hoc} 
way to achieve the consistency with the precision data.
However, when the Top Seesaw was first
proposed in 1998, it lay outside of the $S$-$T$ ellipse by
several standard deviations\,\cite{seesaw},
and the model was thus DOA (dead on arrival).  
Remarkably, in 1999, with  a refined initial
state radiation and $W$-mass determination at LEP-II, the $S$-$T$ 
error ellipse shifted along its major axis toward the upper right.
Since then the theory remains fully consistent
at the $2\sigma$ level, as illustrated in Fig.\,5.
Indeed the theory lies within the $S$-$T$ plot for 
natural values of its parameters.  One might say
that, with the theoretically expected scale for the seesaw partner
mass of $M_\chi \sim 4$\,TeV, the shift in the
error ellipse was predicted by the theory --- the Top Seesaw has
therefore scored its first predictive phenomenological success!
Or, more conservatively, 
we may view the measured error ellipse as a determination of the heavy seesaw
partner mass, and obtain roughly $M_\chi \sim 4$\,TeV.
In this picture, the high precision
electroweak measurements are therefore probing the mass
of a heavy new particle, the $\chi$ quark, significantly above
the electroweak mass scale.\\

\vspace*{-4cm}
\begin{figure}[H]
\label{fig:diagram00}
\begin{center}
\hspace*{-15mm}
\includegraphics[width=13cm, height=15cm]{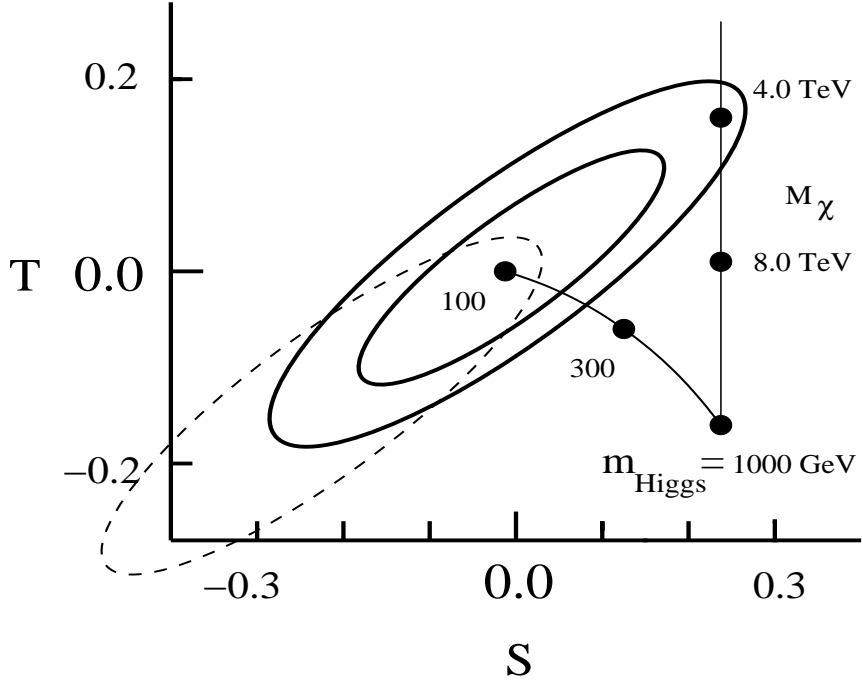} 
%\\[-20mm]  
%\vspace*{7cm}
%\special{psfile=stplothe.ps
%          angle=0 hscale=80 vscale=80 hoffset=70 voffset=-0}
%\vspace*{-4mm}
\end{center}
\vspace*{-3.3cm}
\caption[]{
The $68\%$ and $95\%$\,C.L.  $S$-$T$ contours (solid), superimposing the
Standard Model curve for Higgs mass varying from $100$\,GeV 
up to $1000$\,GeV. The pre-1999 $95\%$ ellipse is
shown with a dashed line.
For the Top Seesaw model with a $1$\,TeV composite Higgs,
we show the $S$-$T$ contributions as a function of $\chi$ mass.  
The data is therefore consistent
with a $\sim\!1$\,TeV Higgs and $M_\chi \sim 4.0$\,TeV.
(The $S$-$T$ ellipses are taken from 1999 precision fit\,\cite{99fit}.)   
}
\end{figure}

Let us briefly summarize the logical path that leads
to the Top Quark Seesaw model, irrespective of
the recent interest in bosonic extra dimensions as a
rationale for this scheme.
Indeed, the observed large top quark mass at Tevatron
is suggestive of new dynamics responsible 
for generating the EWSB involving intimately the top quark.
The ``Top quark condensation'' or ``Top-mode
Standard Model'' \cite{nambu,miransky,BHL,condrev},
is the earliest and simplest idea that
involves a BCS-like pairing $ \VEV{\bar{t}t} $.
It predicts a top quark mass in the SM determined by the
quasi-infrared fixed point\,\cite{hill},
$m_t \sim 220$ GeV, provided the new dynamics scale $\cut$ 
for the condensate generation is chosen to be very large.
The model involves fine-tuning in the gap equation under
the large $\Lambda$ limit, and  the degree of fine-tuning is of 
$\O(m_t^2/\cut^2)$.
%\sim \O(v_{\rm weak}^2/\Lambda^2)$.  
The minimal top condensate model predicts a too heavy top quark
mass, so the simplest scheme is ruled out.

In top condensation, with the fermion-bubble approximation
[omitting the full renormalization group (RG) improvement 
inherent in \cite{hill}], it is conceptually easy to see that
a dynamical mass gap $m_{\rm dyn}$ is generated and related
to the weak scale through the Pagels-Stokar formula\,\cite{PS},
\bea
f_\pi^2 = v_{\rm weak}^2 = \frac{N_c}{16\pi^2}m_{\rm dyn}^2
                       \ln(\Lambda^2/m_{\rm dyn}^2) \,,
\label{eq:PS00}
\eea
where 
$v_{\rm weak} =[2\sqrt{2}G_F]^{-1/2} \simeq 174$\,GeV. 
This relation leads to $m_{\rm dyn} \sim 700$\,GeV
for a typical Topcolor breaking scale
$\cut\sim 3.5$\,TeV. Thus, the degree of
fine-tuning is roughly reduced to the order of 
$\sim\!\(m_{\rm dyn}/\cut\)^2 \!\sim\! (1/4)^2 \!\sim\! 10\%$, 
which is at a reasonable level and is actually ``realistic''
for the Nambu-Jona-Lasinio (NJL) model as an approximation to
the full dynamics. 
[E.g., the NJL model with fermion loops slightly 
exaggerates the degree of fine-tuning, and when it fits to
QCD, one has a degree of fine-tuning, roughly about  
$\({\rm mass\,gap}/\cut\)^2 \!\sim\!\(\f{1}{3}M_p/M_p\)^2 \sim 11\% $, 
where $M_p\sim 1$\,GeV is the mass of proton
and $M_p/3$ the dynamical mass of constituent quarks.]
If the top quark mass had been $\sim 700$ GeV, our problem would
have been solved, and the EWSB would necessarily
be identified with a $\bar{t}t$ condensate. Raising the scale
of $\Lambda$ leads to the aforementioned fine-tuning problem 
and the top quark is too light
to produce the full electroweak condensate.

Topcolor\,\cite{TCATC,Eichten} is gauge
dynamics that can produce a nonzero $\VEV{\bar{t}t}$ condensation.
It involves an imbedding of QCD into
a larger group, which is essentially dictated
by the  quantum numbers of the top quark to be 
$SU(3)\rightarrow SU(3)_1\times SU(3)_2 \times \cdots $
(and possibly also the $U(1)_Y\rightarrow
U(1)_{Y1} \times U(1)_{Y2}\times \cdots $).
While this construction always seemed
{\em ad hoc}, with latticized bosonic extra-dimensions as
an organizing principle, 
we have seen that it becomes natural\,\cite{wang0,wang1,wang2}.
Topcolor can directly produce the $\VEV{\bar{t}t}$ condensate,
and the Pagels-Stokar relation (\ref{eq:PS00}) requires 
$\Lambda \sim 10^{14}$\,GeV. Thus
the fine-tuning $\sim\! m_t^2/\Lambda^2 \!\sim\! 10^{-23}$ becomes a
severe problem in the simplest realization.  Alternatively, Topcolor
can produce a light top mass 
at the natural scale $\Lambda \sim \O(1)$\,TeV, 
and then another strong dynamics, e.g., Technicolor, is required
to provide the majority strength of the EWSB. 
This is known as Topcolor Assisted Technicolor (TC2)\,\cite{TCATC},
and it frees one from the requirement
that the top quark condensate generates all of the observed 
$v_{\rm weak}$. It also largely solves the problematic 
constraints on the Extended Technicolor (ETC), which
prohibits the generation of a large mass $m_t\sim v_{\rm weak}$.  
Many interesting phenomenological
consequences of this TC2 scheme arise\,\cite{Eichten,pheno}.

We can, alternatively, construct a Top Quark Seesaw model in which
the dynamical mass term involving the top
quark is of order $700$\,GeV and thus is associated
with the {\em full} electroweak symmetry breaking.
This involves typically a pairing of the  $t_L$ ($I=\half$) with a new
quark, $\chi_R$ ($I=0$), which has the same quantum numbers
as $t_R$.   We choose, for naturalness sake,
$\Lambda \sim \O({\rm TeV})$, and hence
this mass term is of the order
$\sim\! 700$\,GeV by the Pagels-Stokar formula (\ref{eq:PS00}).
We then incorporate an $I=0$ quark with the
same quantum numbers as $t_R$, $\chi_L$, with
additional mass terms, and we construct a seesaw mechanism.
With the seesaw it is possible to 
adjust the physical mass of the top quark to 
its experimental value of $174$\,GeV\,\cite{seesaw}.  
Hence, the Topcolor Seesaw mechanism can be readily implemented by
introducing a pair of iso-singlet, vector-like
quarks $\chi_L$ and $\chi_R$,  of hypercharge $Y = 4/3$, 
in analogy with the $t_R$.  This model
produces a bound-state Higgs boson, primarily composed
of $\bar{t}_L\chi_R$ with a mass of order $\sim 1$\,TeV or so,
while the $\chi$ mass is at the TeV scale.

Note that the  Top Seesaw model 
{\em does not invoke Technicolor}, but rather 
replaces Technicolor entirely with Topcolor.
In a sense, it is a pure ETC model, where ETC (Topcolor) is 
sufficiently strong to form condensates.
It thus offers new model building possibilities, and
may allow interesting extensions to
solve the flavor problem. The basic dynamics of the model can
be extended to all families if one is willing to 
tolerate more fine-tuning.  Again, extra-dimensions
point the way to a full flavor model extension \cite{wang2}.
While there are the
additional ``$\chi$" quarks involved in the strong dynamics, 
{\em these do not carry
 weak-isospin quantum numbers}. This is an advantage from the
viewpoint of model building, since the constraint of the
$S$ parameter is essentially irrelevant for the
Top Seesaw, since we have only a chiral
top quark condensate in the EWSB channel, and we extend by
including only vector-like fermions.

The Top Seesaw model
makes a robust prediction about the nature of the
electroweak condensate: the left-handed top quark is unambiguously
identified as the electroweak-gauged condensate fermion.  The scheme
demands the presence of Topcolor interactions, 
but beyond the $I=1/2$ component of the EWSB, the remainder of the
structure, e.g., the $\chi$ quarks and the additional strong forces
which they feel, appear to be fairly arbitrary.  
However, as we have seen above, 
a remarkable aspect of the Top Seesaw 
model,   is that the ingredients,  which otherwise appear to be rather
arbitrary, i.e., Topcolor, (tilting $U(1)$'s), vector-like
$\chi$ quarks, etc., are all naturally given by 
theories of extra-dimensions where top and gauge fields 
propagate in the bulk\,\cite{wang0,wang1,wang2}. 
The theory may be depicted
graphically from the latticized bulk in Fig.\,4 as explained above.  
One obtains an effective $1+3$ dimensional
Lagrangian description in which all of the SM
gauge groups are replicated for each Kaluza-Klein (KK) mode,
e.g., for QCD we find $SU(3)\rightarrow SU(3)\times SU(3)\times \cdots$,
with $N$ additional copies for $N$ KK modes.  Moreover,
the vector-like $\chi$ quarks can arise as the KK modes
of fermions in the bulk.

As mentioned at the outset, the Top Seesaw scheme 
implies that, in the absence of the seesaw
mechanism, the top
quark would have a much larger mass, of order $\sim\! 700$ GeV. This 
has the effect of raising the masses of all the colorons and
any additional heavy gauge bosons, permitting the full Topcolor structure to
be moved to somewhat higher mass scales.  This gives more model-building
elbow room, and may reflect the reality of new strong dynamics.
We believe that the Top Seesaw is a sufficiently
significant and novel, but relatively new idea in dynamical
models of EWSB and opens up a large range of new model building
possibilities.

In this work, we perform a systematic analysis of
the dynamical vacuum structure for minimal top seesaw models
by quantitatively solving the gap equations.
The top mass and the full EWSB are generated together. The
inclusion of bottom seesaw is further studied.
We carry out the analysis using an improved broken phase
formulation, in comparison to the traditional gauge-invariant
formalism; the former allows us to treat all the seesaw mixing effects 
in a precise way and thus reliably analyze the model parameter space.
The composite Higgs mass spectrum is computed by 
several independent approaches. 
%the large-$N_c$ fermion-bubble approximation and also by an improved 
%renormalization group approach (including the Higgs
%self-coupling evolution). 
We further study the precision bounds via the
$S$-$T$ oblique corrections and the $\Zbb$ vertex correction, from which we 
derive nontrivial constraints on the parameter space 
and the composite Higgs spectrum.
The effects of Topcolor instantons\,\cite{TCATC} 
are also analyzed, as a source 
to generate part or all of the bottom quark mass.

%%%%%%%%%%%%%%%%%%%%%%%%%%%%%%%%%%%%%%%%%%%%%%%%%%%%%%%%%%%%%%%%%%%%%%%%%
\newpage
\section{\hspace*{-2mm}{\large
Dynamical Top Seesaw Model and the Gap Equations}}
\label{sec:tseesaw}

\subsection{\hspace*{-2mm}{\normalsize
The Minimal Model}}

In the minimal Top Seesaw scheme \cite{seesaw}
the full EWSB occurs via the
condensation of the left-handed top quark $t_L$ with a new, right-handed
weak-singlet quark $\chi_R$. 
The $\chi_R$ quark has hypercharge $Y= 4/3$ and 
is thus indistinguishable from the $t_R$.  
The dynamics which leads to this condensate is Topcolor, as discussed
below, and no tilting $U(1)$'s  are required. 
The fermionic mass scale of this weak-isospin 
$I=1/2$ condensate is  $\sim 700$\,GeV. 
This corresponds to the formation of a dynamical boundstate
weak-doublet Higgs field, 
$H \sim (\overline{\chi_R} t_L, \overline{\chi_R} b_L)^T$.
To leading order in $1/N_c$ this yields, via the Pagels-Stokar formula,
the proper Higgs vacuum expectation value $v_{weak} = 174$\,GeV
and the top quark $I=\half$ dynamical mass term,
\bea
m_{t\chi}\ov{t_L}\,\chi_R \,+\, {\rm h.c.}\,,
\qquad\qquad (\mtx \sim 700\,{\rm GeV})\,.
\eea 
Moreover, the model incorporates a left-handed 
weak-isosinglet $\chi$ quark, with $(I,\,Y)=(0,\,4/3)$.
Thus, $\chi$ quarks have an allowed Dirac mass term,
\bea
\label{dirac1}
\mu_{\chi\chi}\ov{\chi_L}\chi_R \,+\, {\rm h.c.} \,.
\eea
This may be viewed as a dynamical mass through additional new 
dynamics (yet unspecified)  at a still higher mass scale.  
However, since the $\chi_R$ and $\chi_L$ quarks
carry the same $(I,\,Y)$ charges, we prefer to introduce
Eq.\,(\ref{dirac1}) by hand and ignore, 
temporarily, its dynamical origin.
Furthermore, the left-handed $\chi$ quark
can form an allowed weak-singlet Dirac mass term 
with the right-handed top quark, leading to,
\bea
\label{dirac2}
\mu_{\chi t}\ov{\chi_L} \,t_R \,+\, {\rm h.c.} \,,
\eea
which again may be viewed as a dynamical mass term in an enlarged theory. 
There is no direct left-handed top
condensate with the right-handed anti-top in this scheme,
since they do not share the same strong Topcolor dynamics (cf. Sec.\,2.2). 
Thus, the resulting mass matrix for the $t-\chi$ system is,
\bea
\label{eq:ssmatrix}
- \left( \overline{t_L} ~~ \overline{\chi_L} \right)
 \left\lgroup 
\begin{array}{cc}
0 & m_{t \chi} \\[0.2cm]
\mu_{\chi t} & \mu_{\chi \chi} \\
\end{array}
\right\rgroup
\left(\!\!
\begin{array}{c}
t_R \\
\chi_R
\end{array}
\!\!\right)
\,+\, {\rm h.c.} \,.
\eea
This seesaw mass matrix can be exactly diagonalized by rotating the
left- and right-handed fields,
\bea
\label{eq:seesawd1}
\left(
\begin{array}{c}
t_L \\[2mm]
\chi_L
\end{array}
\right)
 = {\mathbf K^t_L}
\left(
\begin{array}{c}
t^\prime_L \\[2mm]
\chi^\prime_L
\end{array} 
\right),
~~~&~~~ 
\left(
\begin{array}{c}
t_R \\[2mm]
\chi_R
\end{array}
\right)
 = {\mathbf K^t_R}
\left(
\begin{array}{c}
t^\prime_R \\[2mm]
\chi^\prime_R
\end{array} 
\right),
\eea
with
\bea
\label{eq:seesawd2}
{\mathbf K^t_L} =
\left\lgroup
\begin{array}{rr}
\!\! c_L & s_L \\
\!\!-s_L & c_L
\end{array} 
\right\rgroup ,
~~~&~~~
{\mathbf K^t_R} =
\left\lgroup
\begin{array}{rr}
\!\!-c_R & s_R \\
\!\! s_R & c_R
\end{array} 
\right\rgroup ,
\eea
which are determined by obtaining the (positive)
mass eigenvalues, $m_t$ and $M_\chi$.  For convenience,
we have used the abbreviation $s_L = \sin \theta_L$, and so forth.
Our parametrization has also implicitly assumed the mass matrix
to be real, and thus orthogonal.
In the absence of further ingredients, this will always be the case
because any stray complex phase in the mass matrix can be absorbed by
redefining the fermion fields.
The (rotated) mass eigenstate
fields are denoted by $t^\prime$ and $\chi^\prime$ to distinguish
them from the interaction eigenstate fields $t$ and $\chi$\,.
The mass eigenvalues and rotation angles are given by,
\bea
\label{eq:mt}
m^2_t   &=& \frac{1}{2}
\left[ \, \mu^2_{\chi \chi} + \mu^2_{\chi t} + m^2_{t \chi} -
\sqrt{ \left( \mu^2_{\chi \chi} + \mu^2_{\chi t} + m^2_{t \chi}
\right)^2
- 4 \, \mu^2_{\chi t} \, m^2_{t \chi} } \, \right] \, , \\[4mm]
& \longrightarrow & \left. \frac{m_{t\chi}^{2}\mu_{\chi t}
^{2}}{\mu_{\chi t}^{2}+\mu_{\chi\chi }^{2}}\right| _{(\mu_{\chi\chi }\gg 
\mu_{\chi t}, \,m_{t\chi})}
, \nonumber \\[5mm]
\label{eq:MX}
m^2_\chi  &=& \frac{1}{2}
\left[ \, \mu^2_{\chi \chi} + \mu^2_{\chi t} + m^2_{t \chi} +
\sqrt{ \left( \mu^2_{\chi \chi} + \mu^2_{\chi t} + m^2_{t \chi}
\right)^2
- 4 \, \mu^2_{\chi t} \, m^2_{t \chi} } \, \right] \, , \\[4mm]
& \longrightarrow & \left. \mu_{\chi\chi}^2 + \mu^2_{\chi t}
\right|_{(\mu_{\chi\chi }\gg \mu_{\chi t},\, m_{t\chi})}, 
\nonumber \\[6mm]
\label{eq:rotL}
\(\!\! \ba{c} s_L  \\ c_L \ea \!\!\) 
&=& \frac{1}{\sqrt{2}} \[ 1
\mp \frac{ \mu^2_{\chi \chi} + \mu^2_{\chi t} - m^2_{t \chi}}
{M^2_\chi - m^2_t} \]^{\half}
\, , \\[4mm]
\label{eq:rotR}
\(\!\! \ba{c} s_R \\ c_R \ea \!\!\)
&=& \frac{1}{\sqrt{2}} \[ 1
\mp \frac{ \mu^2_{\chi \chi} - \mu^2_{\chi t} + m^2_{t \chi}}
{M^2_\chi - m^2_t}\]^{\half}
\, .
\eea
The fermionic mass matrix thus admits
a conventional seesaw mechanism, yielding the physical top quark mass as an
eigenvalue that is $\sim m_{t\chi}\mu_{\chi t}/\mu_{\chi\chi} 
\ll m_{t\chi} \sim 700$\,GeV. 
The top quark mass can be adjusted to its experimental value by
the choice of $ \mu_{\chi t}/\mu_{\chi\chi} $. The
diagonalization of the fermionic mass matrix does not
affect the physical vacuum expectation value (VEV), 
$v_{weak} \simeq 174$\,GeV, of the composite Higgs doublet.
Indeed, the Pagels-Stokar formula is now modified as,
\bea
\label{ps2}
v_{\rm weak}^2 \equiv f_\pi^2 
\simeq \frac{N_c}{16\pi^2} \frac{m_{t}^2}{\sin^2\theta_R} 
\( \ln\frac{\Lambda^2}{\overline{M}^2} + c\)
\eea
where $m_t$ is the physical top mass, 
$\sin\theta_R =s_R\approx \muxt/\muxx$ 
the right-handed seesaw angle, 
$\ov{M}=\sqrt{\mu_{\x t}^2+\mu_{\x\x}^2}$,
and $c$ denotes sub-leading terms, and we expect $c\sim {\cal{O}}(1)$. 

The Pagels-Stokar formula now differs from that obtained (in large-$N_c$
approximation) for pure top quark condensation models, by a large enhancement
factor $1/\sin ^{2}\theta_R$.\, 
This is a direct consequence of the seesaw mechanism. 
The mechanism incorporates $\psi_L= (t_L,\,b_L)$, 
which provides the source of
the weak-isospin $I=1/2$ quantum number 
of the composite Higgs boson, and thus the
origin of the EWSB vacuum condensate. 
Note that we have
separated the problem of EWSB from the weak-isosinglet physics in the 
$\chi_{L,R}$ and $t_R$ sector, which is an advantage of the
seesaw mechanism since the electroweak constraints are not so
restrictive on the isosinglets.

\subsection{\hspace*{-2mm}{\normalsize
Topcolor Dynamics}}

Let us turn to some of the
dynamical questions, e.g.,
how does Topcolor produce the dynamical
$m_{t\chi}$ mass term? 
We proceed by introducing an embedding of QCD into the gauge groups 
$SU(3)_{1}\otimes SU(3)_{2}$, with coupling constants $h_{1}$ and $h_{2}$,
respectively. These symmetry groups are broken down to $SU(3)_{QCD}$ at a
high mass scale. We assign the representations for relevant fermions 
under the full set of gauge groups 
$SU(3)_{1}\otimes SU(3)_{2}\otimes SU(2)_{W}\otimes U(1)_{Y}$ as below,
\begin{equation}
\psi_{L} \!:~~ ({\bf 3,1,2,} \, +1/3) \,,~~~~~
\chi_{R} \!:~~ ({\bf 3,1,1,} \, +4/3) \,,~~~~~
t_{R}, \,\chi_{L}\!:~~ ({\bf 1,3,1,} \, +4/3) \,.
\end{equation}
This set of fermions is incomplete; the representation specified has
$[SU(3)_1]^3$, $[SU(3)_2]^3$, and $U(1)_Y [SU(3)_{1,2}]^2$ gauge anomalies.
These anomalies will be canceled by fermions associated with either the
dynamical breaking of $SU(3)_1 \otimes SU(3)_2$, or with the
$b$ quark mass generation
(an explicit realization of the latter case will be given in 
Sec.~\ref{sec:bseesaw}). 
The crucial dynamics of the EWSB and 
top quark mass generation will not depend on the details of these
additional fermions. Schematically, the picture looks like:\\

\begin{center}
\underline{$\ \ \ \ \ \ \ \ ~~SU(3)_{1}$ \ \ \ \ \ \ \ \ \ $\ \ \ \ \ \ \,\,
SU(3)_{2}$ \ \ \ \ \ \ \ \ \ }
\vspace*{2mm}

$\left\lgroup
\begin{array}{c}
\,\left( 
\begin{array}{c}
t_L \\ 
b_L
\end{array}
\right)\, 
\\[4mm]
\chi_{R} 
\\[2mm] 
\cdots
\end{array}
\right\rgroup 
${\small \ \ \ \ \ }~~$
~\left\lgroup
\begin{array}{c}
\left( 
\begin{array}{c}
t_R \\ 
b_R
\end{array}
\right)\,
\\[4mm]
\chi_{L} 
\\[2mm]
\cdots
\end{array}
\right\rgroup
$
\end{center}
This can be viewed as a two lattice-brane approximation
to a higher dimensional model with localized
chiral fermions \cite{wang2}.

We further introduce a scalar field, ${\Sigma}$, transforming as
$({\bf \overline{3},3,1,} \, 0)$,  with a negative mass
$M_{\Sigma }^{2}$ and an 
associated quartic potential such that $\Sigma $ develops a diagonal VEV,
\begin{equation}
\langle {\Sigma}^i_j \rangle = V \; \delta_{j}^{i}\,,
%\label{}
\end{equation}
and Topcolor group is broken down to the usual QCD, 
\begin{equation}
SU(3)_{1}\otimes
SU(3)_{2}\longrightarrow SU(3)_{QCD} \,,
%\label{}
\end{equation}
yielding massless gluons and an octet of degenerate colorons 
with mass $\cut$ given by
\begin{equation}
\cut^{2}= (h_{1}^{2}+h_{2}^{2})\,{V}^{2}\,.
%\label{}
\end{equation}
$\Sigma$ is just the Wilson link connecting the
two branes in the $4+1$ picture,
and $V$ the inverse compactification scale.  
Alternatively, from a pure $3+1$ perspective
this symmetry breaking can arise dynamically, which
is akin to dimensional deconstruction \cite{acg}.
We will describe $\Sigma$ as
a fundamental field in
the present model for the sake of simplicity.

The scalar $\Sigma$ also has the correct quantum numbers to form a 
Yukawa interaction with the singlet seesaw quarks $\chi_{L,R}$ and thus
provides the requisite mass term $\mu_{\chi \chi}$,
\begin{equation}
-y_{\chi}\,\overline{\chi _{R}}\,\Sigma\, \chi _{L}+{\rm h.c.}
\longrightarrow -\mu_{\chi\chi}\,
\overline{\chi }\chi  \,.
\end{equation}
This also happens automatically in the latticized
extra-dimension scheme where this term plays the
role of the fermion (hopping) kinetic term. 
We stress that this is an electroweak singlet mass term.
In this scheme $y_\chi$ is a perturbative 
coupling constant so that ${V} \gg \mu_{\chi \chi }$.
Finally, as both $t_{R}$ and $\chi_{L}$ carry identical Topcolor and
$U(1)_{Y}$ quantum numbers, 
we should also include the explicit weak-singlet mass term,
of the form, $\mu_{\chi t}\overline{\chi _{L}}\,t_{R}+{\rm h.c.}$\,.

At energy scales below the coloron mass,
the effective Lagrangian of this minimal model is 
$SU(3)_C\otimes SU(2)_W\otimes U(1)$ invariant and can be written as,
\begin{equation}
{\cal {L}}_{0}= {\cal {L}}_{\rm kinetic}-(\mu_{\chi\chi}\,\overline{\chi _{L}}
\,\chi_{R}
+\mu_{\chi t}\,\overline{\chi _{L}}\,t_{R}+{\rm h.c.})+{\cal {L}}_{\rm int} \,.
\label{eq:xmass}
\end{equation}
${\cal {L}}_{\rm int}$ contains the residual Topcolor interactions from the
exchange of the massive colorons, and can be written as
an operator product expansion,
\begin{equation}
{\cal {L}}_{\rm int}=  -\frac{h^{2}_{1}}{\cut^{2}}\left( \overline{\psi _{L}}\,
\gamma^{\mu }\frac{T^{a}}{2}\psi _{L}\right)
\left( \overline{\chi _{R}}\,
\gamma_{\mu }\frac{T^{a}}{2}\chi _{R}\right) +LL+RR+\cdots \,,
\label{njl0}
\end{equation}
where $LL$ $(RR)$ refers to left-handed (right-handed) current-current
interactions and $T^a$'s are the broken $SU(3)$ generators.
Since the Topcolor interactions are strongly coupled, forming 
boundstates, higher dimensional operators might have a significant 
effect on the low energy theory. However, if the full Topcolor dynamics 
induces chiral symmetry breaking through a second order (or weakly first order)
phase transition, then one can analyze the theory using the 
fundamental degrees of freedom,
namely the quarks, at scales significantly lower than the Topcolor scale. 
We will assume that this is the case, which implies that 
the effects of the higher dimensional operators are suppressed by powers of the
Topcolor scale, and it is sufficient to keep in the low energy theory 
only the effects of the operators shown in Eq.\,(\ref{njl0}).
Furthermore, the $LL$ and $(RR)$ interactions do not affect the low-energy
effective potential in the large-$N_c$ limit, so we will ignore them. 
(One should keep in mind that these interactions may have other effects,
such as contributions to the custodial symmetry violation parameter $T$,
but these effects 
are negligible if the Topcolor scale is in the multi-TeV range).

To leading order in $1/N_c$ and upon performing the
familiar Fierz rearrangement, we obtain the following 
scalar-type NJL\,\cite{NJL}  interaction,
\begin{equation}
{\cal {L}}_{\rm int}= \frac{h_{1}^{2}}{\Lambda^{2}}(\overline{\psi_{L}}\,
\chi_{R})\,(\overline{\chi_{R}}\,\psi _{L}) ~.
\label{njl1}
\end{equation}
It is convenient to pass to a partial mass eigenbasis with the following
transformations for right-handed fields,
\begin{equation}
\label{eq:partialrotation}
\chi _{R} \rightarrow \cos \theta \;\chi _{R} - \sin \theta\;
t_{R} \,,~~~~~
\quad t_{R} \rightarrow \cos \theta \;t_{R} + \sin \theta \;\chi _{R} \,,
\end{equation}
where
\begin{equation}
\label{eq:thetaRapp}
\tan \theta = \frac{\mu_{\chi t}}{ \mu_{\chi \chi}}  \,.
\end{equation}
In this basis, the NJL Lagrangian takes the form,
\bea
{\cal {L}}_{0} & = & {\cal {L}}_{\rm kinetic}-
\overline{M}\,\overline{\chi_{R}}\,\chi_{L}+{\rm h.c.}
\nonumber \\ [2mm]
& & + \frac{h_{1}^{2}}{\Lambda^{2}}
\left[ \overline{\psi _{L}}\, \left(\cos \theta \;\chi_{R}
-\sin \theta \;t_{R}\right)\right] 
\left[ \left(\cos\theta \; \overline{\chi_{R}}
-\sin\theta \; \overline{t_{R}} \right) \,\psi _{L}\right] \,, 
\label{njl2}
\eea
with 
\bea
\overline{M}=\sqrt{\mu_{\chi\chi}^{2}+ \mu_{\chi t}^{2}} ~.
\label{barM}
\eea

\subsection{\hspace*{-2mm}{\normalsize
Gap Equation Analysis}}

At this stage we have the choice of using
the renormalization group (RG), or to study the mass gap equation for
$\mtx$. Ultimately these should be equivalent.  
The RG approach requires the 
construction of the effective potential of the
composite Higgs boson, and its minimization. 
The gap equations get us there directly.  
A further rationale for studying the gap equations is that they
in principle allow one to explore the limits, such as 
$\overline{M} > \Lambda$ which are conceptually more difficult
with the renormalization group. (The dimension-6 operator
makes no sense above the scale $\cut$ in the RG, but
the cut-off theory can still be expressed in the gap equation
language.) In the following, we will start with the 
gap equation analysis, and we find it
instructive to begin by treating $\mtx$ as a mass-insertion and
examine its dependence on the parameters $\overline{M}$ and $\theta$.
An improved derivation of the seesaw gap equation without mass-insertion 
will be given in Appendix-A1 and Sec.\,2.4.

To derive gap equations, we expand the NJL vertex in Eq.\,(\ref{njl2})
and find that the four individual vertices,
$\(\ov{t_L}\x_R\)\(\ov{\x_R}t_L\)$,\,
$\(\ov{t_L}t_R\)\(\ov{t_R}t_L\)$,\,
$\(\ov{t_L}t_R\)\(\ov{\x_R}t_L\)$,\, and
$\(\ov{t_R}t_L\)\(\ov{t_L}\x_R\)$,\,
can form two types of dynamical condensates, 
$\langle \ov{t_L}\x_R\rangle$ and 
$\langle \ov{t_L} t_R\rangle$. Correspondingly, we have two
mass-gap terms, 
\bea
-\,\mtx\overline{t}_L \chi_R \,-\,m_{tt}\overline{t}_L t_R \,,
\eea
where the diagonal mass $m_{tt}$ can be conveniently put into the 
top propagator while the off-diagonal mass $\mtx$ will be included 
up to $\O(\mtx^3)$ in the present analysis.
We can then write down the two gap equations for $\mtx$ and $\mtt$, as
graphically shown in Fig.\,6. %\ref{fig:txtt}.
It is clear that these are the large-$N_c$ Schwinger-Dyson equations
[expanded up to $\O(\mtx^3)$]  for the NJL-Lagrangian (\ref{njl2}).
From Fig.\,6, we derive,
\beq
\label{eq:gapeqtxtt}
\ba{c}
\mtx = \dis -sc\f{h_1^2}{\cut^2}\sum_{j=1}^4\Delta_j\,,~~~~~
\mtt = \dis -s^2\f{h_1^2}{\cut^2}\sum_{j=1}^4\Delta_j\,,  \\[8mm]
\longrightarrow  \dis\f{\mtx}{\mtt} =\f{s}{c} =\f{\muxt}{\muxx} \,,
\ea
\ee

%\vspace*{40mm}
\begin{figure}[H]
\label{fig:txtt}
\begin{center}
%\vspace*{8cm}
%\special{psfile=gap1.ps
%          angle=0 hscale=70 vscale=70 hoffset=40 voffset=-20}
%\vspace*{-30mm}
\vspace*{-8mm}
\includegraphics[width=13cm,height=9.5cm]{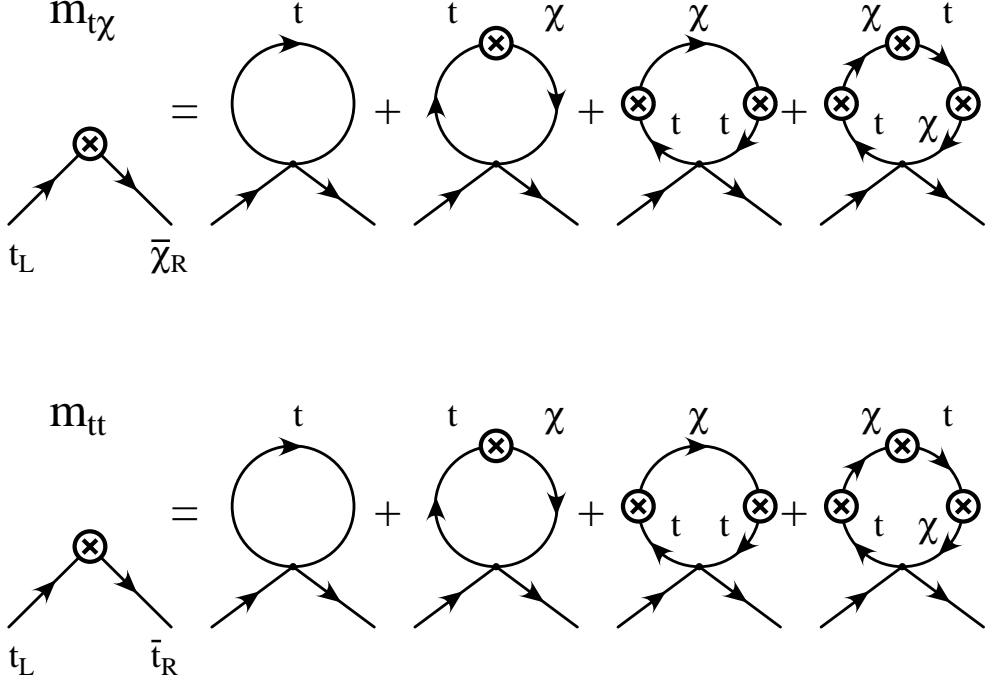} %\\[-20mm]  
\caption[]{Top seesaw gap equations for $\mtx$ and $\mtt$.}
\end{center}
\end{figure}

\noi
where $(s,\,c)=(\sin\theta,\,\cos\theta)$ and the term
$\sum_{j=1}^4\Delta_j$ represents the sum of four loop-integrals
on the right-hand side of each gap equation in Fig.\,6. It is important
to note that the {\em same} loop graphs appear in both gap equations
for $\mtx$ and $\mtt$ so that we have the relation $\mtt/\mtx = s/c$
as above. This means that the two coupled gap equations
are actually reduced to one independent gap
equation, say, for $\mtx$. 
By explicit calculation of the loop integrals, we write
this gap equation in the following form, up to $\O(\mtx^3)$,
\beq
\label{eq:gapeqtx}
\ba{l}
\mtx \,=\, \mtx\dis\f{\k}{\k_c}
\[
1 -\f{\mtx^2}{\cut^2} \((1+s^2)\ln\(\f{\cut^2}{\ov{M}^2}+1\)
  -\f{c^2\cut^2}{\cut^2+\MB^2} 
  +\f{s^4}{c^2}\ln\(\f{c^2}{s^2}\f{\cut^2}{\mtx^2}+1\) 
                      \)  \right.
\\[6mm]
\left. \dis
\hspace*{30mm}
  \,-\,c^2\f{\MB^2}{\cut^2}\ln\(\f{\cut^2}{\MB^2}+1\)
\]     +\O(\mtx^4)  \,,
\ea
\eeq
where for convenience
we have used the definitions, $\k= h_1^2/(4\pi)$ and 
$\k_c=2\pi/N_c$.  
There are several ways to see that these reproduce normal
top condensation in the decoupling limit. For instance,
taking $\overline{M}\rightarrow \infty$ for fixed $\cut$
and using the relation $\mtx = \mtt (s/c)$, we find,
\begin{eqnarray}
\label{gapts}
\mtt \,=\, \mtt\dis\f{\k}{\k_c}
\[
1-\f{\mtt^2}{\cut^2} \ln\(\f{\cut^2}{\mtt^2}+1\)
\] \,,
\end{eqnarray}
which is just the familiar top condensation gap equation, with
$\mtt$ the dynamical top quark mass.
Here we have decoupled $\chi_L$
and $\chi_R$ with $\overline{M}\rightarrow \infty$.  We can also
obtain top condensation by setting $\sin^2\theta = s^2 = 0$ and
$\overline{M}\rightarrow 0$, which decouples $\chi_L$ and  $t_R$, and causes
$\chi_R$ to play the role of $t_R$.
A main advantage of this mass-insertion gap equation (\ref{eq:gapeqtx}) 
is that it allows us to analytically solve for $\mtx$
(ignoring a small $\O(s^4)$ term),
\beq
\label{eq:gapeqtx2}
\mtx ~\simeq~ \dis\cut
\[
\f{\dis 1-\kBI-c^2\f{\MB^2}{\cut^2}\ln\(\f{\cut^2}{\MB^2}+1\)}
  {\dis (1+s^2)\ln\(\f{\cut^2}{\MB^2}+1\)-\f{c^2\cut^2}{\cut^2+\MB^2}}
\]^{\half} ,
\eeq
where we have discarded the trivial solution $\mtx=0$.

\begin{figure}[H]
%\vspace*{7.5cm}
%\special{psfile=fig_keffcrit.eps
%          angle=0 hscale=84 vscale=84 hoffset=90 voffset=-20}
%\vspace{5mm}
\begin{center}
\vspace*{-2cm}
\includegraphics[width=13cm,height=13cm]{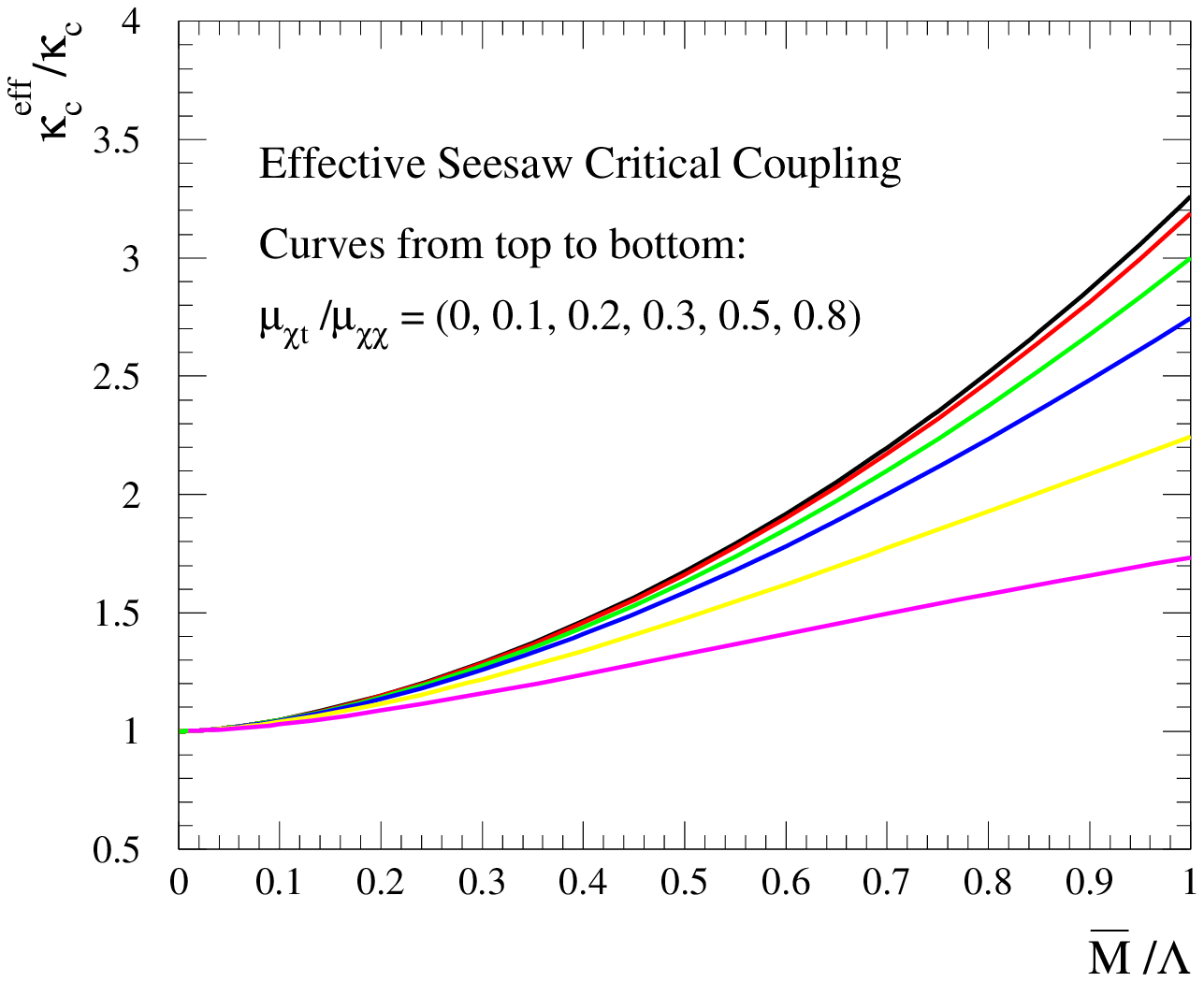}
\end{center}
\vspace*{-1cm}
\caption[]{Effective seesaw critical coupling $\k_c^{\rm eff}$
(scaled by constant $\k_c \equiv 2\pi/3$) 
as a function of $\overline{M}/\cut $, for
${\tan\theta}=\muxt/\muxx =(0,\,0.1,\,0.2,\,0.3,\,0.5,\,0.8)$.
}
\label{fig:keff}
\end{figure}

This clearly shows that for the fixed $\k/\k_c > 1 $,
the condensate turns off like a second order phase transition 
as we raise the scale $\overline{M}$. This is essentially to
compensate the decoupling of the heavy fermion in the loop
of mass $\overline{M}$.
The gap equation (\ref{eq:gapeqtx}) or (\ref{gapts})
also shows that we require super-critical 
coupling as the mass $\overline{M}$ becomes large.
We can further derive the  effective seesaw critical coupling
$\k_c^{\rm eff}$ from the gap equation (\ref{eq:gapeqtx}) or 
(\ref{eq:gapeqtx2}) by setting $\mtx =0$, i.e., we have,
\beq
\label{eq:kceff}
\dis \f{~\k_c^{\rm eff}}{\,\k_c} ~=~ 
\f{1}{\dis 1-c^2\f{\MB^2}{\cut^2}\ln\(\f{\cut^2}{\MB^2}+1\)} \,\,,
\eeq
which is displayed in Fig.\,\ref{fig:keff} as a function 
of $\MB/\cut$. For $\k > \k_c^{\rm eff} $, we have $\mtx > 0$.
We see that $\k_c^{\rm eff} =\k_c$ for $\MB=0$, and
as $\MB$ increases the effective seesaw critical coupling
$\k_c^{\rm eff}$ moves above $\k_c(=2\pi/N_c)$
implying that stronger Topcolor force is required
compared to the non-seesaw case.
Finally, we note that using the complete seesaw diagonalization
(\ref{eq:seesawd1})-(\ref{eq:seesawd2}) 
and the NJL-vertex (\ref{njl1}),
we can derive the exact large-$N_c$ seesaw gap
equation without using a mass-insertion approximation (cf.
Appendix-A1). This will allow us to reliably analyze the
full seesaw parameter space. 

%\vspace*{12mm}

The electroweak structure of the low energy theory is best read off 
from the effective Lagrangian, which may be derived from the 
traditional gauge-invariant renormalization group analysis as below.
We proceed by rewriting the NJL interaction (\ref{njl1})
with the introduction of an auxiliary
color-singlet field, $\Phi_0$, which becomes 
the {\it unrenormalized} composite Higgs doublet,
\be
\label{eq:RG1}
{\cal {L}}_{0} =  {\cal {L}}_{\rm kinetic} -
\left[\,\overline{M}\,\overline{\chi_R} \,\chi_{L}
+ h_{\rm 1}\,\overline{\psi_L} \,(\cos \theta \;\chi_{R} -\sin\theta
\;t_{R} )\,\Phi_0 + {\rm h.c.} \right]
-\Lambda^{2}\Phi_0^{\dagger}\Phi_0 ~.
\end{equation}
To derive the effective Lagrangian at a low energy scale $\mu$, we
integrate out the modes of momenta $M\geq |k| \geq \mu $. For 
$\mu < \MB < \Lambda $, the heavy field $\chi $ decouples, so
that we have,
\begin{equation}
\label{eq:RG1x}
{\cal {L}}_{\mu < \overline{M}} = {\cal {L}}_{\rm kinetic} -
h_1 \sin\theta \left( \overline{\psi _{L}}
t_{R} \,\Phi_0 +{\rm h.c.} \right) + 
Z_{\Phi}\left| D\Phi_0\right|^{2} -
\widetilde{M}_{\Phi}^{2}(\mu) \Phi_0^{\dagger }\Phi_0 -
\widetilde{\lambda} \left(\Phi_0^{\dagger }\Phi_0\right)^{2} ,
\end{equation}
where the effective scalar wave-function renormalization,  mass
term and quartic coupling are given by,
\bea
\label{eq:ZMCRG1}
Z_{\Phi}(\mu) & = & \dis\half\kB \left[
\ln \left(\frac{\Lambda^{2}}{\overline{M}^{2}}\right) +
 \sin^{2}\!\theta\,
\ln \left( \frac{\overline{M}^{2}}{\mu ^{2}}\right) +
\O(1) \right] \,,
\nonumber 
\\ [4mm]
\widetilde{M}_{\Phi}^{2}(\mu) & = & \dis
\Lambda^{2}-\dis\kB
\[ \cut^{2}
-\cos^2\!\theta \,\MB^2\ln\(\f{\cut^2}{\MB^2}\) +\O(\MB^2,\,\mu^2)
\] \,,
\\ [4mm]
\widetilde{\lambda}(\mu) & = & \dis 2\pi\f{\k^2}{\k_c} \left[
\ln \left( \frac{\cut^{2}}{\overline{M}^{2}}\right) + 
 \sin^{4}\!\theta \,
\ln \left( \frac{\overline{M}^{2}}{\mu^{2}}\right) +\O(1)
\right] \,,
\nonumber 
\eea
where $(\k,\,\k_c) = (h_1^2/4\pi,\,2\pi/N_c)$\,.\,
These relations hold for $\mu < \overline{M} $ in the large-$N_{c}$
approximation, and illustrate the decoupling of the $\chi $ field 
at the scale $\mu < \overline{M}$.
In the limit $\sin \!\theta \ll 1$, 
the induced couplings are those of the usual NJL model; 
but the Higgs doublet is predominantly a boundstate 
of $\overline{\chi_R}\psi_L$, 
and the corresponding fermion loop, with 
loop-momentum ranging over $\MB < |k| < \cut$,
controls most of the renormalization group evolution 
of the effective Lagrangian.

In order for the composite Higgs doublet to develop a
VEV, the Topcolor $SU(3)_{1}$  gauge force  must be super-critical,
as indicated by the preceeding gap equation analysis.
Once $\k(=h_1^2/4\pi)$ is super-critical, we are
free to tune the renormalized Higgs boson mass, $M_{\Phi}^{2}(\mu) =
\widetilde{M}_{\Phi}^{2}(\mu) /Z_{\Phi}$, to any desired value. 
This implies that we are free
to adjust the renormalized VEV of the Higgs doublet to the electroweak value,
$\langle \Phi \rangle = v/\sqrt{2} \simeq 174$ GeV. The renormalized
effective Lagrangian at $\mu <\MB$ takes the form,
\begin{equation}
\label{eq:LRG1mu}
{\cal {L}}_{\mu < \overline{M} }= {\cal {L}}_{\rm kinetic}
- g_t \sin\!\theta \left(\overline{\psi_{L}}
t_{R}\,\Phi+{\rm h.c.} \right) + \left| D\Phi\right|^{2}-
M_{\Phi}^{2}(\mu) \, \Phi^{\dagger}\Phi
- \lambda (\Phi^{\dagger}\Phi)^{2}
\end{equation}
where,
\begin{equation}
\Phi = \Phi_0\sqrt{Z_\Phi}\,,~~~~
g_t = \frac{h_{\rm 1}}{\sqrt{Z_{\Phi}}} \,,~~~~
M_{\Phi}^{2}(\mu) 
= \frac{\widetilde{M}_{\Phi}^{2}(\mu)}{Z_{\Phi}} \,,~~~~
\lambda = \frac{\widetilde{\lambda }}{Z_{\Phi}^{2}}\,.
\end{equation}
\vspace*{2mm}

When the Topcolor interaction is super-critical,  $\Phi$ becomes
tachyonic at low energy scales,
$\widetilde{M}^2_{\Phi}(\mu\rightarrow 0) < 0$
and a dynamical condensate will be induced.  
This condensate breaks the electroweak symmetry 
$\SU{2}{L} \otimes {U(1)}_Y \ra {U(1)}_{EM}$ and induces mixing 
between the top and $\chi$ fields.
In the minimal top seesaw model the physical particle spectrum 
can be readily seen by writing the Higgs doublet in the unitary
gauge,
$\,
%\label{eq:UG}
{\Phi} = \frac{1}{\sqrt{2}}\left(
%\begin{array}{c}
v + h, %\\[3mm]
0
%\end{array}
\right)^T ,\,
$
where $h$ is the neutral Higgs boson of the theory. 
The resulting top quark mass can be read off from the renormalized Lagrangian,
\begin{equation}
\label{eq:PS-RG1}
m_{t} \,=\, \dis\frac{g_t \,v}{\sqrt{2}}\,
             \sin \theta \,,
\end{equation}
which corresponds to the  Pagels-Stokar formula in the form
of Eq.\,(\ref{ps2}).

Finally, by minimizing the effective Higgs potential 
in Eq.\,(\ref{eq:LRG1mu}) and 
using the results in Eq.\,(\ref{eq:ZMCRG1}), we can
derive the approximate formula for the physical Higgs mass by keeping
the leading logarithmic terms,
\beq
\label{eq:mHapp}
M_h \,\approx\, 2\mtx \,,
\eeq
which shows that the physical Higgs mass is about two times of the
dynamical mass gap, as expected from the usual
large-$N_c$ bubble approximation\,\cite{BHL,NJL}. 
In the subsection\,2.6, we will derive a
more precise $M_h$ using two improved analyses.

\subsection{\hspace*{-2mm}{\normalsize
Tadpole Condition and Improved Analysis in the Broken Phase}}
\label{sec:tadpole}

\begin{figure}[H]
%\vspace{7cm}
%\special{psfile=fig_tadpole.eps
%          angle=0 hscale=80 vscale=80 hoffset=90 voffset=-20}
\begin{center}
\includegraphics[width=12cm,%height=5cm
]{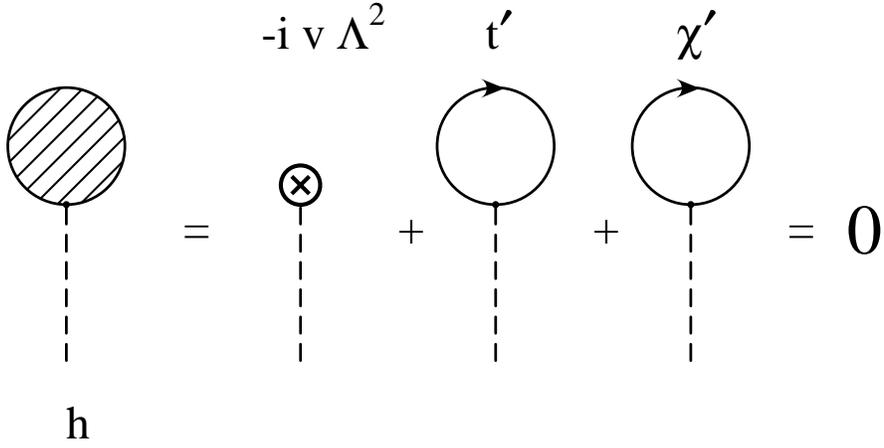}
\end{center}
\caption[]{The large-$N_c$ tadpole condition 
           for minimizing the Higgs potential.
}
%\vspace*{2mm}
\label{fig:tadpole}
\end{figure}

Before proceeding to perform the numerical analysis for gap equations, 
we consider an alternative (yet equivalent) derivation of the gap equation
based on the Higgs tadpole condition in the broken phase 
of the effective theory. (For a simpler example 
of a broken phase analysis in NJL, see \cite{salopek}).
We also present the improved RG analysis in the broken phase of the
low energy theory, which allows us to precisely treat the seesaw
mass diagonalization and the mixing effects in Higgs Lagrangian.
[This is unlike the usual gauge-invariant RG analysis
around Eq.\,(\ref{eq:RG1}) where the Higgs vacuum is unshifted and thus
the exact seesaw mass diagonalization is not allowed.]
As a consequence, the Higgs mass and its
Yukawa coupling can be more precisely analyzed 
in the present broken phase formalism.  
We begin by choosing the unitary gauge of the Higgs doublet and
shifting the bare field $\Phi_0$ to the broken phase vacuum, 
\bea
\label{eq:htshift}
{\Phi}_0 &=& \frac{1}{\sqrt{2}}\left(\!
\begin{array}{c}
{v}_0 + {h}_0 \\[3mm]
0
\end{array}
\!\right) ,
\eea
which results in the fermionic seesaw mass matrix 
given in Eq.\,(\ref{eq:ssmatrix}). Thus, the effective Lagrangian
at the scale $\mu=\cut$ can be written as,

\beq
\label{eq:Leffcut}
\ba{ll}
{\cal L}_{\cut} & 
\!\!\! 
=
- \left( \overline{t_L} ~~ \overline{\chi_L} \right)
 \left\lgroup 
\begin{array}{cc}
0 & m_{t \chi} \\[0.2cm]
\mu_{\chi t} & \mu_{\chi \chi} \\
\end{array}
\right\rgroup
\left(\!\!
\begin{array}{c}
t_R \\
\chi_R
\end{array}
\!\!\right)
-\dis\f{h_1}{\sqrt{2}}\ov{t_L}\x_Rh_0+ {\rm h.c.} 
-\half\cut^2 h_0^2 - \cut^2 v_0h_0  \\[5mm]
& 
\!\!\!
=
-m_t\ov{t^\prime}t^\prime - M_\x\ov{\x^\prime}\x^\prime -\half\cut^2 h_0^2 - \cut^2 v_0h_0 -
\dis\f{h_1}{\sqrt{2}} 
\[ c_L \, \overline{t_L'} + s_L \, \overline{\chi'}_L \] 
\left[ s_R \, t_R' + c_R \, \chi_R' \right] h_0
 + {\rm h.c.} \,,
\ea
\eeq
where we have performed the exact seesaw diagonalization according to
Eqs.\,(\ref{eq:seesawd1})-(\ref{eq:seesawd2}).
Now, we evolve the Lagrangian down to the scale $\mu \,(< \MX \leq \cut)$
by integrating out the momenta $k\in (\mu,\,\cut)$. The heavy quark
$\x$ decouples and we arrive at the renormalized broken phase 
Lagrangian,
\beq
\label{eq:Leffmu}
\ba{l}
{\cal  L}_{\mu < \MX}  
= \dis
-m_t\ov{t^\prime}t^\prime -
\dis\f{~g_t}{\sqrt{2}} c_L s_R   \ov{t^\prime}t^\prime h
+\half\(\partial_\mu h\)^2 
-\Delta\widetilde{T} h  \dis -\half M_h^2 h^2
- V_{\rm int}\(Z_h^{-1}h\)
\ea
\eeq
where 
$\,g_t = h_1/\sqrt{Z_h}\,$ and \,$V_{\rm int}\(Z_h^{-1}h\)$\, contains 
the effective Higgs self-interactions. The Higgs tadpole 
term $\Delta\widetilde{T}$ and mass term $M_h^2$ are defined by,
\beq
\label{eq:TMh}
\Delta\widetilde{T}= 
\( Z_h^{-\half}v\cut^2+\delta\widetilde{T} \) Z_h^{-\half}\,,~~~~
M_h^2 = (\cut^2+\delta\widetilde{M}^2_h)/Z_h \,,
\eeq
with $\delta\widetilde{T}$ and $\delta\widetilde{M}^2_h$ computed from the
one-loop Higgs tadpole and self-energy corrections, respectively. 
The Higgs tadpole condition,  \,$\Delta\widetilde{T} = 0$,\, results in,
\beq
\label{eq:tadpole}
v_0\cut^2+\delta\widetilde{T} = 0 \,,
\eeq
where $\delta\widetilde{T}$ comes from one-loop tadpole diagrams
(cf. Fig.\,\ref{fig:tadpole}). Note that the tadpole loops in
$\delta\widetilde{T}$ will be integrated from zero momentum to 
the cutoff $\cut$ (independent of the renormalization scale $\mu$) 
as they are really vacuum graphs with vanishing external momentum.
The equation (\ref{eq:tadpole}) is just the
minimization condition of the Higgs potential in its broken
phase, and is equivalent to the gap equation derived from the
NJL formalism in Sec.\,2.3 and Appendix-A1, as will be clear shortly.
Fig.\,\ref{fig:tadpole} shows that the condition in Eq.\,(\ref{eq:tadpole})
actually represents the exact large-$N_c$ gap equation without 
mass-insertion. [The mass-insertion tadpole condition, fully equivalent
to gap equation (\ref{eq:gapeqtx}) in Sec.\,2.3, 
will be given in Appendix-A2.]
Now, using the relation $\mtx = h_1v_0/\sq2$, we can explicitly derive, 
from Eq.\,(\ref{eq:tadpole}), a single gap equation for $\mtx$,
\beq
\label{eq:tadgap1}
\vspace*{3.5mm}
%\hspace*{-1cm}
m_{t \chi} ~=~
 \frac{\kappa}{\kappa_c} \left\{ c_L s_R \left( m_t
- \frac{m_t^3}{\cut^2} 
\ln \left[ \frac{\cut^2 + m_t^2}{ m_t^2} \right] \right) 
%\right. \\ & & \: \left. 
+ s_L c_R \left( m_\chi
- \frac{m^3_\chi}{\cut^2} 
\ln \left[ \frac{\cut^2 + m_\chi^2}{m_\chi^2} \right] \right) 
\right\} \, ,
%\nonumber
%\vspace*{3.5mm}
\eeq
where $(\k,\,\k_c) = (h_1^2/4\pi,\,2\pi/N_c)$.
Eq.\,(\ref{eq:tadgap1}) is the same as the exact large-$N_c$ NJL gap equation 
derived in Appendix-A1. 
It also reduces back to the approximate
mass-insertion gap equation (\ref{eq:gapeqtx}) 
(cf. Sec.\,2.3 and Appendix-A2)
after expanding the seesaw rotation angles and mass eigenvalues up
to $\O(\mtx^3)$, as we have verified. 
This provides a consistency check of our analysis.
Since the right-hand side of Eq.\,(\ref{eq:tadgap1})
contains the mass gap $\mtx$ in an implicit way, it is less transparent
than the approximate mass-insertion gap equation (\ref{eq:gapeqtx}) 
presented earlier. But, the precise treatment of all seesaw mixing
effects in Eq.\,(\ref{eq:tadgap1})
has an advantage of allowing us to reliably explore the full 
seesaw parameter space, and is particularly useful in
our later quantitative numerical analysis.

We proceed by computing the
wave-function  renormalization constant of the Higgs field, $Z_{h}$,
and obtain,
\bea
\label{eq:Zh}
Z_{h} &=& 
\dis\half \kB
\left\{
c^2_L s^2_R \ln \left[ \frac{\cut^2 + m_t^2}{m_t^2} \right] +
(c^2_R + s^2_L s^2_R) \ln \left[ \frac{\cut^2 + m_\chi^2}{m_\chi^2}
\right]
\right\} ,
\eea
where we have dropped  the small
${\cal O}(1)$ constant terms (which are not logarithmically enhanced)
together with the tiny $\O(m_t^2/\MX^2)$ terms.
The renormalized $t^\prime$-$\ov{t}^\prime$-$h$ vertex has Yukawa coupling 
$c_L s_R g_t/\sqrt{2}$ with  
$~
g_t = h_1/\sqrt{Z_h}\,.
$
The dynamical mass $\mtx$ in the seesaw matrix takes the form,
$\mtx = h_1v_0/\sq2 = g_tv/\sq2$,\, 
which, with Eq.\,(\ref{eq:Zh}), results in a more precise form of 
the seesaw Pagels-Stokar formula,
\bea
\label{eq:PS-RG2}
v^2 &=& \frac{m^2_{t \chi}}{4 \, \pi \, \kappa_c} \left\{
c^2_L s^2_R \ln \left[ \frac{\cut^2 + m^2_t}{m^2_t} \right]
+ \left( c^2_R + s^2_L s^2_R \right) \ln \left[
\frac{\cut^2 + M^2_\chi}{M^2_\chi} \right] \right\} \, .
\eea
This equation is an improvement over the
previous formula (\ref{ps2}) [or (\ref{eq:PS-RG1})]
in that the exact seesaw mixing effects
associated with the leading logarithmic terms are included. 
To check the consistency, we
note that Eq.\,(\ref{eq:PS-RG2}) reduces back to Eq.\,(\ref{ps2})  
under the limit\,
$(s_L^2,\,s_R^2)\ll 1$ and 
$\mtx \approx m_t/(\muxt/\muxx)
      \approx m_t/\sin\theta_R$
(where $\sin\theta_R\approx \muxt/\muxx$).
Finally, we note that the above Pagels-Stokar formula is derived
under the large-$N_c$ fermion bubble approximation, 
which, for the low scale cutoff $\cut \lesssim 10^{4-5}$\,GeV, 
is found to work well in comparison with the 
full RG evolution (including non-large-$N_c$ terms)\,\cite{Hill90}.

\subsection{\hspace*{-2mm}{\normalsize
Solutions to the Top Seesaw Gap Equation }} 
\label{sec:tsolution}

\begin{figure}[H]
\begin{center}
\vspace*{-18mm}
\hspace*{-1cm}
%\begin{tabular}{cc}
%  \includegraphics[width=13cm,height=13cm]{fig_k2.eps} &
%  \hspace*{-1.5cm}
\includegraphics[width=14.5cm,height=13cm]{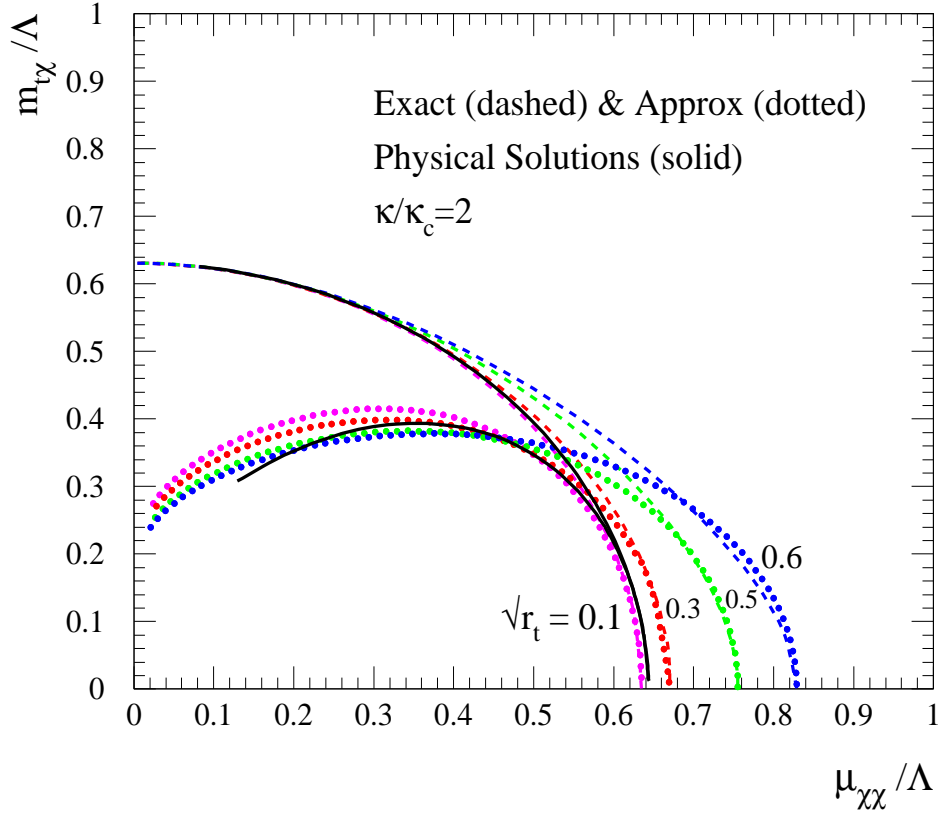}
%\end{tabular}
\end{center}
\vspace*{-9mm}
\caption{
Solutions for top seesaw gap $m_{t\chi}$ with $\k/\k_c=2$ 
and $\sqrt{r_t}=\muxt/\muxx=(0.1,\,0.3,\,0.5,\,0.6)$, respectively.
The physical seesaw solutions (satisfying $m_t= 174$\,GeV and
$v=\sqrt{2}v_{\rm weak}=246$\,GeV) are plotted as solid curves,  extracted from 
Fig.(\,\ref{fig:tssphysrefapp}).  The upper set of curves (dashed curves)
are derived from the
exact large-$N_c$ gap equation  (\ref{eq:tadgap1}) 
and the lower ones (dotted curves)
from the mass-insertion gap equation (\ref{eq:gapeqtx2}).
}
\label{fig:mtxph}
\end{figure}

In this subsection we present a systematic numerical 
analysis of the top seesaw gap equations.
From the approximate or exact gap equation
[cf. Eq.\,(\ref{eq:gapeqtx2}) or Eq.\,(\ref{eq:tadgap1})], 
we can see that the seesaw mass gap $\mtx/\cut$
(scaled by $\cut$) can be solved as
a function of the $\x$-mass parameter $\muxx/\cut$
(scaled by $\cut$), for each given $\k/\k_c$ (the strength of
Topcolor gauge force) and the seesaw parameter $r_t=(\muxt/\muxx)^2$.
Exploring such a relation between $\mtx/\cut$
and $\muxx/\cut$  will allow us to explicitly examine
the behavior of the second order phase transition of the mass
gap $\mtx$ as the $\x$ quark mass scale $\muxx$
becomes large. This is shown in Fig.\,\ref{fig:mtxph} 
for a typical input of $\k/\k_c=2$ and a wide range of
$r_t$ values. We have plotted seesaw solutions using
both the approximate mass-insertion gap equation (\ref{eq:gapeqtx2})
and the exact gap equation (\ref{eq:tadgap1}), 
depicted  as dotted and dashed curves
in Fig.\,\ref{fig:mtxph}. We see that the two type of solutions indeed
converge in the small $\mtx/\cut$ region as expected, and deviate more
from each other for larger $\mtx/\cut$ values.
As the ratio $\muxx/\cut$ moves beyond $\sim\!0.63$, 
the mass gap $\mtx$ smoothly turns off, indicating
a second order phase transition has occurred.
In another limit, $\muxx/\cut\to 0$, the difference 
between the two sets of curves becomes
the largest as the approximate curves of $\mtx/\cut$
all fall into zero while 
the exact ones  smoothly approach to about $0.63$,  
a particular solution of the reduced gap equation,
$\,1 - \kappa_c / \kappa = ({m_{t \chi}}/{\cut})^2 
\ln \( 1 + {\cut^2}/{m^2_{t \chi}} \),\,
$
(with $\k/\k_c=2$),  derived from Eq.\,(\ref{eq:tadgap1}) in the
limit $\muxx/\cut \ra 0$.\\

\vspace*{-4mm}

We now turn to the physical solutions in which we
superimpose the requirements of the top mass, $m_t = 174.3$\,GeV, and 
the full EWSB VEV, $v = 246$\,GeV. 
Our strategy is to fix the coloron mass $\cut$
(characterizing the Topcolor breaking scale), and the
Topcolor gauge coupling at that scale ($h_1$, or
equivalently, $\kappa / \kappa_c$). Then, we are left with
three seesaw parameters $(\mtx,\, \muxt,\, \muxx)$
[or, equivalently,  $(\mtx,\, r_t,\, \muxx)$] to be determined.
Indeed, we have three coupled equations to make this
determination completely feasible: the gap equation (\ref{eq:tadgap1}) 
[or (\ref{eq:gapeqtx2})], the top mass eigenvalue equation 
(\ref{eq:mt}), and the Pagels-Stokar formula 
(\ref{eq:PS-RG2}) [or, (\ref{eq:PS-RG1})].
From this set of solutions, all other physical quantities, such as 
the seesaw mixing angles, the mass of $\x$ quark, and the Higgs 
mass and Yukawa couplings, can be predicted as functions of 
$\cut$ for each given $\k/\k_c$.

\begin{figure}[H]
\begin{center}
\vspace*{-10mm}
\includegraphics[width=16cm,height=20cm]{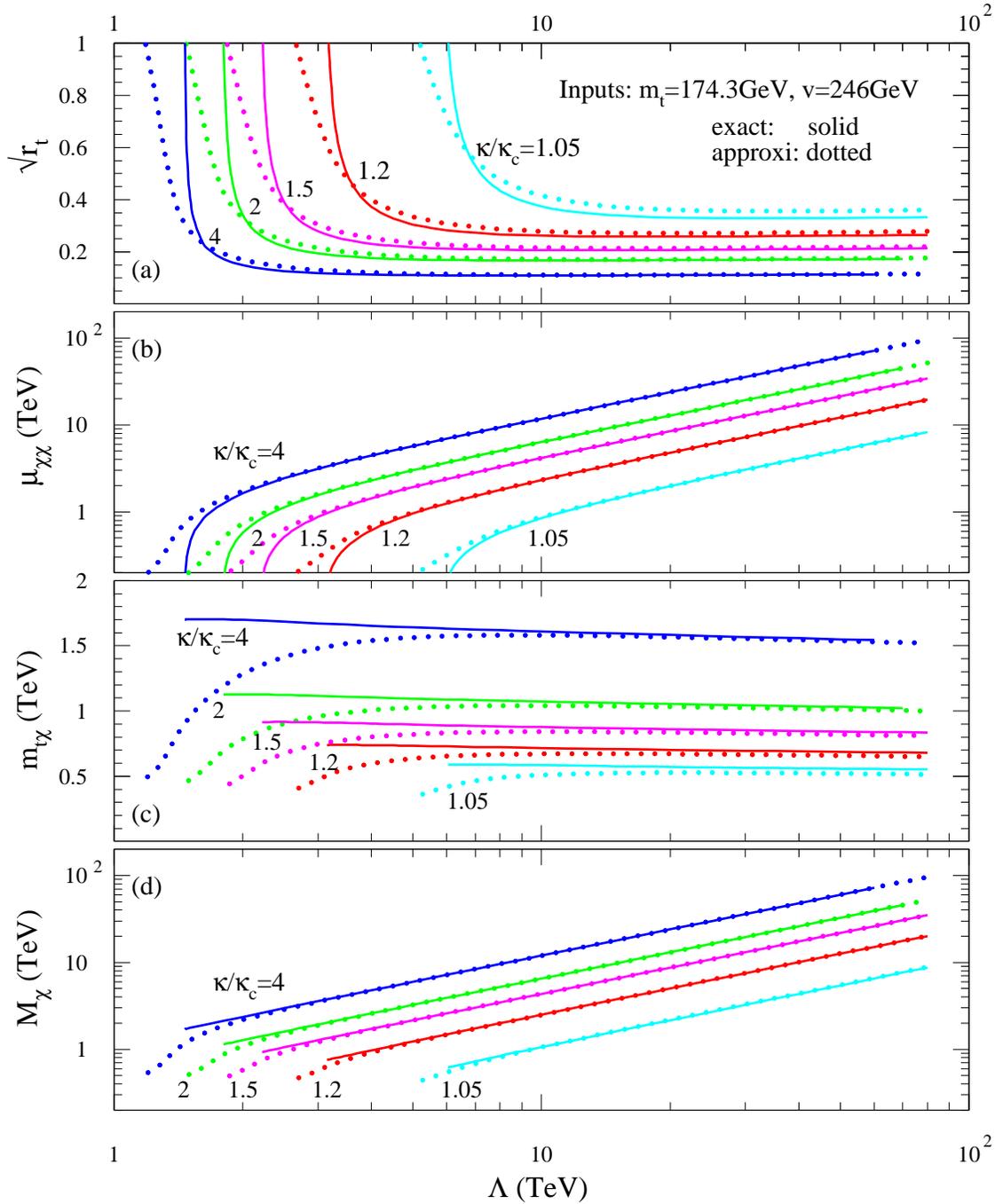}
\end{center}
\vspace*{-13mm}
\caption{Solutions of the top seesaw gap equations are shown in
plots (a)-(c) for $\k/\k_c=(1.2,\,1.5,\,2,\,4)$, with 
$m_t=174.3$\,GeV and $v=\sqrt{2}v_{\rm weak}=246$\,GeV superimposed.
The solid curves are derived from the exact gap equation
(\ref{eq:tadgap1}) 
while the dotted curves from the mass-insertion gap equation
(\ref{eq:gapeqtx2}).
The predicted physical mass-eigenvalue of $\x$ quark is also shown 
in the plot (d).
}
\label{fig:tssphysrefapp}
\end{figure}

In Fig.\,\ref{fig:tssphysrefapp}(a)-(c), we 
present our complete  physical seesaw solutions 
as functions of $\cut$ and for various inputs of  $\kappa / \kappa_c$. 
For completeness, we also show the 
prediction of the $\x$ mass ($\MX$) in Fig.\,\ref{fig:tssphysrefapp}(d).
Fig.\,\ref{fig:tssphysrefapp}(c) shows that the mass gap $\mtx$ 
ranges from $\sim\!700$\,GeV up to $\sim\!1.7$\,TeV for 
$1.05 \leq \k/\k_c \leq 4$, and  is quite
flat in the entire region of $\cut$.
There is also a lower limit
on the allowed region of $\cut$ for each fixed $\k / \k_c$.
For instance, $\cut$  has to be greater than 1.8\,TeV for
$\k / \k_c =2$. 
Furthermore, it is instructive to map our solutions into the plane
of $\mtx/\cut$ vs $\muxx/\cut$ in Fig.\,\ref{fig:mtxph}.
Since the seesaw parameters $(\mtx,\, r_t,\, \muxx)$ are determined
as in  Fig.\,\ref{fig:tssphysrefapp}(a-c)
for each given $\cut$ and $\k/\k_c$,  we see that the physical solution 
for $\k/\k_c=2$ (solid curves) indeed take a unique trajectory
in the $\mtx/\cut - \muxx/\cut$ plane of Fig.\,\ref{fig:mtxph}.
For $\cut$ varying from 1.8\,TeV to 80\,TeV,
the (exact and approximate)
physical solutions move from left to right along the two solid curves
and fall into good agreement for \,$\mu_{\chi\chi}/\cut \gtrsim 0.56$\,.

\begin{figure}[H]
\begin{center}
\vspace*{10mm}
\includegraphics[width=16cm,height=17cm]{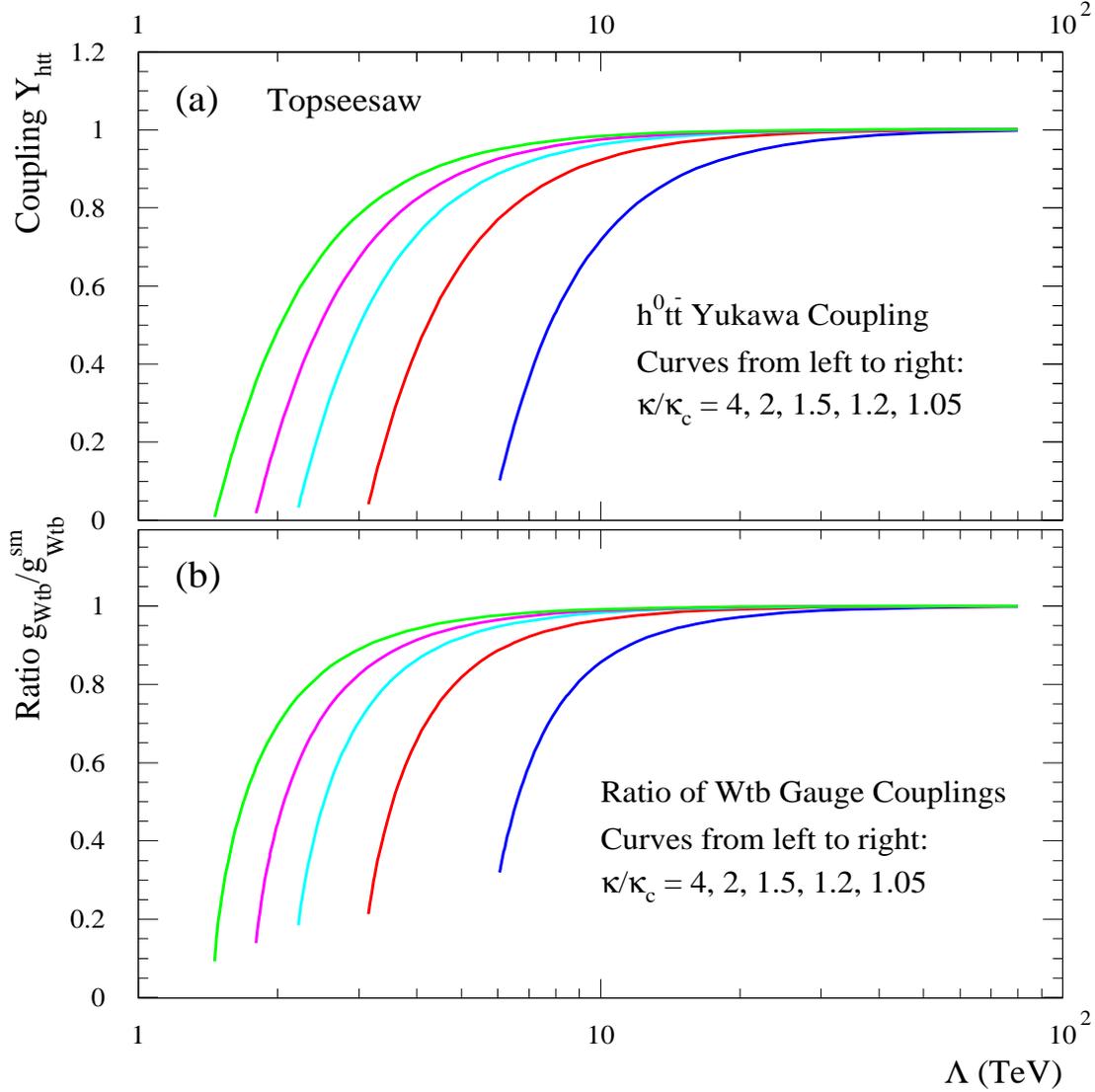}
\end{center}
\vspace*{-13mm}
\caption{The predicted effective Higgs Yukawa coupling $Y_{htt}$ in
(a), and $W$-$t$-$b$ gauge coupling $g_{Wtb}$ shown as the ratio over its
SM value in (b). 
}
\label{fig:htt-Wtb}
\end{figure}

With these solutions we are ready to predict physical observables.
We first consider the effective  $t^\prime$-$\ov{t^\prime}$-$h$ Yukawa coupling,
which can be extracted from Eq.\,(\ref{eq:Leffmu}),
\beq
\label{eq:htt}
Y_{htt} ~=~
g_t\,c_L s_R  ~=~ \dis\f{h_1}{\sqrt{Z_h}} c_Ls_R  \,.
\eeq
In the limit of $r_t=(\muxt/\muxx)^2 \ll 1$ and
$x_t=(\mtx/\muxx)^2 \ll 1$,  (\ref{eq:htt})
can be approximated as, 
$Y_{htt} \approx  \sqrt{r_t}(h_1/\sqrt{Z_h})$. With the leading
order seesaw mass relation, 
$m_t\approx \sqrt{r_t}\mtx
= \sqrt{r_t} (h_1/\sqrt{Z_h})(v/\sq2 )
$,\,
we arrive at an approximate equation, 
$Y_{htt}\approx  \sq2 m_t/v \approx 1$,\,
as in the SM. 
Now, we can understand the gross behavior of  $Y_{htt}$ in 
Fig.\,\ref{fig:htt-Wtb}(a). Namely, for the low $\cut$ region, the
seesaw solutions of $r_t$ and $\mtx/\muxx$ are quite  
sizable [cf. Fig.\,\ref{fig:tssphysrefapp}(a-c)]
so that the above limit $(r_t,\,x_t) \ll 1$ is
not good and the deviation $Y_{htt}-1$ is large; also
smaller $\k/\k_c$ values have larger $r_t$, suggesting larger
deviation of $Y_{htt}$ from unity.  But, when
$\cut$ increases, the ratios $(r_t,\,x_t)$ drop off quickly
and thus $Y_{htt}$ approaches $Y_{htt}=1$.

Other important couplings include the effective $W$-$t^\prime$-$b$
and $Z$-$t^\prime$-$t^\prime$ gauge couplings, which are now modified by the
seesaw rotations of $t$ and $\chi$ [cf. Eqs.\,(\ref{eq:rotL})-(\ref{eq:rotR})]. 
The  $W$-$t^\prime$-$b$
coupling $g^{~~}_{Wtb}$, for instance, involves only the
left-handed fields $(t^\prime_L,\,b_L)$  and we derive,
\beq
\dis
\f{g^{~~}_{Wtb}}{g^{\rm SM}_{Wtb}}
~=~  c_L  \,=\,   %\sqrt{1 -s_L^2}
1-\f{x_t}{2(1+r_t)^2}\[1+\f{8r_t-3}{4(1+r_t)^2}x_t\] + \O(x_t^3) \,,
\eeq
where $\,(\sqrt{r_t},\,\sqrt{x_t}) 
  \equiv (\muxt,\, \mtx )/\muxx \,<\, 1$.\,
We see that the effective coupling $g^{~~}_{Wtb}$ is reduced
from its SM value, and the deviation becomes small in
the limit $(r_t,\,x_t) \ll 1$ (valid in the
large $\cut$ region, cf., Fig.\,\ref{fig:tssphysrefapp}).
This picture is quantitatively shown in Fig.\,\ref{fig:htt-Wtb}(b).
Such deviations are important for precision experimental tests at
various colliders before the seesaw quark $\x$ can be directly produced.

Finally, we remark that, using the freedom to adjust $r_t$ 
[or equivalently, $\sin\theta$ in Eq.\,(\ref{eq:partialrotation})],
we can apparently accommodate any fermion mass 
lighter than $\sim\!700$\,GeV.
However, this requires some fine-tuning.   This freedom
may be useful in constructing more complete models involving all three
generations. The top quark is unique, however, in that its large mass is 
very difficult to accommodate in any other way,
and there is less apparent fine-tuning. 
We therefore believe it is generic, in any model of this kind, 
that the top quark receives the bulk of its mass through this seesaw mechanism.
%With the seesaw mechanism embedded in a more general theory, 
%there are more composite scalars, and one of the neutral Higgs bosons may be 
%as light as ${\cal O}(100)$\,GeV \cite{seesaw,bulk,dob}.

\newpage
%%%%%%%%%%%%%%%%%%%%%%%%%%%%%%%%%%%%%%%%%%%%%%%%%%%%%%%%%%%%%%
\subsection{\hspace*{-2mm}{\normalsize
The Composite Higgs Boson Mass  }}
\label{sec:mH}

\begin{figure}[H]
\begin{center}
\vspace*{-10mm}
\includegraphics[width=16cm,height=17cm]{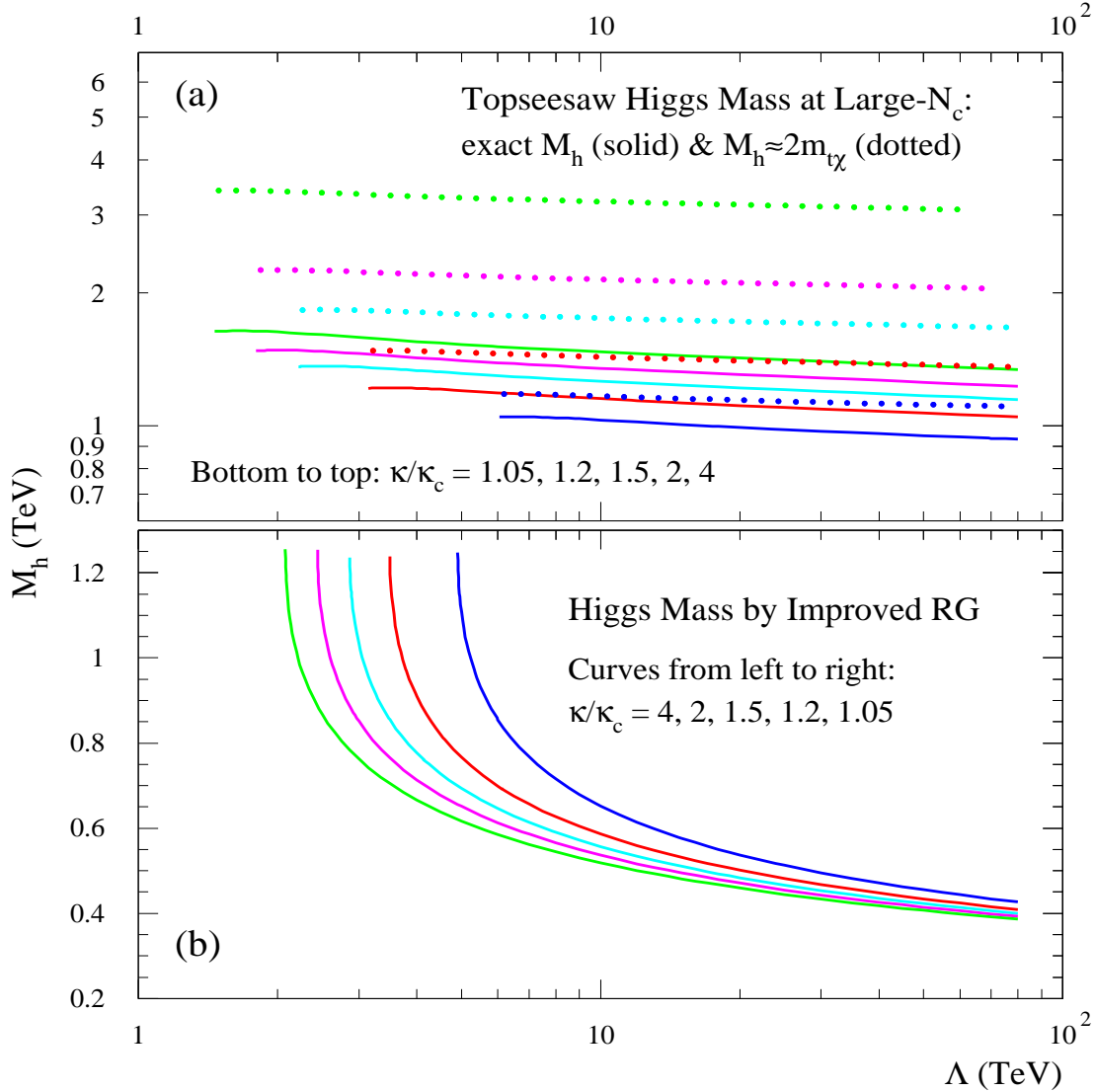}
\end{center}
\vspace*{-13mm}
\caption{The predicted mass spectrum of the top seesaw Higgs
boson: (a) by the large-$N_c$ fermion-bubble calculation;
and (b) by an improved RG analysis 
including the Higgs self-coupling evolution.
}
\label{fig:Mhnew}
\end{figure}

With the seesaw gap equation solved in the previous subsection,
we can proceed to analyze the mass spectrum of the composite Higgs
boson.
From Eqs.\,(\ref{eq:Leffcut}), (\ref{eq:TMh}) and (\ref{eq:Zh}),
and taking the usual large-$N_c$ fermion-bubble approximation,
we can straightforwardly compute the physical Higgs boson mass $M_h$.  
A lengthy calculation gives,
\beq
\ba{ll}
\label{eq:Mh}
M_h^2 = & \!\!\!\dis \f{1}{Z_h}
\left\{
\(1-\kB\)\cut^2  \right. \\[6mm] 
& \dis
+\kB
\[
\(
3s_L^2c_R^2 + \(c_L^2c_R^2+s_L^2s_R^2\)\f{\MX^2}{\MX^2-m_t^2}+
2c_Lc_Rs_Ls_R\f{\MX m_t}{\MX^2-m_t^2}
\)\MX^2\ln\(\f{\cut^2}{\MX^2}+1\)    \right. \\[6mm]
& \dis  \hspace*{7mm}
+\(
3c_L^2s_R^2 - \(c_L^2c_R^2+s_L^2s_R^2\)\f{m_t^2}{\MX^2-m_t^2}-
2c_Lc_Rs_Ls_R\f{\MX m_t}{\MX^2-m_t^2}
\)m_t^2\ln\(\f{\cut^2}{m_t^2}+1\)  \\[6mm]
& \dis
\left.\left.
\hspace*{7mm}
-2s_L^2c_R^2\f{\cut^2\MX^2}{\cut^2+\MX^2}
-2c_L^2s_R^2\f{\cut^2m_t^2}{\cut^2+m_t^2}
\]
\right\}  \,.
\ea
\eeq
To compare with Eq.~(\ref{eq:mHapp}), we consider
the limit  $r_t\equiv (\muxt/\muxx)^2\to 0$ and expand all quantities
in terms of the small parameter $x_t\equiv (\mtx/\muxx)^2$, so that,
$(m_t,\,\MX)\approx \(0,\, \muxx\sqrt{1+x_t}\) $ and
$(s_L, s_R)\approx (x_t,\,0)$.\, With these, we verify that
the $M_h^2$ formula (\ref{eq:Mh}) 
reduces to $\,M_h\approx 2\mtx$,\, in agreement with
the approximate mass relation 
(\ref{eq:mHapp}) derived by the gauge-invariant RG analysis.
Using the physical seesaw solutions [cf. Fig.\,\ref{fig:tssphysrefapp}(a)-(c)],
we can plot the predicted Higgs mass from Eq.\,(\ref{eq:Mh})
[Eq.\,(\ref{eq:mHapp})] as the solid [dotted] curves
in Fig.\,\ref{fig:Mhnew}(a).
It is important to note that our current large-$N_c$ 
fermion-bubble approximation predicts a heavy Higgs mass, typically
around $1$\,TeV\,\footnote{
With the seesaw mechanism embedded in a more general theory, 
there are more composite scalars with mixings, and one of the neutral 
Higgs bosons may be as light as 
${\cal O}(100)$\,GeV\,\cite{seesaw,dob}.},
saturating the SM unitarity bound.

When the ratio $\k/\k_c$ becomes closer to one (i.e., $\k$ becomes
more critical),  the Higgs mass becomes
lighter, as expected from the mass formula (\ref{eq:Mh}).
Also, the approximate relation
$M_h\approx 2\mtx$ in (\ref{eq:mHapp}) holds better for smaller 
$\k/\k_c \sim 1$ (to about $30\%$) and becomes less reliable
for larger $\k/\k_c$ value with an overestimate factor up to
$\sim\!2$.  This shows that the current improved broken phase calculation
of $M_h$ (including exact seesaw mixings) already works 
better than the usual approach which
results in $M_h\approx 2\mtx$ \cite{seesaw,georgi,dob}.

Finally, we note that the above calculation
of the Higgs mass $M_h$ includes only the large-$N_c$
fermion-bubble contributions, but ignores the non-large-$N_c$
Higgs propagation in the loop. 
For the leading logarithmic terms in $M_h$, this
corresponds to solving the RG equations (RGEs)
for top Yukawa coupling ($y_t$)
and Higgs self-coupling ($\la$) by keeping the 
fermion-bubble terms.
This approach also applies to the calculation of top mass $m_t$
and results in the Pagels-Stokar formula which,
in the case of a low cutoff scale $\cut \sim 10^{4-5}$\,GeV,  
is found to agree well with the full RG evolution.
In the minimal top-condensate model\,\cite{BHL}, 
the large-$N_c$ fermion-bubble calculation
of $m_t$ agrees with full RG analysis to 
$5-14\%\,(34\%)$ level for $\cut=10^{5-4}\,(10^{19})\,$GeV,
while for the Higgs mass prediction, the former tends to overestimate
$M_h$ by a factor of $1.8-2\,(1.2)$ for 
$\cut=10^{5-4}\,(10^{19})\,$GeV\,\cite{Hill90}.
This is due to the fact that for a high scale $\cut$, $m_t$
is controlled by the infrared quasi-fixed point\,\cite{hill};
for a low scale $\cut$, the  
infrared fixed point is not reached and the $m_t$ value is
mainly determined by the dominant large-$N_c$ RG running so that
the fermion-bubble calculation works well\,\cite{Hill90}.

The Higgs mass in the case of a high scale $\cut$ is again
controlled by the infrared quasi-fixed point (where the $y_t$-term
and $\la$-term tend to cancel in the $\beta$-function of $\lambda$);
however, the situation with a low scale $\cut$ is different as the
infrared fixed point is not reached and 
the positive (non-large-$N_c$) $\lambda^2$-term in the $\beta$-function
of $\la$ has a sizable numerical coefficient compared to the negative 
large-$N_c$ $y_t^4$-term.
This $\lambda^2$-term can drive $\lambda$ (and thus $M_h$) to
lower value and corrects the usual fermion-bubble calculation by 
a factor $\sim\!1.8-2$ for $\cut=10^{4-5}$\,GeV\,\cite{BHL,Hill90},
but, the uncertainties
of the one-loop RG predictions (from the unknown non-perturbative
dynamics associated with the compositeness condition at
$\mu=\cut$) also become much larger, 
of $\O(100-200)$\,GeV \cite{BHL}, as the infrared fixed point
is not so relevant. 
Hence, the one-loop full RG analysis 
(with compositeness conditions)\,\cite{BHL} 
may not be more reliable than the usual fermion-bubble calculation
for theories with a {\it low scale $\cut$}.  
Similar features should hold for the
$M_h$ analysis in the top seesaw model 
[except a complication by the new mass scale $\MX$ between 
$(m_t,\,M_h)$ and $\cut$].  
Nevertheless, we feel it is useful to  
implement such an improved one-loop RG analysis of $M_h$ below
(in the spirit of Ref.\,\cite{BHL}), as a comparison.

Using the mass-independent $\overline{\rm MS}$ scheme\,\cite{BHL},
we consider the top-seesaw RG evolution in two steps:
(i).  for the range $\cut \geq \mu \geq \MX$;
(ii). for the range $\MX \geq \mu \geq (M_h,\,m_t)$.
We start with the gauge-invariant effective Lagrangian (\ref{eq:RG1})
at $\mu=\cut$, 
\be
\label{eq:Lcut0}
{\cal {L}}_{\cut} =  {\cal {L}}_{\rm kinetic} -
\left[\,
\muxt\,\ov{\x_L}t_R+
\muxx\,\overline{\chi_{R}} \,\chi_{L}
+ h_{\rm 1}\,\overline{\psi_L} \,\x_R\,\Phi_0 + {\rm h.c.} \right]
-\Lambda^{2}\Phi_0^{\dagger}\Phi_0 \,,
\end{equation}   
where for simplicity 
the partial rotation (\ref{eq:partialrotation}) is not taken  
since we will use a mass-independent RG scheme\,\cite{BHL}
and consider $\MX \simeq \muxx$.
For $\cut > \mu \geq \MX$, the effective Lagrangian
${\cal {L}}_{\MX\leq \mu<\cut}$ contains,
\be
\label{eq:Lmux}
\ba{l}
-\(\,
\muxt\,\ov{\x_L}t_R+
\muxx\,\overline{\chi_{R}} \,\chi_{L}
+ h_{\rm 1}\,\overline{\psi_L} \,\x_R\,\Phi_0 + {\rm h.c.} \)
+Z_{\Phi}\left| D\Phi_0\right|^{2} -
\widetilde{M}_{\Phi}^{2}(\mu) \Phi_0^{\dagger }\Phi_0 -
\widetilde{\lambda} \left(\Phi_0^{\dagger }\Phi_0\right)^{2}  
\\[4mm]
=
-\(\, \muxt\,\ov{\x_L}t_R+
\muxx\,\overline{\chi_{R}} \,\chi_{L}
+ g_t\,\overline{\psi_L} \,\x_R\,\Phi + {\rm h.c.} \)
+\left| D\Phi\right|^{2} -
{M}_{\Phi}^{2}(\mu) \Phi^{\dagger }\Phi -
{\lambda} \left(\Phi^{\dagger }\Phi\right)^{2} \,, 
\ea
\ee
where $g_t=h_1/Z_\Phi^{1/2}$, 
$M_\Phi^2=\widetilde{M}_\Phi^2/Z_\Phi$, and
$\la =\widetilde{\la}/Z_\Phi^2$, with
$Z_\Phi (\mu) \simeq (\k/\k_c)\ln\(\cut/\mu\)\,$
and
$\widetilde{\la} (\mu) \simeq 4\pi^2(\k^2/\k_c)\ln\(\cut/\mu\)\,$
in the $\ov{\rm MS}$ scheme.
%Our focus here is to examine how the inclusion
%of (non-large-$N_c$) $\lambda$ running may affect $M_h$
%since the large-$N_c$ RG running works well for $m_t$ analysis
%(\`a la Pagels-Stokar formula) for low scale $\cut\lesssim 10^{4-5}$\,GeV.
The SM gauge couplings are negligible for the current analysis and
we can write the RGE of $\la$ in the region $\cut \geq \mu \geq \MX$,
\be
\beta (\la) = \dis\f{d\la}{d\ln\mu}
\simeq \dis\f{1}{8\pi^2}\[-N_cg_{t}^4
+2N_c\la g_{t}^2+12\la^2\] ,
\label{eq:la-RG1}
\ee         
where the $\la$-terms on the right-hand side 
tend to decrease $\la$ (and $M_h$) 
and are ignored in the usual
fermion-buble calculation (which is justified for 
$g_{t}^2 \gg \la$ and $N_c\gg 1$). 
The large-$N_c$ relation $g_t=h_1/Z_\Phi^{1/2}$ gives
the $\psi$-$\chi$-$\Phi$ Yukawa coupling,
\be
\label{eq:ytxNc}
\dis
g_t^2(\mu) \,\simeq\, \f{8\pi^2/N_c}{\ln(\cut/\mu)} \gg 1 \,,
\ee
which suggests the compositeness boundary condition
$g_t^2(\cut)=\infty$. 
The complete large-$N_c$ RGE for $g_t$ is, 
\be
\label{eq:ytxNc2}
\dis
\f{dg^2_t}{d\ln\mu} =\f{1}{8\pi^2} \[N_cg_{t}^2-3\f{N_c^2-1}{N_c}g_3^2\]g_t^2\,, 
\ee
where the effect of the QCD coupling $g_3$ is found to be numerically
negligible for the current analysis, so that $g_t^2$ may be solved analytically,
\be
\label{eq:ytx}
\dis
g^2_{t}(\mu) \simeq \[g_{t}^{-2}(\cut) +\f{N_c}{8\pi^2}\ln\f{\cut}{\mu}\]^{-1} . 
\ee
The boundary value $\la(\cut)$ may be estimated using the above large-$N_c$
fermion-bubble relation $\la=\widetilde{\la}/Z_\Phi^2$, corresponding to 
keeping the first term on the right-hand side of the RGE (\ref{eq:la-RG1}), i.e.,
\be
\label{eq:la-Nc}
\dis
\la (\mu) \simeq \[g_{t}^{-2}(\cut) +\f{N_c}{8\pi^2}\ln\f{\cut}{\mu}\]^{-1}  \,,
\ee
from which, we define the compositeness conditions at $\mu=\cut$,
\be
\label{eq:BC}
\la (\cut) = g_t^2(\cut)=\infty \,.
\ee
Using this and (\ref{eq:ytx}),
we can solve the complete RGE (\ref{eq:la-RG1}) and 
deduce $\la(\MX)$. 
As $\mu$ approaches the scale $\MX$, we perform the partial
diagonalization (\ref{eq:partialrotation}) to the mass terms in
Eq.\,(\ref{eq:Lmux}) and then decouple $\x$ at $\mu \leq \MX$.
This gives the effective Lagrangian (\ref{eq:LRG1mu}) 
derived earlier, with $\ov{M}\simeq \MX$ and
the renormalized $t$-$t$-$\Phi$ Yukawa coupling
$y_t(\mu) = g_t(\mu)\sin\theta$ for $\mu \leq \MX$.
The on-shell condition $m_t(m_t)=y_t(m_t)v/\sqrt{2}=174$\,GeV
requires $y_t(m_t)\simeq 1$, so that 
$y_t(\mu)$ is constrained to be small, close to $1$,
\be
\label{eq:ytNc}
\dis
y^2_{t}(\mu) \simeq \[1 -\f{N_c}{8\pi^2}\ln\f{\mu}{m_t}\]^{-1} 
             \gtrsim 1 \,,~~~~~~(\mu \leq \MX)\,. 
\ee
The numerical effect of  $y_t(\mu)$
on the relevant $\la$ running is found to be small
for  $\mu\in \(M_h,\,\MX\)$.
Thus, the step-(ii) RG evolution of $\la$  in the region
$\MX\geq\mu\geq M_h$  is essentially controlled 
by the simplified RGE,  $d\la/d\ln\mu \simeq 3\la^2/2\pi^2$. 
The physical Higgs mass is then numerically solved from the on-shell
condition, 
\be
M_h^2 =2v^2\la(M_h)
\simeq \dis
{2}v^2
\[{\dis \f{1}{\la(\MX)}+\f{3}{4\pi^2}\ln\f{\MX^2}{M_h^2} }
\,\]^{-1}  ,
\label{eq:MH-RG2}        
\ee
and is plotted in Fig.\,\ref{fig:Mhnew}(b).

Since the $\x$ mass $\MX$ is determined from solving the seesaw gap equation
for each given $\cut$ and  $\k/\k_c$ in Sec.\,2.5,
Fig.\,\ref{fig:Mhnew}(b)
shows different Higgs mass spectrua as $\k/\k_c$ varies.
We see that for $\cut < 10$\,TeV, $M_h$ ranges around
$(0.7-1.25)\,{\rm TeV}\sim \!1$\,TeV, while 
for $\cut \gtrsim 10$\,TeV the $\la$ running becomes more significant,
bringing $M_h$ down to  $\sim\!650-400$\,GeV which is about
a factor 2 below the large-$N_c$ fermion-bubble calculation in
Fig.\,\ref{fig:Mhnew}(a), as also expected from the analysis of 
the minimal top-condensate model\,\cite{BHL,Hill90}.
However, we must note that for dynamical symmetry breaking theories
with a low scale cutoff $\cut \sim 10-100$\,TeV, 
the infrared fixed point becomes less relevant and the uncertainties
in $M_h$ associated with the compositeness condition (\ref{eq:BC})  
are large, around $\O(100-200)$\,GeV, so that the naive one-loop RG
running is not so reliable 
and higher loop corrections could be important as well.
Furthermore, the simplest mass-independent $\ov{\rm MS}$ RG scheme may have
its drawback in treating such low scale dynamical theories,
in comparison with the mass-dependent renormalization\,\cite{GP}
which suggests that the large Higgs mass nearby $\cut$ will suppress
$\lambda$ running and result in higher 
$M_h$ values\,\cite{Bando,condrev}.
Hence, the RG improved spectrum in Fig.\,\ref{fig:Mhnew}(b) 
only serves as a reference to show how the traditional large-$N_c$ 
fermion-bubble calculation in the top seesaw model might be improved 
when including the perturbative Higgs self-coupling evolution.

%\input seesaw_2.tex

%%%%%%%%%%%%%%%%%%%%%%%%%%%%%%%%%%%%%%%%%%%%%%%%%%%%%%%%%%%%%%%%%%%%
\newpage
\section{\hspace*{-2mm}{\large
Extensions with Bottom Quark}}
\label{sec:bseesaw}

\subsection{\hspace*{-2mm}{\normalsize 
The Mechanism for Bottom Quark Mass}}

As things stand, we have not addressed the 
issue of the bottom quark mass.
The simplest way of producing the $b$ quark mass is to
include additional weak-singlet fermionic fields 
$\omega _{L}$ and $\omega _{R}$  together with $b_{R}$, 
which are charged under the gauge group
$SU(3)_1\otimes SU(3)_2\otimes SU(2)_W\otimes U(1)_Y$,
\begin{equation}
\dis  
b_{R}, \omega _{L} \!:~\, \({\bf 1,3,1,} -\f{2}{3}\) \, ,~~~~~~ 
\omega _{R}        :~\, \({\bf 3,1,1,} -\f{2}{3}\) \,.
\label{omega}
\end{equation}
Such assignments for the $b-\w$ sector nicely
cancel the unwanted gauge-anomalies from the Top Seesaw sector 
(cf. Sec.\,2.1),
so that we can regard their presence as a generic part of the standard
Topcolor picture.
We further allow $\overline{\omega}_{L}\omega _{R}$ and  
$\overline{\omega}_{L} b_{R}$ mass terms, in addition to the 
$\chi -t$ mass terms [cf. the Eq.\,(\ref{eq:xmass}) in Sec.\,2.2],
\begin{equation}
\label{eq:xwmass}
{\cal {L}}_{\rm mass} ~=\, 
-\(\mu_{\chi\chi}\,\overline{\chi _{L}}\,\chi_{R}
   \,+\,\mu_{\chi t}\,\overline{\chi _{L}}\,t_{R} \)
-\(\mu_{\omega \omega}\,\overline{\omega_L} \omega_{R} 
 \,+\, \mu_{\omega b}\,\overline{\omega_L} b_{R} \)
\,+\,{\rm h.c.} \,.
\end{equation}
With the previous assignments
for the $\chi$ quarks,
the extended model can be schematically represented as below, 
\noindent
%{\bf Inclusion of b quark I:$\protect\bigskip $}
\begin{center}
\underline{$\ \ \ \ \ \ \ \ \ \ \ \ 
~\,SU(3)_{1}$~\, \ \ \ \ \ \ \ \ \ \ \ \ \ \ \
$ \ \ \ \ \ SU(3)_{2}$ \ \ \ \ \ \ \ \ \ \ \ }
\vspace*{2.2mm}

$\left\lgroup
\begin{array}{c}
\left(\! 
\begin{array}{c}
t_L \\ 
b_L
\end{array}
\!\right)  ~~I=\half%
%TCIMACRO{\UNICODE[m]{0xbd} }%
%BeginExpansion
%{\frac12}%
%EndExpansion
\\ [7mm]
\left(\! 
\begin{array}{c}
\chi_R \\ 
\omega_R
\end{array}
\!\right) ~~I=0
\end{array}
\right\rgroup 
${\small \ \ \ \ \ }$\left\lgroup 
\begin{array}{c}
\left(\!
\begin{array}{c}
t_R \\ 
b_R
\end{array}
\!\right) ~~I=0 
\\ [7mm]
\left(\! 
\begin{array}{c}
\chi_L \\ 
\omega_L
\end{array}
\!\right) ~~I=0
\end{array}
\right\rgroup $
\end{center}
\vspace*{7mm}

We see that the additional quark $\w_R$ joins  the strong
Topcolor $SU(3)_1$ like $\x_R$. After the Topcolor breaking and
integrating out massive colorons, we have following
NJL interactions, 
\bea
\label{eq:Lxw0}
{\cal L}_{\rm int} 
&= & \frac{ h_1^2 }{\cut^2} \; \left[ \left( \bar{\psi}_L \chi_R \right)
    \left( \bar{\chi}_R \psi_L \right)
   + \left( \bar{\psi}_L \omega_R \right) \left( \bar{\omega}_R \psi_L \right)
   \right] 
\nonumber \\[2mm]
& \to & -\cut^2 \(\Phi_{t0}^{\dagger} \; \Phi_{t0} +
                  \Phi_{b0}^{\dagger} \;\Phi_{b0} \) 
 - h_1 \, \( \bar{\psi}_L \; \Phi_{t0} \; \chi_R  +
             \bar{\psi}_L \; \Phi_{b0} \; \omega_R\) 
 + {\rm h.c.} \, ,
\eea
which contains two scalar doublets 
$\Phi_{t0}$ and $\Phi_{b0}$ after the bosonization of the NJL vertices.
The Lagrangian ${\cal L}_{\rm mass}+{\cal L}_{\rm int}$,
however, poses a global $U(1)$  symmetry 
under which the fields transform as,
\beq
\label{eq:U1PQ}
\ba{llll}
\psi_L     \ra \psi_L \,,                     &                       
t_R        \ra e^{i \alpha} \, t_R \,,        &
b_R        \ra e^{i \alpha} \, b_R \,,        &  \\[2mm]
\x_{L(R)}  \ra e^{i \alpha} \, \x_{L(R)} \,,  &
\w_{L(R)}  \ra e^{i \alpha} \,\w_{L(R)} \,,   &  
{\Phi}_{t0} \ra e^{-i \alpha} \,{\Phi}_{t0} \,, & 
{\Phi}_{b0} \ra e^{-i \alpha} \,{\Phi}_{b0} \,.
\ea
\eeq
If this  symmetry were exact, the  dynamical condensates
$\langle \ov{t_L}\x_R \rangle$ and  $\langle \ov{b_L}\w_R \rangle$
(or, equivalently, the scalar VEVs 
$\langle \Phi_{t0} \rangle$ and
$\langle \Phi_{b0} \rangle$) would spontaneously break it
and generate a problematic massless Goldstone boson (the Peccei-Quinn axion).  
Fortunately, the symmetry is anomalous, and the Topcolor 
instanton effect\,\cite{TCATC} induces an effective  
Peccei-Quinn breaking term via the 't~Hooft flavor 
determinant\,\cite{thooft} with the form, 
\beq
\label{eq:Hdet}
\dis\f{c_0}{\cut^2}\det\[\ov{\psi_L}
\(\ba{c} \x_R \\[2mm] \w_R \ea\)\]  \,+\,{\rm h.c.}
~=~
\dis\f{c_0}{\cut^2} \ep^{\alpha\beta} 
\(\ov{\psi_L}^{\alpha}\x_R\)\(\ov{\psi_L}^{\beta}\w_R\)\,+\,{\rm h.c.}\,,
\eeq
where $c_0$ is a (complex) constant depending on details of the Topcolor
strong dynamics and from experience with QCD we expect,
$c_0\sim \O(0.1-1)$.  
In analogy with the $\eta^\prime$ in QCD, this
effective interaction will provide a non-zero  
mass for the axionic pseudo-Goldstone boson.
It is also possible that such an effective Peccei-Quinn breaking term
may also arise from additional flavor dynamics at a scale much above
the Topcolor breaking scale\,\cite{georgi}. In general, we parametrize
the Peccei-Quinn breaking interaction as,
\bea
\label{eq:LPQ}
{\cal L}_{{\rm PQB}} 
& = & \frac{\xi \, h^2_{1}}{\cut^2} \epsilon^{\alpha \beta}
\left[ \left( \overline{\chi_R} \;\psi_L^{\alpha} \right)
       \left( \overline{\omega_R} \;\psi_L^{\beta} \right)
+       \left( \overline{\psi_L}^{\alpha} \chi_R \right)
       \left( \overline{\psi_L}^{\beta} \omega_R \right)
\right] \nonumber \\[4mm]
& \ra & - \xi\; \epsilon^{\alpha \beta} 
\[ \cut^2
          \Phi_{t0}^{\alpha} \;\Phi_{b0}^{\beta} 
         + h_1 \(\overline{\chi_R} \;\psi^\alpha_L \;\Phi_{b0}^{\beta}
                 + \overline{\omega_R} \;\psi^\beta_L \;\Phi_{t0}^{\alpha}
               \)\]
                 \,+\, {\rm h.c.}  \,.
\eea
where we ignore a possible phase in the parameter $\xi$ and let
it be real for the purpose of the current study.  
With the Topcolor instantons as the origin
of this effective interaction, we can estimate the typical size of $\xi$,
\beq
\label{eq:xi}
\dis
\xi=c_0/h_1^2 = c_0\[\dis\f{8\pi^2}{3}\kB\]^{-1} 
    \sim \O(10^{-2}-10^{-3})\,,
\eeq
where $c_0\sim \O(0.1-1)$ and  $\k/\k_c\sim 2-4 $.
Since the relevant values of $\xi$ are tiny, it is justified to
treat it as a perturbation and only include the corrections up to
$\O(\xi^1)$.  We note that,
in addition to generating an explicit axion mass, the above 
interaction (\ref{eq:LPQ}) also provides  a  correction
to the mass terms 
\,$\mtx\ov{t_L}\x_R$\,  and
\,$\mbw\ov{b_L}\w_R$,\, 
i.e., we generally have, 
from (\ref{eq:Lxw0}) and  (\ref{eq:LPQ}),
\beq
\label{eq:mtxmbw}
\mtx \,=\, \dis h_1\(    \langle\Phi_{t0}\rangle
                   +\xi\langle\Phi_{b0}\rangle  \) \,,~~~~~
\mbw \,=\, \dis h_1\(    \langle\Phi_{b0}\rangle
                   +\xi\langle\Phi_{t0}\rangle  \) \,.
\eeq
The second equation gives the physical $b$ mass, $m_b\approx \mbw \muwb/\muww$,
via the following seesaw matrix,
\bea
\label{eq:bseesaw}
- \left( \overline{b_L} ~~ \overline{\w_L} \right)
 \left\lgroup 
\begin{array}{cc}
0 & m_{b \w} \\[0.2cm]
\muwb & \muww \\
\end{array}
\right\rgroup
\left(\!\!
\begin{array}{c}
b_R \\
\w_R
\end{array}
\!\!\right)
\,+\, {\rm h.c.} \,.
\eea   
For $\xi\gtrsim 10^{-2}$,
there is the interesting possibility that the $b$ mass may completely
originate from ${\cal L}_{\rm PQB}$ (for example, from Topcolor instanton
effects).
This requires $\langle\Phi_{b0}\rangle =0$, implying the 
leading order Lagrangian (\ref{eq:Lxw0}) to have a zero mass-gap in
the $\(\ov{b_L}\w_R\)$ channel which can be realized 
when $\w$ becomes very heavy ($\muww \gg \cut$)
and decouples.  In this special case,
the whole model reduces back to our minimal top seesaw
model studied in Sec.\,2, except that now the $b$ quark 
acquires its mass from Topcolor instantons,
\beq
m_b\approx \mbw \muwb/\muww \,, ~~~~~
\(\;{\rm with}~ 
\mbw = \xi h_1\langle\Phi_{t0}\rangle =\xi\mtx\;\) \,.
\eeq
Consequently, the Higgs doublet $\Phi_{b0}$ is also removed
from the low energy theory and the remaining analysis of this
decoupling limit becomes identical to Sec.\,2.
However, in the more general cases where $\w$ does not decouple
from the theory (\,$\muww \!\lesssim \!\cut$\,), 
the $b$ quark can acquire its mass
from both terms in the second relation of (\ref{eq:mtxmbw});  
and furthermore, for
$\xi\lesssim 10^{-3}$ and $\muwb/\muww \lesssim 1$,
the mass $m_b$  predominantly comes from the leading order term.
Such non-decoupling scenarios also have a rich physical Higgs spectrum as both
Higgs doublets (including the massive axion) will be accessible
in our low energy theory.  These will be systematically studied
below.

\subsection{\hspace*{-2mm}{\normalsize
Gap Equations for Top and Bottom Seesaws and 
the Physical Solutions 
}}

\begin{figure}[H]
\begin{center}
\vspace*{-3mm}
\includegraphics[width=14.5cm]{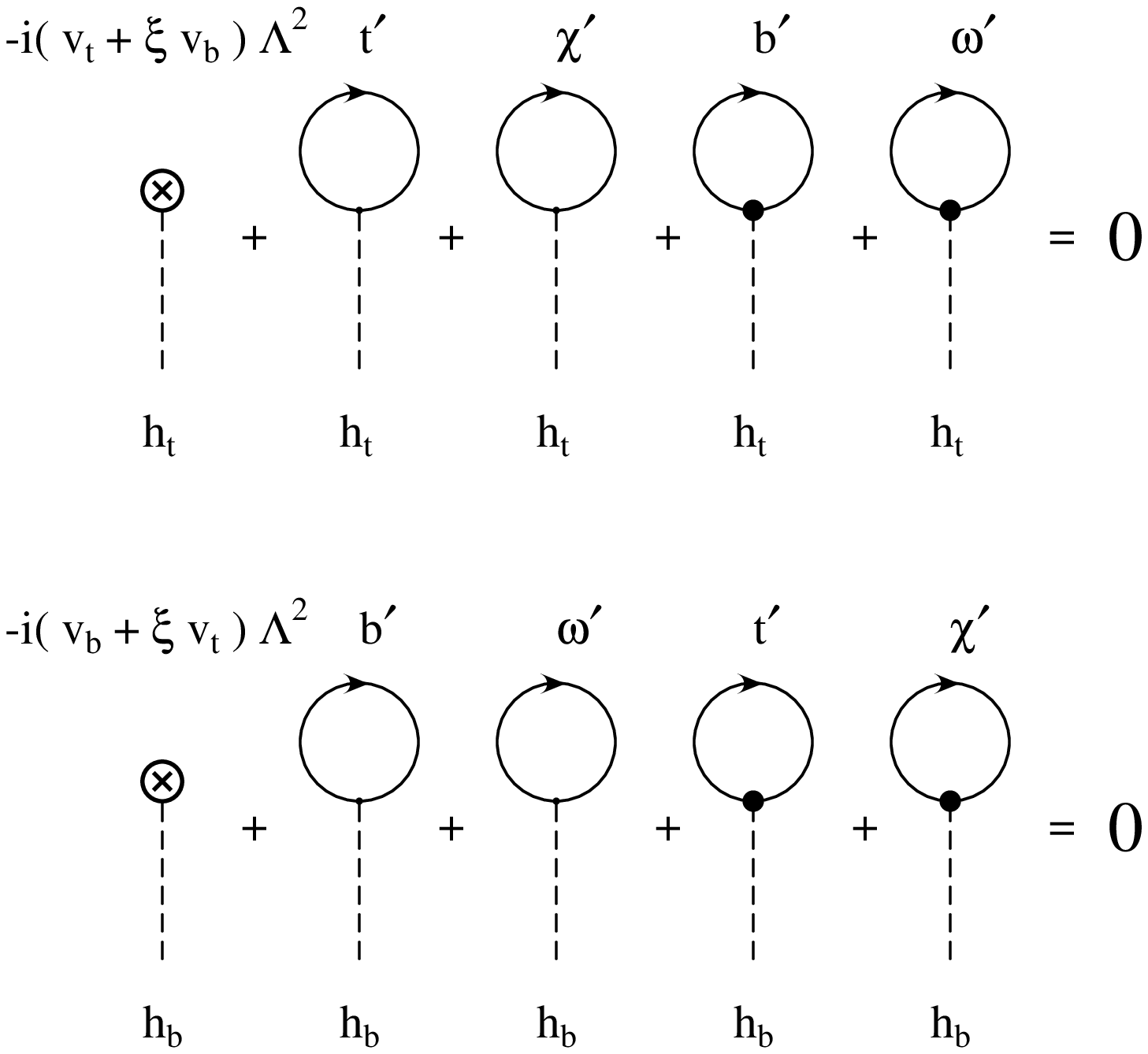}
\end{center}
\vspace*{-5mm}
\caption{
Coupled gap equations for top and bottom seesaws up to $\O(\xi)$.
The black dots denote the vertices associated with small $\xi$ couplings.
}
\label{fig:tb-tadpole}
\end{figure}

In this subsection, we derive the gap equations
for both top and bottom seesaws up to $\O(\xi)$ and analyze
their physical solutions. This is in analogy with Sec.\,\ref{sec:tadpole},
but with the 
$b$ seesaw mass gap and $\O(\xi)$ corrections included.
We start by explicitly defining the bare fields of the
two Higgs doublets $\Phi_{t0}$ and $\Phi_{b0}$
in the shifted vacuum,
\beq
\label{eq:hthb}
{\Phi}_{t0} \,=\,
\(\!
\ba{c}
\(v_{t0} + h_{t0}^0 + i\pi^0_{t0}\)/\sq2 \\[3mm]
\pi^-_{t0}
\ea
\!\) ,~~~~~~
{\Phi}_{b0}  \,=\,
\(\!
\ba{c}
\pi^+_{b0}   \\[3mm]
\(v_{b0} + h_{b0}^0 + i\pi^0_{b0}\)/\sq2 
\ea
\!\) ,
\eeq
where, as in usual 2-Higgs-doublet model (2HDM) and upon renormalization, 
the rotations of $h_{t}^0$ and $h_{b}^0$
give the mass-eigenstates of neutral Higgs bosons $(h^0,\,H^0)$, while
the combinations of other six scalars 
$\pi_{t}^{0,\pm}$ and $\pi_{b}^{0,\pm}$ 
result in three would-be Goldstone bosons (eaten by $W^\pm ,Z^0$) and three
physical Higgs states $(A^0,\,H^\pm)$. 
Now, we can explicitly write the two seesaw mass-gaps in (\ref{eq:mtxmbw}) as,
\bea
\label{eq:newmtxmbw}
\mtx = \frac{h_1}{\sqrt{2}} \left( v_{t0} + \xi v_{b0} \right)\,, ~~~~~
\mbw = \frac{h_1}{\sqrt{2}} \left( v_{b0} + \xi v_{t0} \right)\, .
\eea
In the same spirit of Sec.\,2.4 and using the Lagrangian
$\,{\cal L}_{\rm mass}+{\cal L}_{\rm int}+{\cal L}_{\rm PQB}$,\,
we obtain two coupled gap equations up to $O(\xi)$ 
from the tadpole conditions of the neutral Higgs fields 
$h_{t0}^0$ and  $h_{b0}^0$, as shown in 
Fig.\,\ref{fig:tb-tadpole}. Thus, we can derive them as,
\beq
\label{eq:gaptb}
\mtx \,=\, \dis\frac{\kappa}{\kappa_c} \left[ F_t + \xi F_b \right]\,,
~~~~~~
\mbw \,=\, \dis\frac{\kappa}{\kappa_c} \left[ F_b + \xi F_t \right] \, ,
\eeq
or, equivalently, up to $\O(\xi)$,
\beq
\dis\kB (\mtx-\xi \mbw)  \,=\, F_t \,,
~~~~~~
\dis\kB (\mbw-\xi \mtx)  \,=\, F_b \,,
\label{eq:gaptb2}
\eeq      
where
\beq
\label{eq:Ftb}
\ba{lll}
F_t &=& \dis c^t_L s^t_R \left( m_t
- \frac{m_t^3}{\cut^2} 
\ln \left[ \frac{\cut^2 + m_t^2}{ m_t^2} \right] \right)
 + s^t_L c^t_R \left( \MX
- \frac{\MX^3}{\cut^2} 
\ln \left[ \frac{\cut^2 + \MX^2}{\MX^2} \right] \right) 
\,,
\\[5mm]
F_b &=& \dis c^b_L s^b_R \left( m_b
- \frac{m_b^3}{\cut^2} 
\ln \left[ \frac{\cut^2 + m_b^2}{ m_b^2} \right] \right)
 + s^b_L c^b_R \left( \MW
- \frac{\MW^3}{\cut^2} 
\ln \left[ \frac{\cut^2 + \MW^2}{\MW^2} \right] \right) 
\,,
\ea
\eeq
and the seesaw rotation angles $s_{L,R}^{t,b}$ and $c_{L,R}^{t,b}$ are
similarly defined as in Eqs.\,(\ref{eq:rotL})-(\ref{eq:rotR}).
We see that the two gap equations decouple from each other at the leading order
$\O(\xi^0)$ and the correlations appear at $\O(\xi)$
which are generally small. The $\O(\xi)$ terms become important only for
very large $\tanb = v_t/v_b$ and sizable $\xi \gtrsim 10^{-2}$.
For instance, a typical case with $\tanb=40$ and $\xi=2\times10^{-2}$
gives the ratio $(\xi v_t)/v_b = 80\%$, implying that the $\xi$-term 
makes up about $80\%$ of the mass-gap $\mbw$ and
thus the $b$ mass.
Another important role of the $\O(\xi)$
interactions is their contributions to the Higgs masses,
especially, the mass of the pseudo-scalar $A^0$.

Similar to the RG analysis in Sec.\,2.4, we can further evolve the Higgs
Lagrangian $\,{\cal L}_{\rm mass}+{\cal L}_{\rm int}+{\cal L}_{\rm PQB}$,\,
from the scale $\cut$ down to $\mu \(< \mu_{\x\x,\w\w}\leq \cut\)$ 
by integrating out loops with the heavy fermions $(\x,\,\w)$. 
The Higgs fields get renormalized, e.g., 
$h_{t0}^0=Z_{ht}^{1/2}h_t^0,\,  h_{b0}^0=Z_{hb}^{1/2}h_b^0,\,$ and
so on.
We can write down the renormalized Higgs VEVs,
$v_t = Z_{h_t}^{1/2} v^0_t$ and $v_b = Z_{h_b}^{1/2} v^0_b$,
and define their ratio,  $\tanb=v_t/v_b$, as usual.
Here, the two neutral Higgs wave-function renormalization constants
are computed as,
\beq
\label{eq:Zhtb}
\ba{lll}
Z_{h_t} & = & \dis\half\kB \left\{
c^{t \, 2}_L s^{t \, 2}_R \ln \left[ \frac{\cut^2 + m_t^2}{m_t^2} \right] +
(c^{t \, 2}_R + s^{t \, 2}_L s^{t \, 2}_R) 
\ln \left[ \frac{\cut^2 + m_\chi^2}{m_\chi^2}
\right]
\right\} +\O(\xi^2) \,, \\[5mm]
Z_{h_b} & = & \dis\half\kB   \left\{
c^{b \, 2}_L s^{b \, 2}_R \ln \left[ \frac{\cut^2 + m_b^2}{m_b^2} \right] +
(c^{b \, 2}_R + s^{b \, 2}_L s^{b \, 2}_R) 
\ln \left[ \frac{\cut^2 + m_\omega^2}{m_\omega^2} \right] \right\} +\O(\xi^2) \, ,
\ea
\eeq
in which the $\xi$-corrections appear only at $\O(\xi^2)$ as can be seen
from the interaction Lagrangian ${\cal L}_{\rm int}+{\cal L}_{\rm PQB}$.
Then, from Eqs.\,(\ref{eq:mtxmbw}) and (\ref{eq:Zhtb}), we derive two new
Pagels-Stokar formulae, 
\beq  
\label{eq:PS-vtvb}
\ba{lll}
v_t^2 &=&\dis 
\frac{\(\mtx-\xi\mbw\)^2}{4 \, \pi \, \kappa_c} \left\{
c^{t\;2}_L s^{t\;2}_R \ln \left[ \frac{\cut^2 + m^2_t}{m^2_t} \right]
+ \left( c^{t\;2}_R + s^{t\;2}_L s^{t\;2}_R \right) \ln \left[
\frac{\cut^2 + M^2_\chi}{M^2_\chi} \right] \right\} + \O(\xi^2) \,, 
\\ [6mm]
v_b^2 &=&\dis 
\frac{\(\mbw-\xi\mtx\)^2}{4 \, \pi \, \kappa_c} \left\{
c^{b\;2}_L s^{b\;2}_R \ln \left[ \frac{\cut^2 + m^2_b}{m^2_b} \right]
+ \left( c^{b\;2}_R + s^{b\;2}_L s^{b\;2}_R \right) \ln \left[
\frac{\cut^2 + \MW^2}{\MW^2} \right] \right\} + \O(\xi^2) \,,
\ea
\eeq
with a physical constraint from the EWSB, 
$(v_t^2+v_b^2)^{1/2}=v\simeq 246$\,GeV.
Again, we see that the $\xi$-correction may be important only for the 
second equation of $v_b$ when $\tanb$ is very large
and $\xi$ is sizable. 
Since typically $\mtx\lesssim 1$\,TeV and $\mbw \gtrsim 10-20$\,GeV, 
we see that the effects of $\xi$ 
in Eq.\,(\ref{eq:PS-vtvb}) is negligible  for $\xi \lesssim 10^{-3}$.

\begin{figure}[H]
\begin{center}
\vspace*{-1.5cm}
\includegraphics[width=16cm,height=20cm]{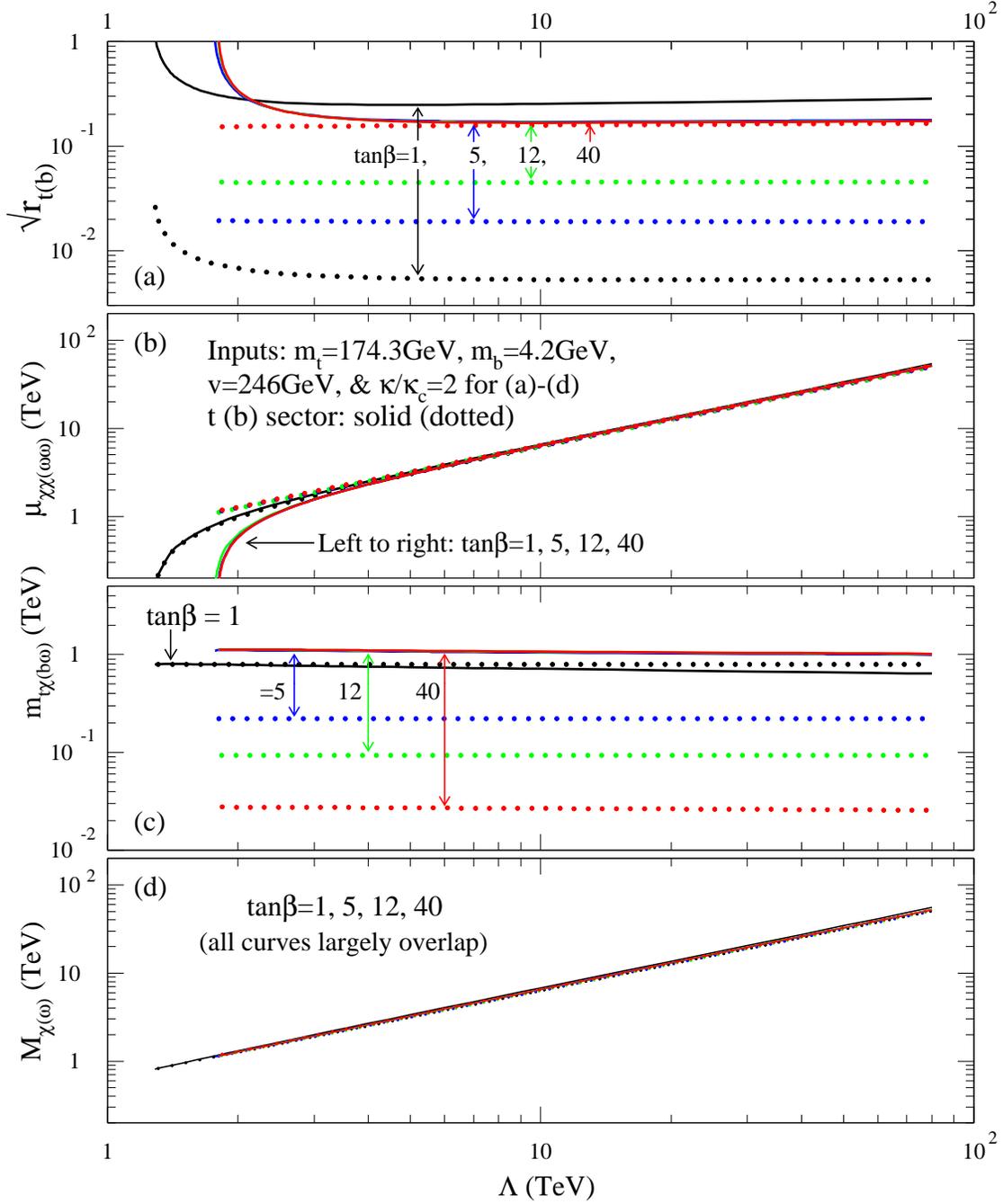}
\end{center}
\vspace*{-13mm}
\caption{
Solutions of the top and bottom seesaw gap equations 
with $\kappa / \kappa_c =2$ and $\tanb \in (1,\,5,\,12,\,40)$,
where we have superimposed the physical constraints,
$(m_t,\,m_b)=(174.3,\,4.2)$\,GeV and $v=246$\,GeV.
The solid curves are for the top sector while
the dotted curves for the bottom sector.
Here we have chosen the region $\xi\lesssim 10^{-3}$ in which the 
$\xi$-effects are negligible (invisible).
}
\label{fig:tbss_soall}
\end{figure}

\begin{figure}[H]
\begin{center}
\vspace*{-1.5cm}
\includegraphics[width=16cm,height=18cm]{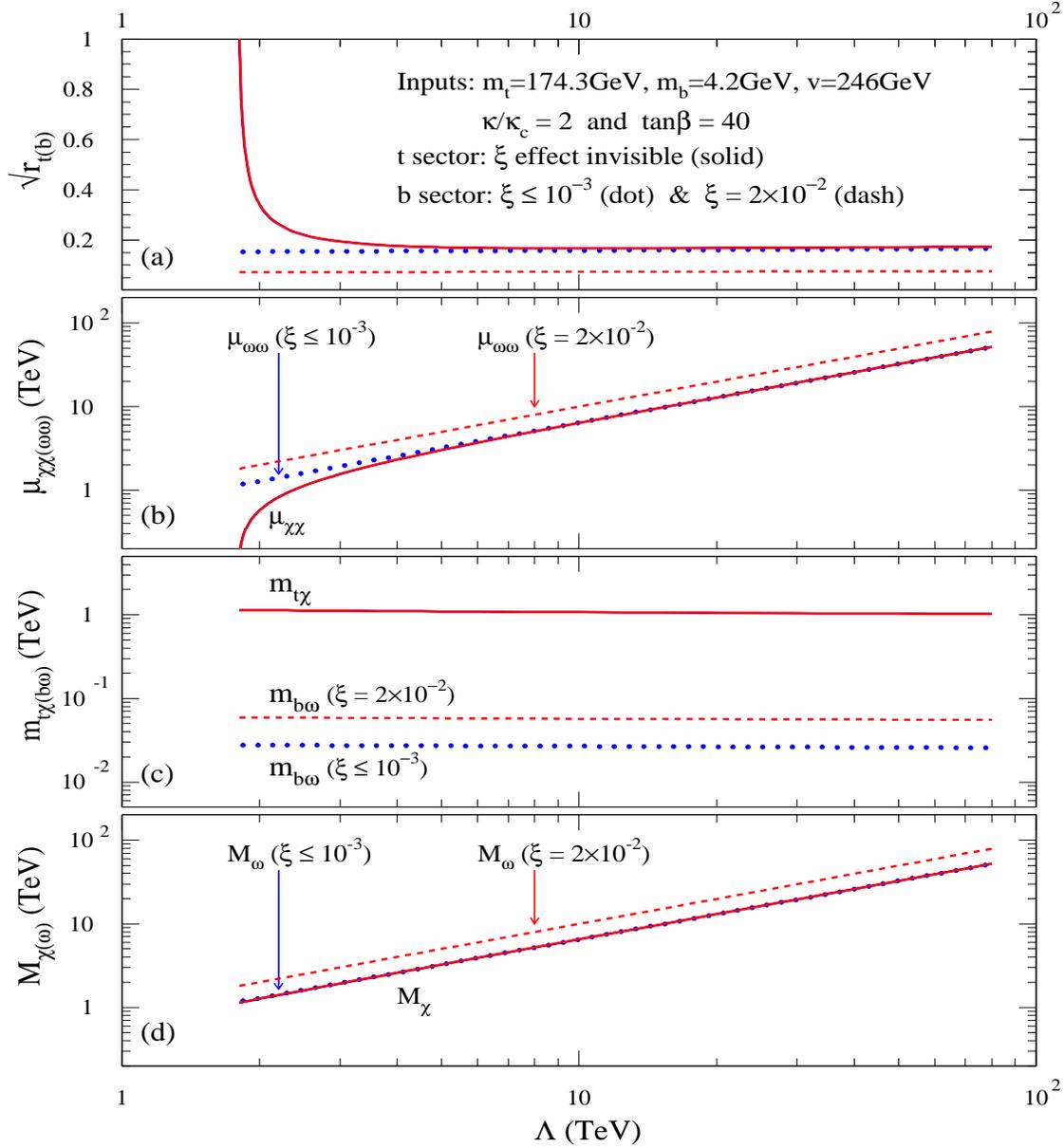}
\end{center}
\vspace*{-13mm}
\caption{
The effect of the $\O(\xi)$ corrections on the physical seesaw solutions,
for $\kappa / \kappa_c =2$ and $\tanb = 40$.
The solid curves are for the top sector while
the bottom sector is depicted by dotted curves [$\xi\lesssim \O(10^{-3})$]
and dashed curves [$\xi= 2\times 10^{-2}$]. A sizable value of 
$\xi =2\times 10^{-2}$ (representing typical instanton effect) 
can provide about $80\%$ of the $b$ mass for $\tanb = 40$; while
for $\xi\lesssim \O(10^{-3})$, the $b$ mass is almost fully given
by $\O(\xi^0)$  seesaw corrections.
} 
\label{fig:tbss40xi2m2}
\end{figure}

Now, we are ready to solve the gap equations for the top-bottom seesaw system.
We note that our extended model has three input parameters
$\(\cut,\,\k/\k_c,\,\tanb\)$, and three extra
unknown parameters $(\mbw,\, r_b,\, \muww)$ (with $r_b\equiv \muwb/\muww$) from
the $b$-seesaw sector, in addition to $(\mtx,\, r_t,\, \muxx)$ from the
$t$-seesaw sector. On the other hand, we have six
physical conditions in total: two seesaw gap equations
[in Eq.\,(\ref{eq:gaptb})], 
two Pagels-Stokar formulae [in Eq.\,(\ref{eq:PS-vtvb})], 
and two mass-eigenvalue equations 
[in Eq.\,(\ref{eq:mt}) for $m_t$ and a similar one for $m_b$].
Thus, all six seesaw parameters can be completely solved as functions of
$\cut$ for each given  $\(\k/\k_c,\,\tanb\)$.
Consequently, the masses of $\x$ and $\w$ are also predicted, 
together with all seesaw mixing angles. 
We display our systematic numerical solutions for a wide range of
$\tanb$ values in Fig.\,\ref{fig:tbss_soall}, where we have chosen 
$\xi \lesssim 10^{-3}$ and found that the $\xi$-corrections are negligible
and the difference from $\xi=0$ case is invisible in the plots.
From this figure, we also see that the $\chi$ and $\omega$ are highly
degenerate for all solutions;
the same feature holds for the parameters $(\muxx,\,\muww)$ when 
$\cut\gtrsim 2-3$\,TeV. 
This fact can be understood by noting that 
the real difference between the top and bottom sectors is
controlled by the experimental ratio $m_t/m_b\approx 40 \gg 1$ 
and the input ratio $\tanb = v_t/v_b$. The former is connected to
seesaw parameters via,
\beq
\dis \f{m_t}{m_b}  \approx \f{\mtx\muxt/\muxx}{\mbw\muwb/\muww} 
                   =       \f{\mtx}{\mbw}\sqrt{\f{r_t}{r_b}}  \,,
\eeq
while the latter can be deduced from the
Pagels-Stokar formula (\ref{eq:PS-vtvb}) after ignoring the $\O(\xi)$
corrections and the insensitive logarithmic factors, i.e., 
$\tanb =v_t/v_b \sim \mtx/\mbw$, where we 
have also expanded the right-hand sides of (\ref{eq:PS-vtvb}) like 
Eq.\,(\ref{ps2}) in which we can see the heavy masses $(\MX,\,\MW)$
[or $(\muxx,\,\muww)$]
of the vector-like fermions $(\x,\,w)$ have only logarithmic dependence,
obeying the decoupling theorem\,\cite{App,vectorF}. 
Similar decoupling behavior appears on the right-hand sides of the
gap equations (\ref{eq:gaptb})-(\ref{eq:Ftb}).
Indeed, it is this
decoupling nature that makes the right-hand side of (\ref{eq:PS-vtvb})
insensitive to $(\MX,\,\MW)$. Thus, we arrive at two approximate relations
below, which control the qualitative features of the two sectors,
\beq
\label{eq:appeq}
\dis
\sqrt{\f{r_t}{r_b}} \sim \f{m_t/m_b}{\tanb} \sim \f{40}{\tanb}\,,
~~~~~~
\f{\mtx}{\mbw}\sim \tanb \,.
\eeq
Using these, we can now understand, in Fig.\,\ref{fig:tbss_soall},
why the main difference between 
$t$ and $b$ sectors are reflected in the ratios
$(r_t,\,r_b)$ for small $\tanb$ values, but manifest in the mass-gaps
$(\mtx,\,\mbw)$ for large $\tanb$ values. Finally,
because of their vector-like decoupling nature,
the heavy masses $(\MX,\,\MW)$ or $(\muxx,\,\muww)$
remain highly degenerate, and numerically they are located 
at around $(0.63-0.65)\cut$ for $\k/\k_c=2$,
as shown in  Fig.\,\ref{fig:tbss_soall}.
However, we  expect such a picture for the $b$-sector to be
modified when $\xi$-correction to the mass gap $\mbw$ becomes significant in the
very high $\tanb$ region. As a typical case, we may consider $\tanb=40$
and $\xi=2\times 10^{-2}$ [which is a generic size of the Topcolor instanton
contribution with $c_0=\O(1)$ and $\k/\k_c=2$ in Eq.\,(\ref{eq:xi})].
In this case, we deduce a ratio $(\xi v_t)/v_b =\xi\tanb = 80\%$
for the mass-gap $\mbw$ in Eq.\,(\ref{eq:newmtxmbw}),
implying that the $\xi$-term
makes up about $80\%$ of $\mbw$ and the $b$ mass.
Consequently, the Eq.\,(\ref{eq:PS-vtvb}) no longer gives the relation
$v_t/v_b \sim \mtx/\mbw$ [and thus (\ref{eq:appeq})]
because in the second formula of Eq.\,(\ref{eq:PS-vtvb}) 
the $\xi\mtx$ term is non-negligible on the right-hand side.
But, the $t$-sector remains essentially the same as before 
since in the mass-gap $\mtx$ [cf. Eq.\,(\ref{eq:newmtxmbw})]  the ratio 
$(\xi v_b)/v_t=\xi/\tanb \lll 1$ and is completely
negligible even for small $\tanb$.
Our numerical solutions for this large $(\tanb,\,\xi)$ scheme are
shown as dashed curves in Fig.\,\ref{fig:tbss40xi2m2}, in comparison with 
the small or zero $\xi$ cases ($\xi \lesssim 10^{-3}$) shown as dotted curves.
Indeed, we see sizable modifications for the seesaw parameters in
the $b$-sector, i.e., the gap $\mbw$ is lifted up by a factor of $\sim\!2$
while the ratio $r_b=\muwb/\muww$ is shifted down by about one-half.
As a consequence, the mass scale $\MW$ (or $\muww$) for $\w$ is also pushed up 
somewhat, closer to the scale $\cut$. 
This gives an interesting example in which the effects of
Topcolor instantons\,\cite{TCATC}  are significant and  
provide the dominant contribution to the bottom mass.

\subsection{\hspace*{-2mm}{\normalsize
Mass Spectrum of Composite Higgs Sector
}}

We proceed to analyze the physical Higgs mass spectrum of this extended
model. Starting from the Lagrangian 
${\cal L}_{\rm mass}+{\cal L}_{\rm int}+{\cal L}_{\rm PQB}$\, at
$\mu =\cut$
and performing the seesaw mass diagonalization,
we evolve it down to the scale $\mu \(< M_{\x,\w} \leq \cut\)$ by
integrating out the loop momenta between $\mu$ and $\cut$ 
and arrive at the renormalized effective Lagrangian with only light quarks
$(t^\prime,\,b^\prime)$ and the two-doublet Higgs bosons,
\beq
\label{eq:Leff2mu}
\ba{l}
{\cal L}_{\mu < M_{\x,\w} \leq \cut} \,=\, 
\dis -m_t\;\ov{t^\prime}t^\prime-m_b\;\ov{b^\prime} b^\prime  \\[3mm]
\hspace*{5mm} 
\dis
-\f{1}{\sq2}\[g_tc_L^ts_R^t\;\ov{t^\prime} t^\prime h_t^0 +
             g_bc_L^bs_R^b\;\ov{b^\prime} b^\prime h_b^0  \] 
-\f{i}{\sq2}\[g_t\cb c_L^ts_R^t\;\ov{t^\prime}\gamma_5 t^\prime h_t^0+
              g_b\sb c_L^bs_R^b\;\ov{b^\prime}\gamma_5 b^\prime h_b^0  
\]A^0  \\[4mm]
\hspace*{5mm} 
\dis
-\;\ov{b^\prime}\[g_t\cb c_L^bs_R^tP_R+g_b\sb s_R^bc_L^tP_L\]t^\prime H^- 
+{\rm h.c.} 
-\[\Delta\widetilde{T}_t\;h_t^0 +\Delta\widetilde{T}_b\;h_b^0\]  \\[4mm]
\hspace*{5mm}
\dis
-\dis\half\[M_{22}^2{h_t^0}^2 +
            M_{11}^2{h_b^0}^2 +
            2\xi M_{12}^2h_t^0h_b^0 +
            M_A^2 {A^0}^2 + 2M_{H^\pm}^2 H^+H^- \]
-V_{\rm int}(h_t^0,h_b^0,A^0,H^\pm) \,,
\ea
\eeq
where 
$(g_t,\,g_b) = (h_1/{Z_{h_t}^{1/2}},\,h_1/{Z_{h_b}^{1/2}})$,
$(\sb ,\,\cb )= (\sin\!\beta,\,\cos\!\beta)$, 
$P_{L,R}=(1\mp\gamma_5)/2$,
and the unitary gauge is chosen so that 
only the physical Higgs scalars $(h_t^0,h_b^0,A^0,H^\pm)$
are relevant. Here, $\Delta\widetilde{T}_t$ and $\Delta\widetilde{T}_b$
are the tadpole terms which we used to derive the gap equations
(\ref{eq:gaptb}) above.
The Higgs mass terms are computed up to $\O(\xi)$ and are expressed
as,
\beq
\label{eq:Mdef}
\ba{lll}
M_{22}^2 = M_{22,0}^2 + \xi\delta M_{22}^2\,,~~ &
M_{11}^2 = M_{11,0}^2 + \xi\delta M_{11}^2\,,~~ &
\xi M_{12}^2 = \xi\delta M_{12}^2\,,\\[4mm]
M_{H^\pm}^2 = M_{\pm,0}^2 + \xi\delta M_{\pm}^2\,,~~ &
M_A^2 = \xi\delta M_A^2 \,, &
\ea 
\eeq
where the leading $O(\xi^0)$ contributions are,
\beq
\label{eq:LO}
\ba{l}
\dis
\hspace*{-1.5mm}
M_{22,0}^2
\!= M_h^2[{\rm Eq.}(\ref{eq:Mh})]\left|_{(s,c)_{L,R}\to (s^t,c^t)_{L,R}}\,,~~
\right.
M_{11,0}^2
\!= M_h^2[{\rm Eq.}(\ref{eq:Mh})]\left|_{(s,c)_{L,R}\to (s^t,c^t)_{L,R}}
^{(m_t,\MX)\to (m_b,\MW)}\,,~~  \right. 
\\[5mm]
\hspace*{-1.5mm}
\dis
M_{\pm,0}^2 \; \!=\dis\f{\k}{\k_c}
\[\cut^2\(Z_{h_t}^{-1}\cb^2+Z_{h_b}^{-1}\sb^2\)\(\f{\k_c}{\k}-1\) 
+\(a_{11}m_t^2+a_{12}\)\ln\(\f{\cut^2}{m_t^2}+1\)  \right.           \\[6mm]
\hspace*{-1.5mm}
\hspace*{23mm}  \dis  \left.
+\(\ov{a}_{21}\MX^2+a_{22}\)\ln\(\f{\cut^2}{\MX^2}+1\)
+\(\ov{a}_{31}\MX^2+a_{32}\)\ln\(\f{\cut^2}{\MW^2}+1\) \] ,  \\[6mm]
\hspace*{-1.5mm}
\(\!\!
\ba{l}
a_{11}\\[2mm]
a_{21}\\[2mm]
a_{31}\\[2mm]
a_{41}
\ea
\!\!\)  \!=\! \dis\f{\cb^2}{Z_{h_t}}\!\(\!\!\ba{l} 
                                (s_R^tc_L^b)^2 \\[2mm] 
                                (c_R^tc_L^b)^2 \\[2mm]
                                (s_R^ts_L^b)^2 \\[2mm]
                                (c_R^tc_L^b)^2
                         \ea\!\!\) \!\!+\!
            \f{\sb^2}{Z_{h_b}}\!\(\!\!\ba{l}
                                (c_L^ts_R^b)^2 \\[2mm] 
                                (s_L^ts_R^b)^2 \\[2mm] 
                                (c_L^tc_R^b)^2 \\[2mm] 
                                (s_L^tc_R^b)^2          
                          \ea\!\!\)\!,~~~
\(\!\!\ba{l}
a_{12}\\[2mm] 
a_{22}\\[2mm] 
a_{32}\\[2mm] 
a_{42}
\ea \!\!\)  \!=\! \f{2\sb\cb}{{Z_{h_t}^{\half}Z_{h_b}^{\half}}}\!\!
\(\!\!  \ba{l}
m_tm_b  c_L^ts_R^tc_L^bs_R^b  \\[2mm] 
\MX m_b s_L^tc_R^tc_L^bs_R^b  \\[2mm] 
\MW m_t s_L^tc_R^tc_L^bs_R^b  \\[2mm]
\MX\MW  s_L^tc_R^ts_L^bc_R^b
\ea\!\!\)\!,
\ea
\eeq
with $\,\ov{a}_{21}=a_{21}+(a_{41}\MX^2+a_{42})/(\MX^2-\MW^2),\,$
and  $\,\ov{a}_{31}=a_{31}+(a_{41}\MW^2+a_{42})/(\MW^2-\MX^2).\,$
The axionic pseudo-scalar $A^0$ is massless at this order due to
the Peccei-Quinn symmetry (\ref{eq:U1PQ}).   
One recovers a simple and intuitive picture under the
approximate limit $(r_t,\,r_b)\ll 1$, i.e.,
\beq
M_{22,0}\approx  2\mtx\,,~~~
M_{11,0}\approx  2\mbw\,,~~~
M_{\pm,0}\approx \sq2 \(\mtx^2+\mbw^2\)^{\half}\,,
\eeq
which are all controlled by the dynamical mass gaps $(\mtx,\,\mbw)$
and become equal in the special case of $\tanb=1$,
as expected.
These approximate formulae agree  with our independent
Higgs potential analysis in Appendix-B.

For the $\O(\xi)$ corrections, we first perform a careful calculation of
the $A^0$ mass, and obtain,
\bea
\label{eq:MA}
M^2_A \!&=&\! \xi\delta M_A^2
\,=\, \frac{2 \xi \Lambda^2}{\dis\sin\!2\beta \sqrt{Z_{h_t}Z_{h_b}}}
\,+\, \O(\xi^2) \,.  
\eea
It is remarkable to notice that the Peccei-Quinn breaking mass
$M_A$ is proportional to $\sqrt{\xi}\cut$ instead of being controlled
by the dynamical mass gaps  $(\mtx,\,\mbw)$. 
As noted above, the essential difference between
$A^0$ and the other Higgs scalars is that $A^0$ is a
massless Goldstone boson at $\O(\xi^0)$ 
and its non-vanishing mass comes from the explicit
Peccei-Quinn breaking $\xi$-term. Hence, it is natural to see
that $M_A$ is not controlled
by the dynamical gaps  $(\mtx,\,\mbw)$, but instead 
scales like\footnote{We have confirmed
Eq.\,(\ref{eq:MA}) by using an independent Higgs potential analysis
(cf. Appendix-B).}\,  $\sqrt{\xi}\cut$\,.  
This results in the
$A^0$ being relatively heavier than naively expected, provided
$\xi \gtrsim 10^{-3}$. 
Such an $\O(\xi\cut^2)$ correction also shows up in other 
Higgs mass formulae at $\O(\xi)$ and is thus a generic
feature of the explicit Peccei-Quinn breaking.  
So, we can express the leading  $\xi$-corrections
to $(h_t^0,\,h_b^0,\,H^\pm)$ masses
in terms of $M_A^2$ while the rest of the $\xi$-terms are of 
$\O(\xi\mtx^2,\,\xi\mbw^2)$ and thus much less significant. 
With this in mind, we compactly summarize the $\O(\xi)$
masses of  $(h_t^0,\,h_b^0,\,H^\pm)$ as, 
\beq
\ba{l}
\xi\delta M_{11}^2 \simeq \sb^2 M_A^2\,,~~~~
\xi\delta M_{22}^2 \simeq \cb^2 M_A^2\,,~~~~
\xi\delta M_{12}^2 \simeq -\sb\cb M_A^2 + 4\xi \(\mtx^2+\mbw^2\)\,,\\[3mm]
\xi\delta M_{\pm}^2 \simeq M_A^2 -\xi\(\cot\!\beta\;\mtx^2+\tanb\;\mbw^2\) \,,
\ea
\eeq
which can be most easily extracted from the Higgs potential analysis
shown in Appendix-B.   
Due to the mixing mass term between $h_t^0$ and $h_b^0$, we
diagonalize them into the  mass-eigenstates $(h^0,\,H^0)$
with the physical masses,
\beq
M_{h,H}^2 =\dis\half\[M_{11}^2+M_{22}^2\pm
           \sqrt{\(M_{11}^2-M_{22}^2\)^2+4M_{12}^4}\;\]\,.
\eeq
The corresponding rotation angle 
$\alpha\in [-\pi/2,\,  0]$ is determined by 
$\tan (2\alpha) = 2M_{12}^2/(M_{11}^2-M_{22}^2)$.\,

\begin{figure}[H]
\begin{center}
%\vspace*{-0.7cm}
%\hspace*{-6mm}
\includegraphics[width=18cm,height=18cm]{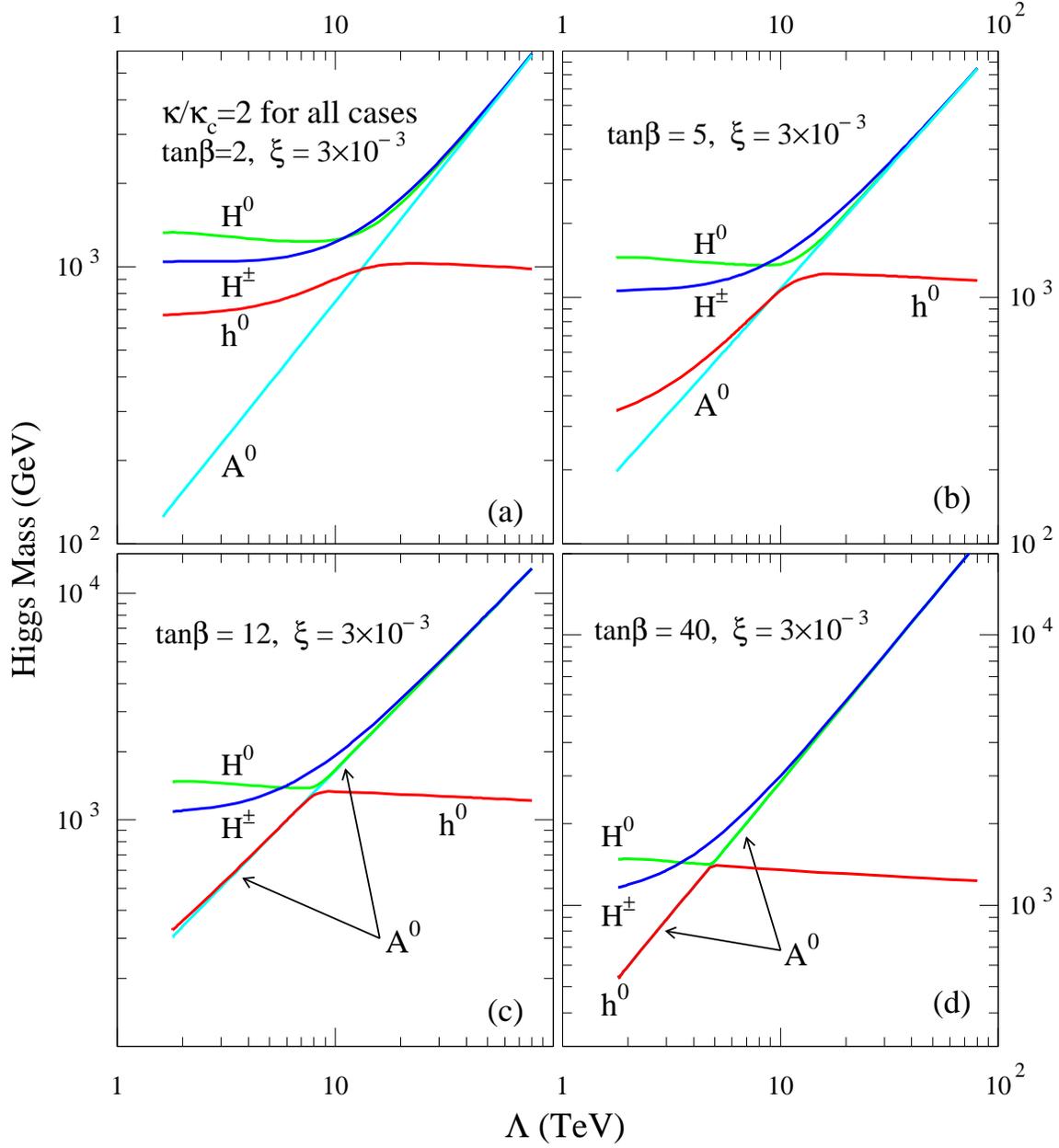}
\end{center}
\vspace*{-5mm}
\caption{
The mass spectrum of the Higgs bosons are plotted for 
$\kappa / \kappa_c = 2$ and $\tanb \in (2,\, 5,\, 12,\, 40)$
in the extended model with bottom seesaw, where
the parameter $\xi$ is representatively chosen as
$\xi = 3 \times 10^{-3}$
}
\label{fig:higgsmasses}
\end{figure}

Based upon these, we can finally analyze the 
Higgs mass spectrum of this model using the physical solutions to
the seesaw gap equations derived in the previous subsection.
We present our numerical results in  Fig.~\ref{fig:higgsmasses},
where we choose
$\kappa / \kappa_c = 2$ and a wide range of $\tanb$ values.
The Peccei-Quinn breaking parameter $\xi$ is set to a representative
value of $\xi =3\times 10^{-3}$ for all plots. 
The proportionality of $M_A$ with $\cut$ can be clearly seen,
and as $M_A$ moves above $1$\,TeV, the Higgs bosons $(H^0,\,H^\pm)$
becomes much more degenerate with $A^0$, while the lightest neutral
Higgs $h^0$ remains around $1$\,TeV, saturating the SM unitarity
bound. This is a quite generic feature of this model unless the 
parameter $\xi$ is much smaller, around $10^{-4}$ or below, which
is unlikely in the Topcolor instanton picture. Also, too small
$\xi ~(\lesssim 10^{-4}-10^{-5})$ will have more significant 
mass-splittings among Higgs bosons $(A^0,\,H^0,\,H^\pm)$ which cause 
large weak-isospin violation in the oblique parameter $T$ (besides
resulting in a very light axion $A^0$). This is  disfavored 
from the experimental viewpoint. Thus, our analysis favors a relatively
heavy axion $A^0$ (together with other Higgs scalars)
and the Topcolor instanton\,\cite{TCATC}
interpretation of the Peccei-Quinn breaking for this model.

%\clearpage

\newpage
\section{\hspace*{-2mm}{\large
Constraints from Precision Observables
}}

After quantitative analyses on the vacuum structure
and composite Higgs spectrum in the dynamical Top Seesaw models,
we proceed to systematically study their experimental constraints
from the electroweak precision data.
The most important bounds come from the radiative corrections
to the oblique parameters $T$ and $S$ \cite{STU} and 
also the corrections to the $\Zbb$ vertex induced by 
the $b$-$\omega$ mixing in the bottom seesaw sector.
It is remarkable that the minimal Top Seesaw model, 
having a typical heavy composite Higgs boson around
$1$\,TeV, is non-trivially compatible with the $S-T$ bounds, 
due to the conspiracy from the large positive seesaw correction
to the $T$ parameter. The case for the extended model
with bottom seesaw is more complex because of the $b$-$\w$ mixing 
and the two Higgs doublets.  In this
extended model the precision $T$ bound requires a certain
degeneracy in the mass spectrum of 
the Higgs scalars and thus favors
a relatively heavy axion $A^0$.  As we will show, with
the Topcolor instanton interpretation of the Peccei-Quinn breaking, the
resulting precision bounds on the heavy $\x$ and $\w$ masses
are at the similar level to that of the minimal Top seesaw model.

\subsection{\hspace*{-2mm}{\normalsize
In the Minimal Top Seesaw Model
}}

\begin{figure}[H]
\begin{center}
%\vspace*{-1.5cm}
\includegraphics[width=17cm,height=17cm]{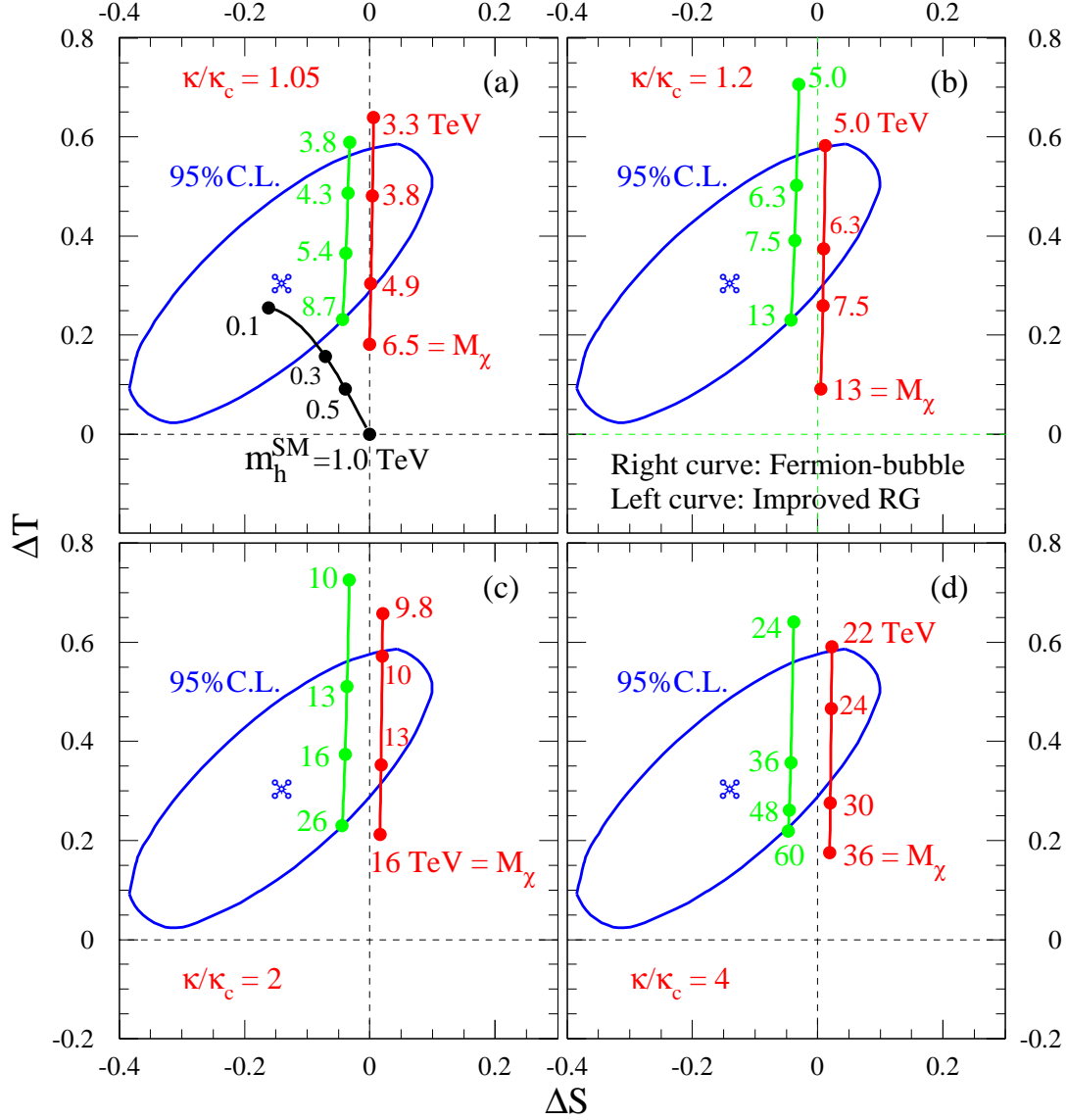}
%fig_tss_stmx.eps}
\end{center}
\vspace*{-7mm}
\caption{
Top seesaw contributions to $\Delta S$ and $\Delta T$ are
compared with the 95\% C.L. error ellipse (with $m_h^{\rm ref} = 1$ TeV)
for $\k/\k_c=1.05,\,1.2,\,2,\,4$.
The $\Delta S$-$\Delta T$ trajectories (including both Higgs and
$\chi$ contributions) are shown as a function of $\MX$.  
In each plot, the curve on the right is derived from
the large-$N_c$ fermion bubble calculation, and as a 
comparison, the curve on the left is deduced by an 
improved RG approach (cf. Sec.\,2.6). 
For reference, the SM Higgs corrections to ($S$,\,$T$), relative to
$m_h^{\rm ref} =1$\,TeV, are given for $m_h^{\rm SM}$ varying from 
100\,GeV up to 1.0\,TeV in plot (a).
}
\label{fig:tssSTmHmx}
\end{figure}

The minimal Top seesaw model has a single composite Higgs boson 
in addition to the singlet seesaw quark $\x$ in the spectrum.
As we have shown in Fig.\,\ref{fig:Mhnew}, this composite Higgs 
scalar has its mass typically around $\sim \!1$\,TeV.
Its contributions to the oblique $S$ and $T$ parameters can be expressed
as,
\beq
\ba{lll}
\Delta S_H & = & \dis +\frac{1}{12 \pi}
\ln \left( \frac{M^2_h}{m_{h,{\rm ref}}^{2}} \right) , \nonumber \\[5mm]
\Delta T_H & = & \dis -\frac{3}{16 \pi \cos^2 \theta_W} 
\ln \left( \frac{M^2_h}{m_{h,{\rm ref}}^2} \right) ,
\ea
\eeq
where $m_{h,{\rm ref}}$ is the reference point of the SM Higgs mass.
Since in the pure SM the current precision data\,\cite{data} favors a 
light Higgs mass around $100$\,GeV, we see that a heavy Higgs scalar with
a $\sim\!1$\,TeV mass will drive  $\Delta T$ in the negative
direction relative to a light SM Higgs and thus is excluded
by the current precision $S-T$ contour shown in 
Fig.\,\ref{fig:tssSTmHmx}(a)\footnote{Our current 
$S-T$ contours are derived using the recent precision data\,\cite{data}
and the global fitting package GAPP\,\cite{GAPP}.}.  
However, the top seesaw sector
has generic weak-isospin violation from the   
$t$-$\chi$  mixing  which will significantly contribute to $\Delta T$ 
in the positive direction, as can be seen from 
the formula,
\bea
\label{eq:ssT}
\Delta T & = & \frac{N_c}{16 \pi^2 v^2 \alpha} \left[
s_L^4 M^2_\chi - s_L^2 (1 + c_L^2) m_t^2 +
2 s_L^2 c_L^2 \frac{M_\chi^2 m_t^2}{M_\chi^2 - m_t^2} \ln 
\frac{M_\chi^2}{m_t^2} 
\right] \nonumber \\[2.5mm]
& = &
\frac{N_c m_t^2}{16 \pi^2 v^2 \alpha}
\left[ 
2\(\ln\frac{\MX^2}{m_t^2} -1\) + \f{1}{r_t}
\right]\f{(m_t/\muxt )^2}{1+r_t}  +\O\(\f{m_t^4}{\muxt^4}\) \,,
\eea
in which $r_t=(\muxt/\muxx)^2\lesssim 1$. Here, 
we have subtracted out the usual SM top contribution
as it was already included in the precision fit.
The expanded formula indeed shows a sizable $\Delta T > 0$;
it also exhibits the decoupling nature of
the vector-like heavy quark $\x$, since the large mass parameters go 
with negative powers (for fixed ratio $r_t$).\footnote{Similar 
features of large $\Delta T>0$ and the decoupling of heavy seesaw
masses were also found in top seesaw models with vector-like
weak doublet seesaw quarks\,\cite{HTY}.}

Next, we compute the 
$\chi - t$ contribution to $\Delta S$, and obtain,
\bea
\label{eq:ssS}
\Delta S & = & \frac{N_c}{36 \pi} s_L^2
\left\{
44 \( \f{\MX^2}{m_z^2} - \f{m_t^2}{m_z^2} \) 
- 2 \ln\frac{M_\chi^2}{m_t^2} 
- 18 c_L^2 G_1 \!\( \frac{m_t^2}{m_z^2}, \frac{M^2_\chi}{m_z^2} \)
\right.
\nonumber \\[2.5mm]
&  &  \hspace*{15.5mm}
\left.
- \(11 \f{m_t^2}{m_z^2}    + 1\)  F_1 \!\left( \frac{m^2_t}{m_z^2}    \right) 
+ \(11 \f{M_\chi^2}{m_z^2} + 1\)  F_1 \!\left( \frac{M^2_\chi}{m_z^2} \right) 
\right\}
\nonumber \\[2.5mm]
& = & 
\frac{N_c}{9\pi }
\[ \dis\ln\f{\MX^2}{m_t^2}-\f{5}{2} +\f{m_z^2}{20m_t^2}
\]\f{(m_t/\muxt )^2}{1+r_t} +\O\(\f{m_t^4}{\muxt^4}\) \,,
\eea
where $m_z$ is the mass of weak gauge boson $Z^0$, and the
relevant functions $F_1(y)$ and $G_1(y_1,y_2)$  are defined as,
\bea
F_1(y) &=& -4 \sqrt{ 4 y - 1 } \: {\rm arctan} 
\frac{1}{\sqrt{ 4 y - 1 } }   \,, \\[2mm]
G_1(y_1,y_2) &=& \frac{ 5 (y_1^2 + y_2^2) - 2 y_1 y_2}{9 ( y_1 - y_2 )^2}
+ \frac{3 y_1 y_2(y_1+y_2) - y_1^3 - y_2^3}{3 (y_1-y_2)^3}
\ln\frac{y_1}{y_2}  \,.
\eea
Now, Keeping the dominant leading logarithmic terms in the above
expanded formulae, we can directly estimate the
relative size of $\Delta S$ versus $\Delta T$,
\beq
\label{eq:STratio}
\dis
\f{\Delta S}{\Delta T} \approx \f{16\pi \alpha}{9} \approx 0.04 \ll 1\,,
\eeq
which shows that $\Delta S$ is only about $4\%$ of $\Delta T$ and thus
negligible in comparison with the typical values of $\Delta T > 0 $, as 
advertised earlier in the introduction.

\begin{figure}[H]
\begin{center}
%\vspace*{-1.5cm}
\includegraphics[width=17cm,height=17cm]{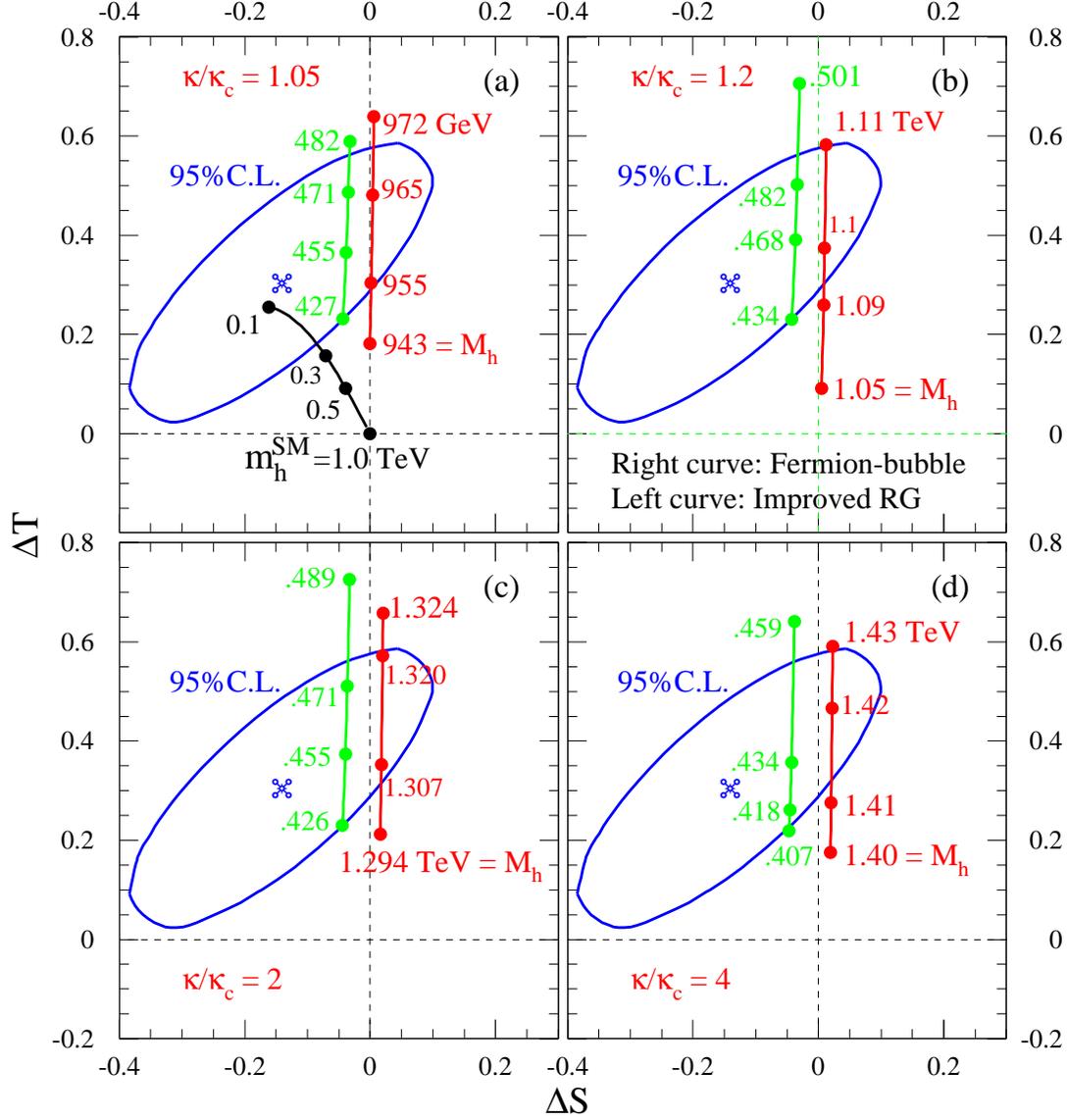}
\end{center}
\vspace*{-7mm}
\caption{
Same as Fig.\,\ref{fig:tssSTmHmx}, 
but with the corresponding $M_h$ values
marked on the  $\Delta S$-$\Delta T$ trajectories instead.
It shows that each trajectory is very insensitive to $M_h$
and the large increase along the positive $\Delta T$ direction is
due to the top seesaw contribution as $\MX$ decreases
(cf. Fig.\,\ref{fig:tssSTmHmx}). 
In each plot, the shift of the left trajectory relative to
the right one is due to the smaller $M_h$ values
(estimated around $400-500$\,GeV), but $  \Delta S\ll \Delta T$ generally
holds so that $\Delta S$ is much less significant.
}
\label{fig:tssSTmHadd}
\end{figure}

In Fig.\,\ref{fig:tssSTmHmx}, 
we assemble the complete $\Delta S$ and $\Delta T$
contributions from the minimal top seesaw model, including both the
corrections from the composite Higgs boson and the seesaw quarks, 
and compare them with
the 95\%\,C.L. contour for $\Delta S-\Delta T$.  
Each figure corresponds
to a different choice of $\kappa / \kappa_c$, and shows the trajectory
in the $\Delta S-\Delta T$ plane as the $\chi$ mass varies.  
As a comparison, we have plotted the results based on both
the large-$N_c$ fermion-bubble calculation 
and the improved RG approach (cf. Sec.\,2.6). 
For the relevant parameter space here, the improved RG approach gives
lower Higgs mass values (around $400-500$\,GeV) so that the
curves are slightly shifted towards the upper left. 
As a consequence, in the improved RG approach
the upper bound on $\MX$ is more relaxed
for $\k/\k_c \gtrsim 1.2$, while the lower bound on $\MX$ remains
at a similar level.  
The figure clearly illustrates that the top seesaw model can be consistent
with the electroweak precision data provided $M_\chi$ is in the appropriate
mass range.
For instance, when the Topcolor force is slightly super-critical, we see that
precision data are effectively probing $M_\chi \sim 4$\,TeV.
A high luminosity Linear Collider at GigaZ
can further improve these indirect precision constraints on the
Top seesaw dynamics with a much smaller $\Delta S-\Delta T$
error ellipse\,\cite{GZ}.
Finally, in Fig.\,\ref{fig:tssSTmHadd}, we display the same 
$\Delta S-\Delta T$ trajectories as in Fig.\,\ref{fig:tssSTmHmx}, 
but with the corresponding Higgs mass ($M_h$) values marked. 
We see that as each trajectory moves up along the $\Delta T$ direction,
the $M_h$ value changes very little and thus the rise of
$\Delta T$ is really due to the decrease of $\MX$ (as marked in 
Fig.\,\ref{fig:tssSTmHmx}). The Fig.\,\ref{fig:tssSTmHadd} further
shows that the relevant Higgs mass is about  $1-1.4$\,TeV in the
large-$N_c$ fermion-bubble calculation and 
$400-500$\,GeV in the improved RG approach. As we explained
in Sec.\,2.6, the large-$N_c$ fermion-bubble calculation
may over-estimate $M_h$ due to the ignorance of non-large-$N_c$
effects of the Higgs propogation in the loop, while the improved
RG approach may under-estimate $M_h$ due to the sizable 
uncertainties associated with the compositeness condition 
at the scale $\cut\sim 10^{4-5}$\,GeV and the use of 
simple mass-independent renormalization in such low cutoff theories.
So, the two approaches are complementary and the real $M_h$
values should lie between these two estimates. 
Actually, the shift between the two trajectories along the $\Delta S$
direction is mainly due to the effect of the Higgs mass.
Thus, taking into account our ignorance of the detailed dynamics
around the scale $\cut$ and above,
we may view the region between the two trajectories inside the  
$\Delta S-\Delta T$ ellips as the viable parameter space allowed
by the precision data.

\subsection{\hspace*{-2mm}{\normalsize
In the Extended Model with Bottom Seesaw
}}

The inclusion of a bottom seesaw generates additional $b$-$\w$ mixing
which have nontrivial contributions to the $S$ and $T$ parameters
and also to the $\Zbb$ vertex. Furthermore, the composite Higgs
sector now contains two doublets and thus
has additional corrections to the precision observables.
We start by calculating the complete set of loop diagrams
[including the mass-eigenstate seesaw quarks 
$(t^\prime,\,b^\prime,\,\x^\prime,\,\w^\prime)$] 
that contribute to the $S$ and $T$ parameters.
The general results can be summarized below, 
\beq
\label{eq:S2exact}
\ba{lll}
\Delta S &=&\dis 
\frac{N_c}{36 \pi} \left\{ s^{t \, 2}_L  
\left[
44 \(\f{\MX^2}{m_z^2} - \f{m_t^2}{m_z^2}\) - 2 \ln\f{M_\chi^2}{m_t^2}
                      - 18 c_L^{t \, 2}
            G_1\!\(\f{m_t^2}{m_z^2}, \frac{M^2_\chi}{m_z^2}\)
\right.\right.  
\\[6mm]
& & \hspace*{23mm}\dis
\left.
-\(11 \f{m_t^2}{m_z^2} + 1\) F_1\!\(\f{m^2_t}{m_z^2}\) 
+\(11 \f{\MX^2}{m_z^2} + 1\) F_1\!\(\f{M^2_\chi}{m_z^2}\) \]
\\[6mm]
& &  \hspace*{8.5mm}  \dis
+ s^{b \, 2}_L \[
28 \(\f{\MW^2}{m_z^2} - \f{m_b^2}{m_z^2}\) 
+2 \ln\f{M_\omega^2}{m_b^2} 
- 18 c_L^{b \, 2} G_1\!\(\f{m_b^2}{m_z^2}, \frac{M^2_\omega}{m_z^2} \)
\right.
\\[6mm]
& &  \left.\left.  \hspace*{23mm}  \dis
- \(7 \f{m_b^2}{m_z^2} - 1\) F_2 \!\( \f{m^2_b}{m_z^2} \) 
+ \(7 \f{M_\omega^2}{m_z^2} - 1\) F_1 \!\( \f{M^2_\omega}{m_z^2} \) 
\right] \right\} ,
\ea
\eeq
\beq
\label{eq:T2exact}
\ba{lll}
\Delta T & = & 
\dis \frac{N_c}{16 \pi s^2_w c^2_w m_z^2} 
\left[
 s^{t \, 2}_L c^{b \, 2}_L F_3\!\(M^2_\x, m^2_b\)
+c^{t \, 2}_L s^{b \, 2}_L F_3\!\(M^2_\omega, m^2_t\)
+s^{t \, 2}_L s^{b \, 2}_L F_3\!\(M^2_\chi, M^2_\omega\) 
\right.
\\[5mm]
& &
\left. \hspace*{20mm}
- s^{t \, 2}_L   c^{t \, 2}_L       F_3\!\(M^2_\chi, m^2_t\)
- s^{b \, 2}_L   c^{b \, 2}_L       F_3\!\(M^2_\omega, m^2_b\)
+ \(c^{t \, 2}_L c^{b \, 2}_L - 1\) F_3\!\(m^2_t, m^2_b\)
\right]  ,
\\[3mm]
\ea
\eeq
where
the functions $F_2(x)$ and $F_3(x_1, x_2)$ are given by,
\bea
F_2(x)           & = & 
\sqrt{1 - 4 x} \ln 
\frac{1 - 2 x - \sqrt{1 - 4 x}}{1 - 2 x + \sqrt{1 - 4 x}}  \,,
\\[2mm]
F_3(x_1, x_2)    & = & 
x_1 + x_2 
- \frac{2 x_1 x_2}{x_1 - x_2} \ln \frac{x_1}{x_2}  \,,
\eea
where $(s_w,\,c_w)=(\sin\!\theta_W,\, \cos\theta_W)$ and $\theta_W$
is the weak angle. 
The above general formulae contain exact seesaw rotation angles
and heavy masses $(\MX,\MW)$ in various places. So,
it is instructive to derive the expanded expressions
in which all large masses exhibit the expected decoupling nature
and the sign of these corrections will become clear.
Thus, we deduce,
\bea
\Delta S  &=&
\frac{N_c}{9\pi }
\[ 
\dis\ln\f{\MX^2}{m_t^2}-\f{5}{2} +\f{m_z^2}{20m_t^2}
\]\f{(m_t/\muxt )^2}{1+r_t}  \nonumber  \\[2mm]
& &
~ +
\dis\[\ln\f{\MW^2}{m_z^2}+\f{7}{3} +9\f{m_b^2}{m_z^2}\ln\f{m_z^2}{m_b^2}
\]\f{(m_b/\muwb )^2}{1+r_b} 
+\O\(\f{m_t^4}{\muxt^4}, \f{m_b^4}{\muwb^4}
   \) \,,
\label{eq:S2app}
\\[4mm]
\Delta T & = &
\f{N_c}{16\pi^2 v^2\alpha}
\left\{
\[2\(\ln\f{\MX^2}{m_t^2}-1\) +\f{1}{r_\x r_t(1+r_t)}
\]\f{(m_t/\muxt)^2}{1+r_t} \right. 
\nonumber 
\\[2mm]
 & &  \hspace*{15.6mm}
-  \left.
\[2\ln\f{\MW^2}{m_t^2} +\f{1}{r_\x r_t(1+r_t)}
\]\f{(m_b/\muwb)^2}{1+r_b}
+\O\(\f{m_t^4}{\muxt^4}, \f{m_b^4}{\muwb^4}, \f{m_t^2m_b^2}{\muxt^2m_z^2}\)
\right\}
\nonumber 
\\[2mm]
& \approx  &
\f{N_c}{16\pi^2 v^2\alpha}
\left\{
\(s_L^{t\,2}-s_L^{b\,2} \)
\[ 2\ln\f{\MX^2}{m_t^2} + \f{1}{r_\x r_t(1+r_t)}
\]  -2s_L^{t\,2}
\right\}  ,
\label{eq:T2app}
\eea
where 
\beq
\label{eq:T2add}
\ba{l}
\dis
r_t\equiv\(\f{\muxt}{\muxx}\)^2\leq 1,~~~~
r_b\equiv\(\f{\muwb}{\muww}\)^2\leq 1,~~~~
r_\x \equiv\(\f{\muxx}{\MX}\)^2\sim 1,~~~~ 
\\[5mm]
\dis
s_L^{t\,2} = \f{(m_t/\muxt)^2}{1+r_t} + \O\(\f{m_t^4}{\muxt^4}\),~~~~
s_L^{b\,2} = \f{(m_b/\muwb)^2}{1+r_b} + \O\(\f{m_b^4}{\muwb^4}\),~~~~
\ea 
\eeq
and in the last line of Eq.\,(\ref{eq:T2app}) we have used the
relation $\MX\simeq \MW$ (cf. Fig.\,\ref{fig:tbss_soall})
to further simplify the expression. 
Now, from Eq.\,(\ref{eq:S2app}) we see that the inclusion of the $b$ seesaw
further adds positive terms to $\Delta S$ which, however, are comparable to
the first term of the $t$ sector only for small $\tanb \sim \O(1-5)$ 
where $r_b\ll r_t$ so that $\muwb \ll \muxt$ 
[cf. Fig.\,\ref{fig:tbss_soall}(a,b)]. For large $\tanb$, 
$\muwb$ becomes closer to $\muxt$ so that
$ (m_t/\muxt )^2 \gg (m_b/\muwb )^2$. Consequently,
$\Delta S$ is dominated by the $t$-seesaw sector 
and thus is very similar to  
the situation in the minimal Top seesaw model where 
$\Delta S\sim 0$ [cf. Eq.\,(\ref{eq:STratio})].

\begin{figure}[H]
\begin{center}
\vspace*{-2.5cm}
\includegraphics[width=16cm,height=15cm]{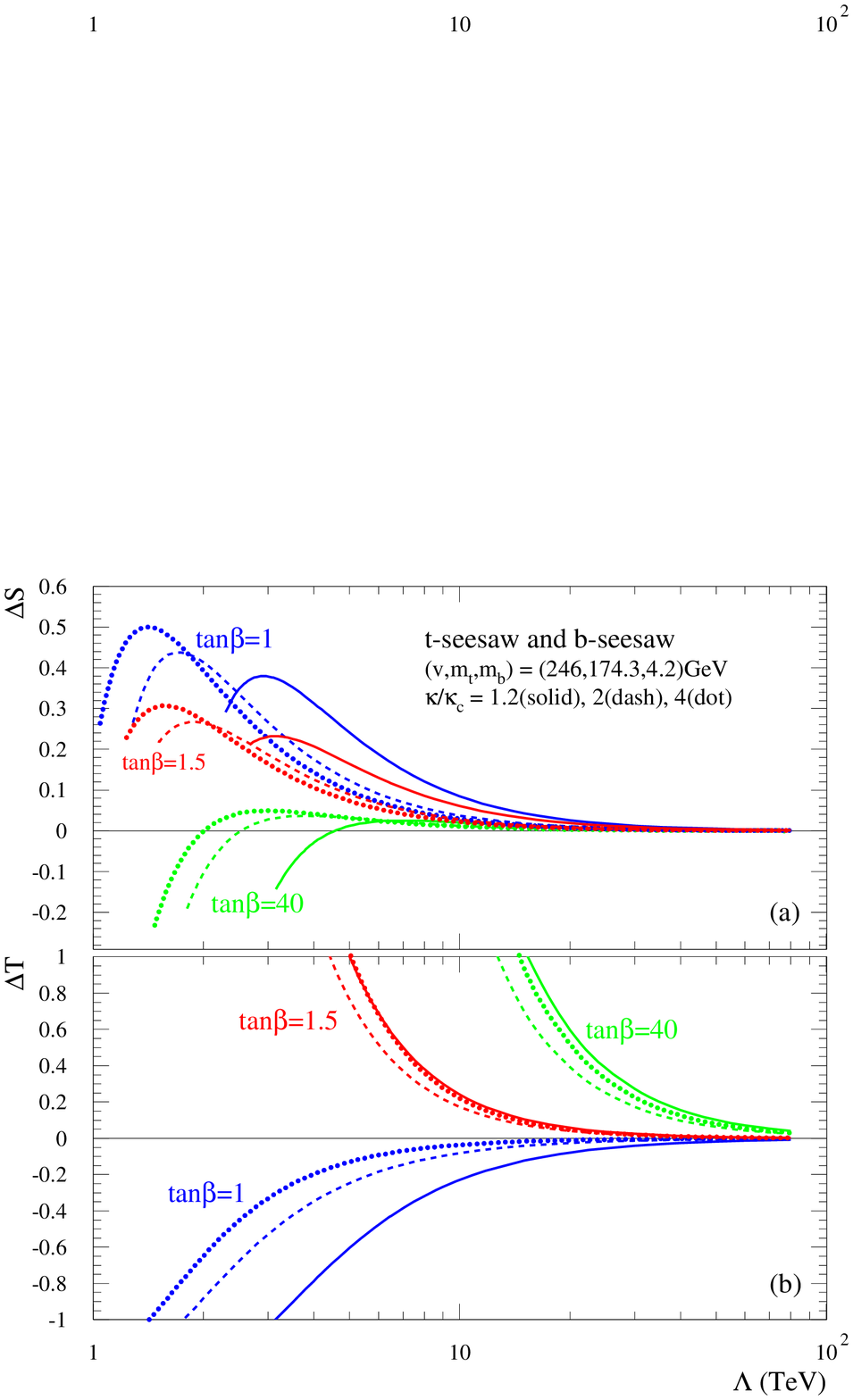}
\end{center}
\vspace*{-7mm}
\caption{
The contributions from the top and bottom seesaws
oblique parameters $S$ (upper plot) and $T$ (lower plot),
based on Eqs.\,(\ref{eq:S2exact})-(\ref{eq:T2exact})
and the physical seesaw solutions 
(cf. Fig.\,\ref{fig:tbss_soall}). 
}
\label{fig:ST2exact}
\end{figure}

With these we can understand the picture shown in 
Fig.\,\ref{fig:ST2exact}(a), based on the exact formulae 
(\ref{eq:S2exact})-(\ref{eq:T2exact})
and the physical seesaw solutions 
(cf. Fig.\,\ref{fig:tbss_soall}).
Next, we examine the more nontrivial features in $\Delta T$ as
shown in  Fig.\,\ref{fig:ST2exact}(b). 
From the last equation in the expanded formula (\ref{eq:T2app}), 
it is instructive to see that the $b$-seesaw sector adds
negative corrections which could cancel the $t$-seesaw
contributions for small $\tanb$ region where we have
$ (m_t/\muxt )^2 \sim (m_b/\muwb )^2$, i.e., $s_L^t\sim s_L^b$, 
as can be understood from
the physical seesaw solutions in  Fig.\,\ref{fig:tbss_soall}(a,b).
Intuitively, we expect that such a cancellation becomes maximal
when $\tanb\to 1$ so that the custodial $SU(2)_c$ symmetry is restored
in the seesaw sector 
aside from the $m_t$-$m_b$ mass difference 
[reflected in the last (negative) constant term on the R.H.S.
of Eq.\,(\ref{eq:T2app})].   This is why we see $\Delta T<0$ for $\tanb=1$
in Fig.\,\ref{fig:ST2exact}(b). However, the $b$-seesaw contribution
in Eq.\,(\ref{eq:T2app}) quickly decreases since $s_L^b$ drops off 
as $\tanb$ moves up,  and when
$\tanb\gtrsim 1.5$ we see that the seesaw contributions become
significantly positive again and $\Delta T$ approaches the values
in the minimal Top seesaw model for $\tanb\gtrsim 40$ where
$r_b\simeq r_t$ ($\muwb\simeq \muxt$) as shown in 
Fig.\,\ref{fig:tbss_soall}(a,b)
so that $ (m_b/\muwb )^2 \ll  (m_t/\muxt )^2 $, making
$b$-seesaw term in $\Delta T$ negligible. 
In summary, for $1.5\lesssim \tanb \lesssim 40$, we still have
sizable positive $\Delta T>0$,
but in the moderate to small $\tanb$ regions $\Delta T$ becomes
smaller than that of the minimal model and thus would help
to weaken the strong constraints from $T$ and lower the bounds on
$(\x,\w)$ masses.
However, the additional positive contributions from the 
two-Higgs-doublet sector
in the extended model tend to shift up $\Delta T$ somewhat, 
this non-trivially renders our final bounds on $M_{\x,\w}$ quite
similar to the situation in the minimal Top seesaw model,
as will be studied below.
\\

Now, we turn to analyze the oblique corrections from the
composite two-doublet-Higgs sector. Since we have derived 
the Higgs mass spectrum in Sec.\,3.3
(cf. Fig.\,\ref{fig:higgsmasses}), 
we can readily compute the corresponding
oblique corrections in our model by using the analytical
formulae below\,\cite{ST2HDM,HPS},
\bea
S_H &=& \frac{1}{12 \pi} \left\{ \cos^2 (\beta-\alpha) \left[
 \ln \frac{M_H^2}{M_h^2}     + G_2(M_h^2, M_A^2) 
-\ln \frac{M^2_{H^\pm}}{M_h M_A}  \right] \right. 
\nonumber \\
& & \left. \: \hspace*{6.5mm}
+ \sin^2 (\beta- \alpha)  \left[ G_2(M_H^2, M_A^2)
- \ln  \frac{M^2_{H^\pm}}{M_H M_A}  \right]
\right\} ,
\\
T_H &=& 
\frac{1}{32 \pi^2 v^2 \alpha} \left\{
\cos^2 (\beta- \alpha) \left[ F_3( M^2_{H^\pm}, M^2_h) 
+ F_3( M^2_{H^\pm}, M_A^2) - F_3( M_A^2, M_h^2 ) \right] 
\right. 
\nonumber \\
& &  \hspace*{14mm}
\left. \: + \sin^2 (\beta- \alpha) \left[ F_3( M^2_{H^\pm}, M^2_H) 
+ F_3( M^2_{H^\pm}, M_A^2) - F_3( M_A^2, M_H^2 ) 
\right]
\right\} ,
\eea
where $(M_h,\,M_H,\,M_A,\,M_{H^\pm})$ are masses of the 
neutral and charged physical Higgs scalars 
$(h^0,\,H^0,\,A^0,\,H^\pm)$ and $\alpha$ is the neutral
Higgs mixing angle (cf. Sec.\,3.3).
The function $G_2(x_1, x_2)$ is given by,
\bea
G_2(x_1, x_2) &=& -\frac{5}{6} + \frac{2 x_1 x_2}{(x_1-x_2)^2}
+ \frac{(x_1+x_2)(x_1^2 - 4 x_1 x_2 + x_2^2)}{2(x_1-x_2)^3}
\ln \frac{x_1}{x_2} \, .
\eea
The above formulae are valid for
$M_{\rm Higgs}^2 \gg m_z^2$ and are well justified for our
model (cf. Fig.\,\ref{fig:higgsmasses}). In the numerical
analysis we have also used more general $S-T$ formulae in
Refs.\,\cite{ST2HDM,HPS} as a consistency check. 
Since $F_3(x_1,x_2)\to 0$ as $x_1 \to x_2$, 
we see that $T_H$ could be much suppressed as long as 
the masses of $(H^0,\,A^0,\,H^\pm)$ have good degeneracy.

\begin{figure}[H]
\begin{center}
%\vspace*{-1.5cm}
\includegraphics[width=17cm,height=18cm]{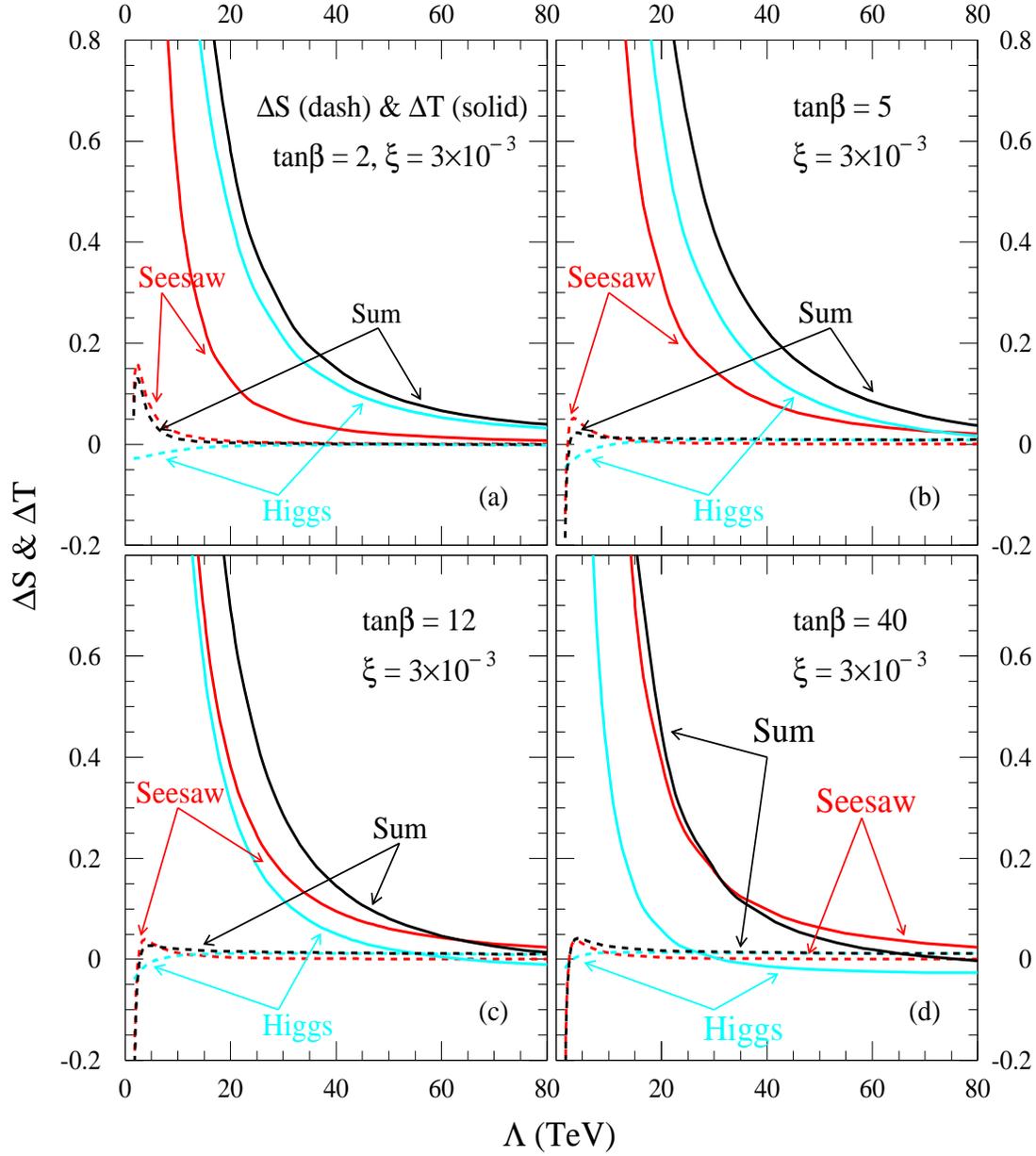}
\end{center}
\vspace*{-7mm}
\caption{
The contributions from the
seesaw sector and the two-doublet-Higgs sector to the 
oblique parameters $T$ (solid curves) and $S$ (dashed curves). 
%\\
}
\label{fig:stk2-list}
\end{figure}

\begin{figure}[H]
\begin{center}
%\vspace*{-1.5cm}
\includegraphics[width=17cm,height=18cm]{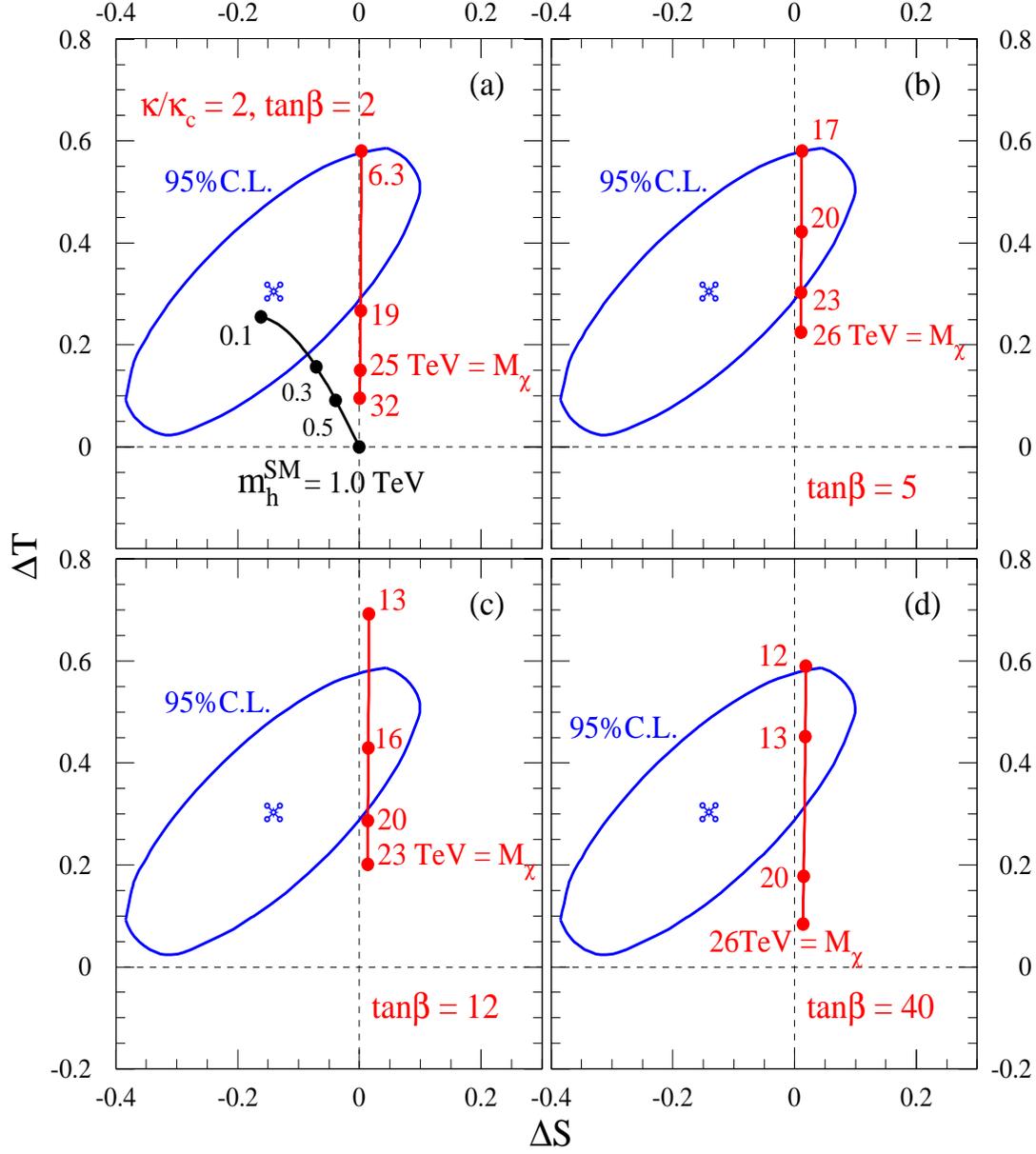}
\end{center}
\vspace*{-7mm}
\caption{
Top and bottom seesaw contributions to $\Delta S$ and $\Delta T$ are
compared with the 95\% C.L. error ellipse (with $m_h^{\rm ref} = 1$ TeV)
for $\k/\k_c=2$ and $\xi = 3 \times 10^{-3}$ 
with a variety of values of $\tanb$.
The $\Delta S$-$\Delta T$ trajectories (including both Higgs and
quark contributions) are shown as a function of $M_\chi$.
For reference, the SM Higgs corrections to ($S$,\,$T$), relative to
$m_h^{\rm ref} =1$\,TeV, are depicted for $m_h^{\rm SM}$ varying from 
100\,GeV up to 1.0\,TeV in plot (a).
}
\label{fig:tbss_stx}
\end{figure}

As shown in Fig.\,\ref{fig:stk2-list},
we find $S_H$ in our model to be generically small 
while $T_H$ can be large and positive for $\cut \lesssim 10$\,TeV
due to the sizable mass-splittings among Higgs scalars
 $(H^0,\,A^0,\,H^\pm)$. However, for larger $\cut$,
the $A^0$ mass increases and becomes more and more degenerate with
$(H^0,\,H^\pm)$ which quickly brings $T_H$ down, as expected.
The seesaw contributions are also plotted in the same figure,
together with the final summed results. We see that the
inclusion of bottom seesaw helps to reduce the total seesaw
contributions in the $T$ parameter, but the two-doublet-Higgs
sector tends to lift it up. 
This non-trivially  brings our final $T$ bounds
in Fig.\,\ref{fig:tbss_stx}
to the  same level as in the minimal Top seesaw model.
For instance, in the case of $\k/\k_c=2$,
Figs.\,\ref{fig:tbss_stx} and \ref{fig:tssSTmHmx}(c)
show that the $\x$ mass 
in the extended model is  bounded into the region around
$6-23$\,TeV for 
$2\lesssim \tanb \lesssim 40$, while in the minimal
Top seesaw model we have  $10\lesssim \MX \lesssim 14$\,TeV.
For the Topcolor force being more critical (i.e.,
smaller  $\k/\k_c$ values below $2$), the seesaw correction  $\Delta T$
is slightly larger (cf. Fig.\,\ref{fig:ST2exact}), 
but at the same time 
the mass $\MX$ ($\MW$) becomes even lower for a given $\cut$
scale [similar to the picture in Fig.\,\ref{fig:tssphysrefapp}(d)] 
and thus the bounds on $\MX$
could be further weakened, in analogy with the 
minimal Top seesaw model.
In summary, the $T$ bound in the extended model restrict
the mass range of $\x$ and $\w$ to be
typically around $3-20$\,TeV, 
depending on the values of $\k/\k_c$ and $\tanb$.

\begin{figure}[H]
\begin{center}
%\vspace*{-15mm}
\includegraphics[width=16cm,height=18cm]{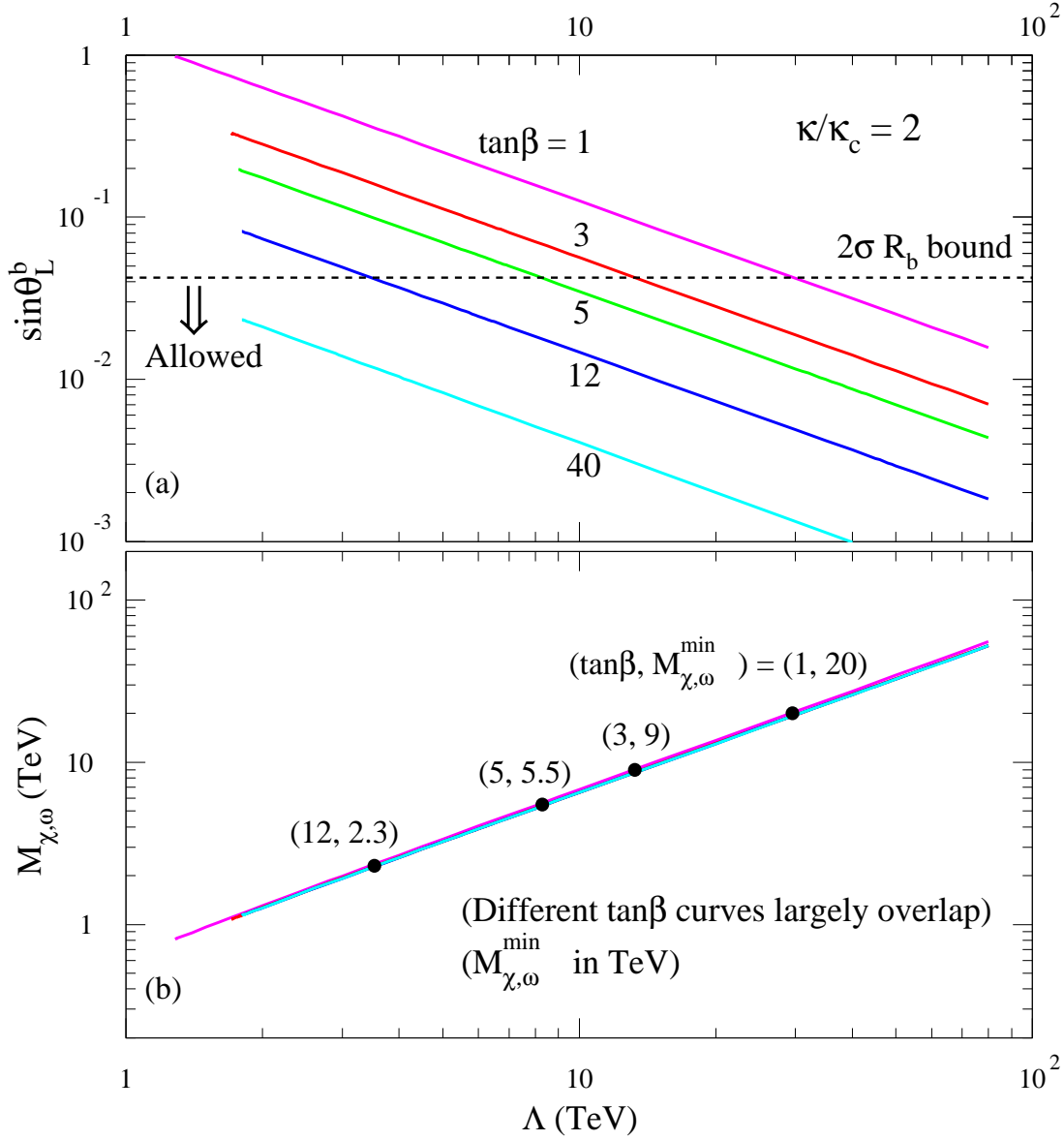}
\end{center}
\vspace*{-13mm}
\caption{
The $R_b$ limits are shown for
the $b$ seesaw angle $s_L^b=\sin\theta_L^b$ in plot (a) and
for the mass $\MW(\simeq \MX)$ in plot (b).
Here, we choose  $\kappa / \kappa_c = 2$ and a wide range of
$\tanb$ values.  
}
\label{fig:tbssZbb}
\end{figure}

Another important bound due to the inclusion of the bottom seesaw
comes from the precision measurement of the 
$Z$-$b$-$\ov{b}$ vertex. 
The seesaw $b-\omega$ mixing induces a positive shift in
the left-handed $Z$-$b$-$\ov{b}$ coupling,
\beq
\delta g_L^b = \dis +\f{e}{2\sin\!\theta_W\cos\!\theta_W}
                      \(s_L^b\)^2  \,,
\eeq
which results in a decrease of 
$R_b=\Gamma[Z\to b\ov{b}]/\Gamma[Z\to {\rm hadrons}]$,
i.e., $R_b \simeq R_b^{\rm SM} - 0.39 \(s^b_L\)^2$, as also
obtained in Ref.\,\cite{georgi}. 
The latest update of $R_b$ data gives\,\cite{Rb},
$R_b=0.21646\pm 0.00065$, which is about $1\sigma$ above the
SM value $R_b=0.2158\pm 0.0002$. 
This puts an upper bound on the $b$-seesaw angle,  
\beq
\dis
s_L^b~\simeq~ \f{m_b/\muwb}{\sqrt{1+r_b}} 
     ~\simeq~ \f{m_b}{\MW\sqrt{r_b}} \,,
\eeq
and correspondingly a lower bound on the mass 
$\MW$ ($\simeq \MX$), as summarized  
in Fig.\,\ref{fig:tbssZbb}.
From the physical seesaw solutions 
[cf. Fig.\,\ref{fig:tbss_soall}(a) in Sec.\,3.2],
we expect that the $R_b$ bound will mainly constrain
the low $\tanb$ region in which $r_b$ is much smaller.
Indeed, the current Fig.\,\ref{fig:tbssZbb}
shows that for larger values of 
$\tanb \;(\gtrsim 15)$, the model is free from the $R_b$ bound,
while for very small values of $\tanb \;(\lesssim 2-3)$, we obtain,
$M_{\x,\w}\gtrsim 10$\,TeV, which is somewhat stronger than the
$T$ bound in Fig.\,\ref{fig:tbss_stx}.
As a final remark, we note that the two-doublet-Higgs sector
can also contribute to the $R_b$, and especially a
charged Higgs boson lighter than about $200-300$\,GeV will
significantly reduce the $R_b$ value\,\cite{Rb-2HDM}. But, in
our model, the relevant Higgs mass spectrum after imposing
the $S-T$ bounds is generically around $\sim\!1$\,TeV or
above (cf. Figs.\,\ref{fig:higgsmasses} and \ref{fig:stk2-list}), 
which renders the Higgs correction to $R_b$ negligible.

\newpage
\section{\hspace*{-2mm}{\large Conclusions
}}

Electroweak symmetry breaking (EWSB) through the Top Quark Seesaw is an 
attractive mechanism that may naturally emerge from theories with 
bosonic extra dimensions.
In this work, we have systematically investigated
the Top Seesaw mechanism for generating the large top mass 
together with the full EWSB. 
We have applied the gap equation analysis to study the seesaw vacuum structure 
and determine the physical parameter space. With the Topcolor breaking
scale ($\cut$) and the Topcolor gauge coupling ($h_1$) as inputs, and
further imposing the physical values of the top mass ($m_t$) and the
full EWSB VEV ($v$), we are able to predict all other seesaw parameters
and thus the physical spectrum of the
model from solving the seesaw gap equation. This includes the masses of
singlet seesaw quark $\x$ and the composite Higgs boson $h^0$. 
The Higgs mass $M_h$ is at the order
of the seesaw mass gap $\mtx$, and typically around $\sim\!0.5-1$\,TeV.
The effective couplings, such as
the Yukawa coupling $h$-$t$-$\ov{t}$ and gauge couplings 
$W$-$t$-$b$ and $Z$-$b$-$\ov{b}$, etc, are also analyzed, 
in comparison with their SM values.

The fermion content of the Top Seesaw is incomplete due to
gauge anomalies, but a minimal choice of additional 
weak-singlet fermions, $\w$, the seesaw
partners for the bottom quark, renders the theory anomaly free
and thus complete. This extended model contains two dynamical
mass gaps $\mtx$ and $\mbw$ in the $(t\x)$ and $(b\w)$ channels,
respectively.
We have performed a complete analysis of the coupled seesaw gap
equations in this extended model. 
The low energy theory contains two composite Higgs doublets.
Topcolor instantons\,\cite{TCATC} are found to 
provide an economic and plausible mechanism for the mass generation 
of the pseudo-scalar $A^0$.
In addition, they may also produce a significant part
of the $b$ mass via the bottom seesaw. 
We have analyzed the resulting Higgs spectrum
in this extended model by using two independent approaches. 
The Higgs mass spectrum typically contains the lightest $h^0$
with a mass around $\sim\!1$\,TeV, and 
three other quite degenerate scalars, 
with masses around one to a few TeV.
We also notice that this model has a particular simple limit,
namely, when the seesaw quark $\w$ becomes heavy enough and 
decouples from the low energy theory, 
it reduces back to the minimal Top seesaw
model with a single Higgs doublet, and in this case, the
bottom mass arises entirely from the Topcolor instanton contribution.

We have further analyzed the electroweak precision bounds on both
the minimal and extended seesaw models.
We find that it is generic in these models to have a small
oblique parameter $S$, but
a significantly positive seesaw contribution to $T$
that largely cancels with the negative $T$ from the heavy 
Higgs boson, in full consistency with the current $S-T$ bounds. 
This makes the dynamical Top seesaw models fully viable, and
as a result, the current precision data is able to indirectly  
confine the heavy $\x$ mass to the natural range of about $\O(3-10)$\,TeV
(for $\k/\k_c\leq 2$)
in the minimal seesaw model. For the extended model with the bottom
seesaw,  the mass of the singlet seesaw quark $\w$ is found to have
good degeneracy with $\x$.
The $b-\w$ mixing tends to reduce the seesaw contribution
in $T$ (especially for the small to moderate $\tanb$ values), 
but the additional correction in the
two-Higgs-doublet sector makes $T$ more positive and thus the final
$S-T$ bounds appear at the similar level to that of the minimal model,
i.e., the allowed seesaw quark masses $M_{\x,\w}$ ranges
from a few TeV up to $\sim\! 30$\,TeV for $1.05\lta \k/\k_c\lta 4$. 
We have also analyzed the correction to the $\Zbb$ gauge
coupling induced by the $b-\w$
mixing and found that the $R_b$ measurement can put stronger bounds
than the $T$ parameter only for very small $\tanb$ region, around
$\tanb \lta 2-3$.

So far, the Top seesaw mechanism,
with necessary ingedients arising automatically
in theories with bosonic extra dimensions, remains
a most natural picture of the dynamical
EWSB scenario, and is consistent with the current experimental
data.  In addition to successfully driving the full EWSB and providing the large
top mass observed at the Tevatron, it has interesting phenomenological
implications, including composite Higgs bosons, additional weak-singlet
quarks in the TeV region, and, ultimately, an entire new layer of strong 
interaction forces at nearby scales to explore.

%\section{\hspace*{-2mm}{\large Acknowledgements
%}}
\vspace*{12mm}
\noi
{\bf  {\large Acknowledgments}}\\[2mm]
We thank many colleagues, especially R. Sekhar Chivukula, 
for discussions and reading the manuscript.
HJH thanks the Fermilab Theory Group
for invitations and support of his summer visits 
during which this collaboration was initiated and
part of his work was performed.   
HJH was supported by the U.S. Department of Energy 
under Grant DE-FG03-93ER40757,   
CTH by the University Research Association for Fermilab 
under contract DE-AC02-76CHO3000 with the U.S. Department of Energy, and 
TT by the U.S. Department of Energy under contract W-31-109-Eng-38.

%\vspace*{15mm}

\newpage
%\pagebreak
\noi
{\bf {\Large Appendices}}
\vspace*{2mm}

\appendix    

\section{\hspace*{-2mm}{\large 
Equivalent Derivations of Top Seesaw Gap Equation 
}}

\subsection{\hspace*{-2mm}{
Exact Large-$N_c$ Gap Equation in the NJL-Formalism
}}

\begin{figure}[H]
\begin{center}
\vspace*{-2mm}
\includegraphics[height=4.5cm]{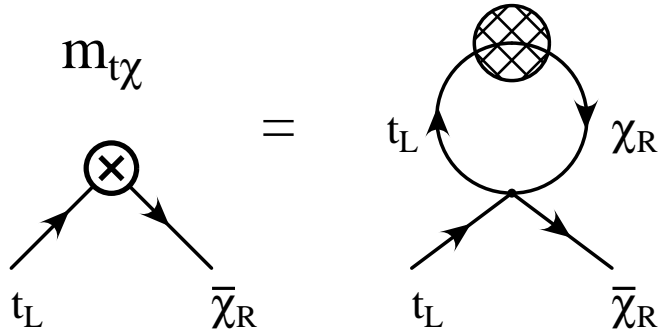}
\end{center}
\vspace*{-7mm}
\caption{
Exact large-$N_c$ Schwinger-Dyson equation for the NJL interaction
in the minimal top seesaw model.
}
\label{fig:NJL-SD}
\end{figure}

In this Appendix, we derive the exact NJL seesaw gap equation
in the  large-$N_c$ limit
based on  the Schwinger-Dyson equation without mass-insertion,
and prove it results in the same equation as the
tadpole condition (\ref{eq:tadgap1})  in Sec.\,2.4.
Starting from the NJL vertex (\ref{njl1}) in Sec.\,2.2, we
can write down the large-$N_c$   Schwinger-Dyson equation
as shown in Fig.\,\ref{fig:NJL-SD}. 

\begin{figure}[H]
\begin{center}
\vspace*{0mm}
\includegraphics[width=14cm]{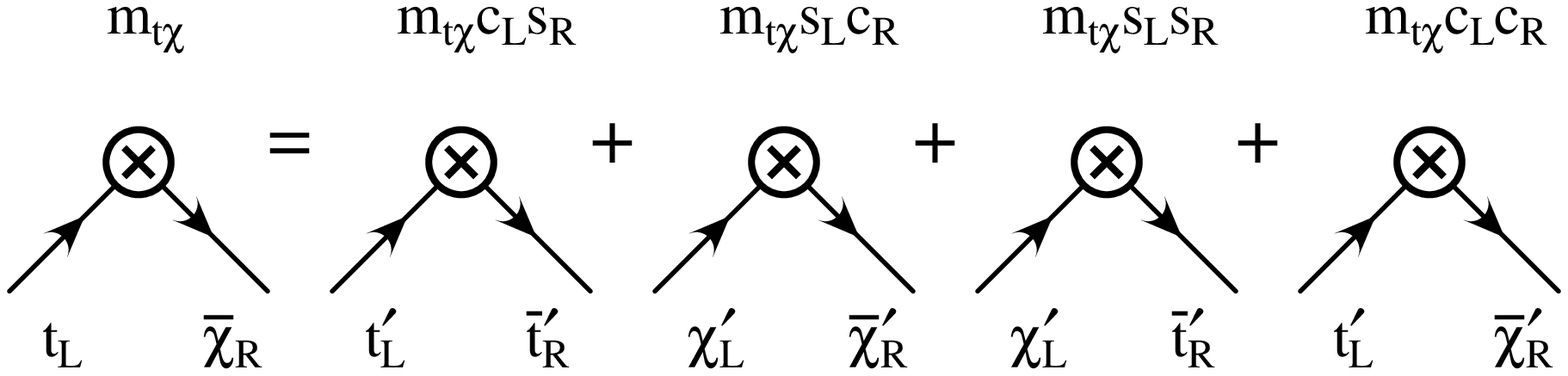}
\\
\vspace*{10mm}
\includegraphics[width=14cm]{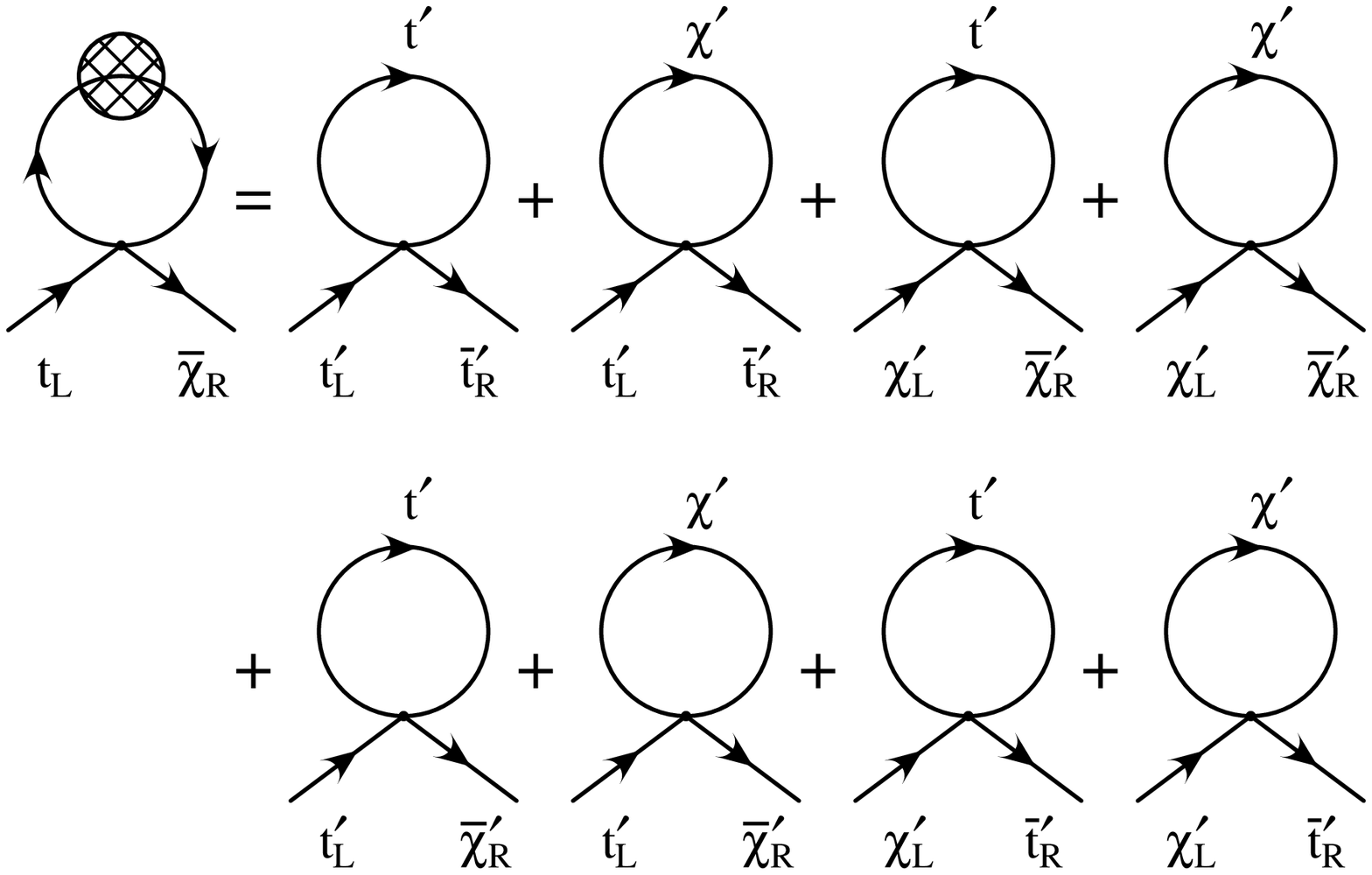}
\end{center}
\vspace*{-1mm}
\caption{
Diagrams expanded from both sides of the large-$N_c$ Schwinger-Dyson equation 
in Fig.\,\ref{fig:NJL-SD}, by using the exact seesaw rotations in
Eqs.\,(\ref{eq:seesawd1})-(\ref{eq:seesawd2}).
}
\label{fig:NJL-SDx}
\end{figure}

Then, we make use of the exact seesaw rotations in 
Eqs.\,(\ref{eq:seesawd1})-(\ref{eq:seesawd2}) to
transform the fields on both sides of the Schwinger-Dyson
equation into the mass eigenbasis. The expanded
diagrams are shown in Fig.\,\ref{fig:NJL-SDx} with proper
rotation angles associated with each graph.
The sums of the expanded diagrams on both sides should be
equal to each other, and in particular, 
each expanded diagram in the upper plot of Fig.\,\ref{fig:NJL-SDx}
must be equal to the sum of the two relevant expanded diagrams
in the lower plot  of Fig.\,\ref{fig:NJL-SDx} which share the
same external lines. (One of two relevant diagrams in the lower
plot of Fig.\,\ref{fig:NJL-SDx} has a $t^\prime$-loop and another
has a $\x^\prime$-loop.)
This leads us to split the Schwinger-Dyson
equation of Fig.\,\ref{fig:NJL-SD} into four
separate equations, which, however, take the following
identical form,
\bea
\label{eq:SDeq}
\mtx &=& -\[ c_Ls_R\Delta_t + c_Rs_L \Delta_\x \] \,,
\eea  
with
\beq
\Delta_t = -\dis\f{h_1^2N_c}{\cut^2}\, {\rm tr}\!\!\int\f{dk^4}{2\pi^4}
\f{i}{~\backslash \!\!\! k -m_t} P_R\,,
~~~~~
\Delta_\x = -\dis\f{h_1^2N_c}{\cut^2}\, {\rm tr}\!\!\int\f{dk^4}{2\pi^4}
\f{i}{~\backslash \!\!\! k -\MX} P_R\,,
\eeq
where $P_{L,R}=\(1\mp\gamma_5\)/2$.
By a direct evaluation of the loop integrals $\Delta_t$
and $\Delta_\x$ with the cutoff $\cut$, we can rewrite
the Schwinger-Dyson equation (\ref{eq:SDeq}) as,
\beq
\label{eq:NJLexactgap}
\vspace*{3.5mm}
%\hspace*{-1cm}
m_{t \chi} ~=~
 \frac{\kappa}{\kappa_c} \left\{ c_L s_R \left( m_t
- \frac{m_t^3}{\cut^2}
\ln \left[ \frac{\cut^2 + m_t^2}{ m_t^2} \right] \right)
%\right. \\ & & \: \left.
+ s_L c_R \left( m_\chi
- \frac{m^3_\chi}{\cut^2}
\ln \left[ \frac{\cut^2 + m_\chi^2}{m_\chi^2} \right] \right)
\right\} \, ,
%\nonumber
\eeq                  
which is just the exact large-$N_c$ seesaw gap equation for
$\mtx$, identical to the tadpole condition (\ref{eq:tadgap1}) in
Sec.\,2.4. This proves the equivalence
between the NJL formalism and the Higgs tadpole formalism
for deriving the seesaw gap equation.
\vspace*{2mm}

\subsection{\hspace*{-2mm}{
Mass-Insertion Gap Equation from Tadpole Condition
}}

\begin{figure}[H]
\begin{center}
\vspace*{-2mm}
\includegraphics[width=13.4cm]{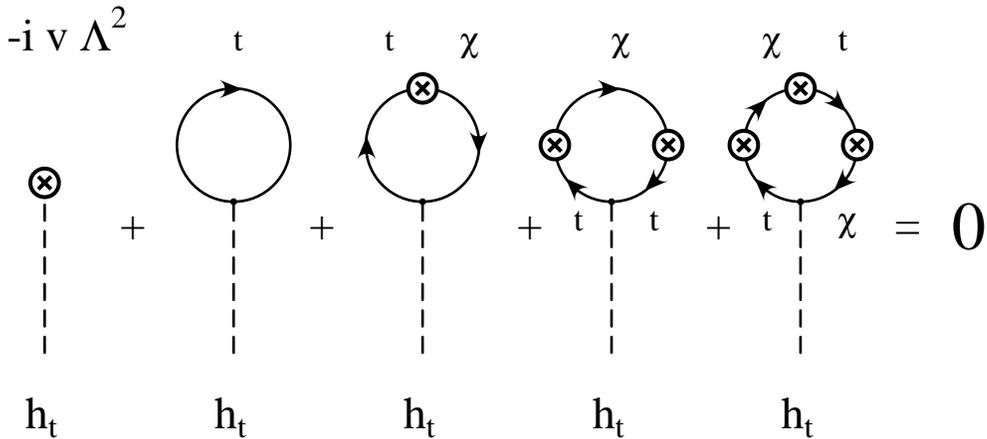}
\end{center}
\vspace*{-3mm}
\caption{
Higgs tadpole condition with mass-insertions to $\O(\mtx^3)$.
}
\label{fig:tadpole-app}
\end{figure}

Here, we derive the tadpole condition with mass-insertion 
up to $\O(\mtx^3)$ and prove it results in the same equation as the
approximate NJL gap equation in Sec.\,2.3.
From the NJL interaction in Eq.\,(\ref{njl2}), we introduce
the auxiliary field $\Phi_0$ which,  in the unitary gauge,
takes the form of (\ref{eq:htshift}) with the 
VEV explicitly shifted.
Then, the effective Lagrangian at the
scale $\mu=\cut$ becomes, 
\beq
\label{eq:LcutMI}
\ba{ll}
{\cal L}_{\cut} &
\!\!\!
=
- \left( \overline{t_L} ~~ \overline{\chi_L} \right)
 \left\lgroup
\begin{array}{cc}
m_{tt} & {\mtx} \\[0.2cm]
0      & \MB        \\
\end{array}
\right\rgroup
\left(\!\!
\begin{array}{c}
t_R \\
\chi_R
\end{array}
\!\!\right)
-\dis\f{h_1}{\sqrt{2}}\ov{t_L}
\(c\x_R+st_R\)h_0+ {\rm h.c.}
-\half\cut^2 h_0^2 - \cut^2 v_0h_0  \\[5mm]
\ea                            
\eeq
where 
\beq
\mtx =\dis c\f{h_1 v_0}{\sq2}\,,~~~~
m_{tt}=\dis s\f{h_1 v_0}{\sq2}\,,~~~~\longrightarrow~~~~
\f{m_{tt}}{\mtx}=\f{s}{c} \,, 
\eeq
and $s/c = \muxt/\muxx$ is the same as in Eq.\,(\ref{eq:thetaRapp}).
The diagonal mass-terms $m_{tt}$ (\,$\MB$\,) will be put into the
$t$ ($\x$) propagator as usual, while the non-diagonal mass-term
$\mtx$ can be included via the mass-insertion order by order. 
It is then straightforward
to derive the tadpole condition $0=v_0\cut^2+\delta\widetilde{T}$
[similar to Eq.\,(\ref{eq:tadpole})] with $\delta\widetilde{T}$
computed from the one-loop Higgs tadpole diagrams. This is shown
in Fig.\,\ref{fig:tadpole-app},
in which we perform the mass-insertion of $\mtx$ up to
the third power. As a result, we derive a single condition
on $\mtx$, which is {\it identical} to the gap equation 
(\ref{eq:gapeqtx}) derived earlier in Sec.\,2.3 by using
the NJL formalism. This shows the equivalence between these
two mass-insertion approaches.

\section{\hspace*{-2mm}{\large 
Potential Analysis for Higgs Mass Spectrum \\
\hspace*{-1.6mm}with Bottom Seesaw
}}

\begin{figure}[H]
\begin{center}
\vspace*{-2mm}
\includegraphics[height=8cm]{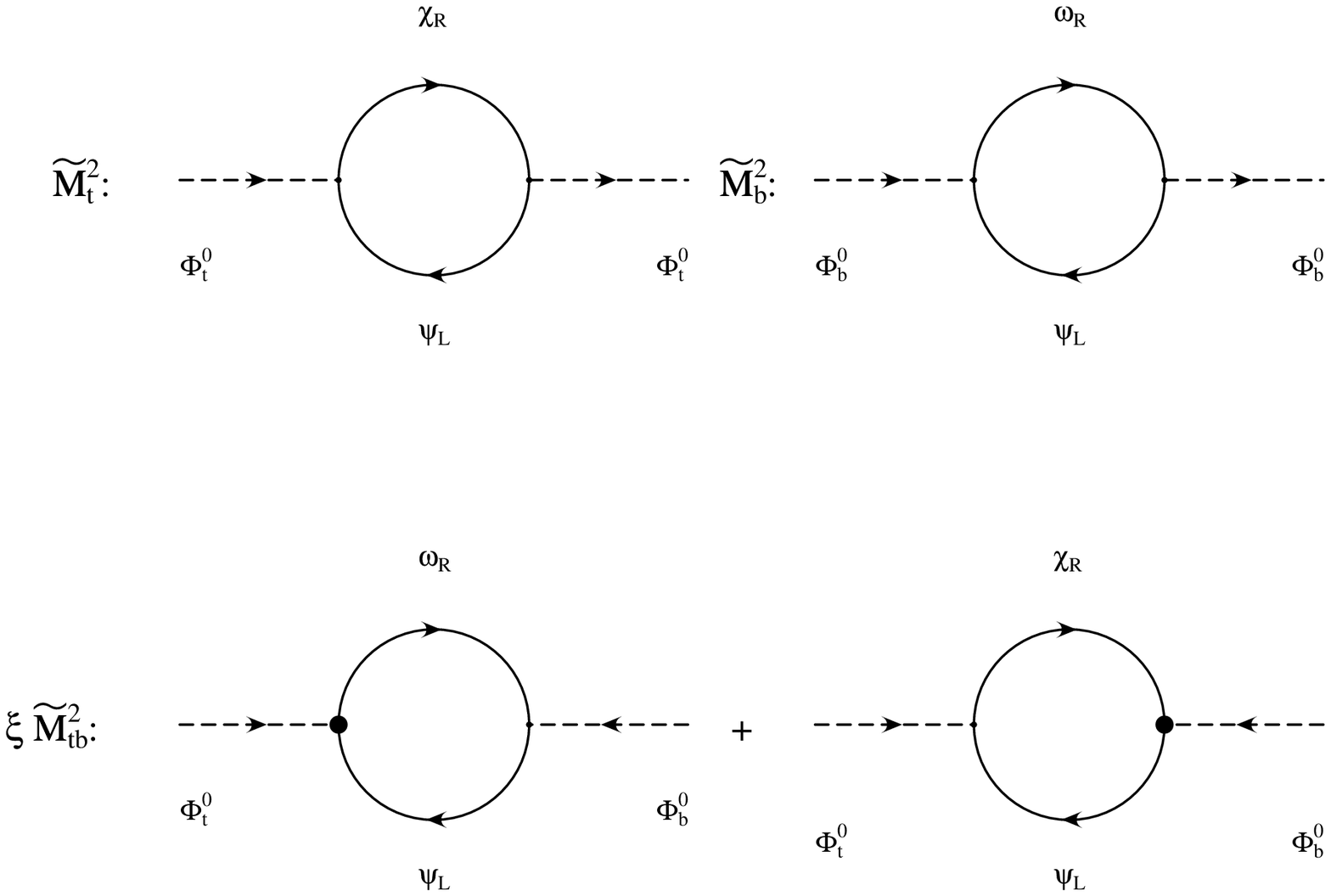}
\\
\vspace*{18mm}
\hspace*{5mm}
\includegraphics[height=9.5cm]{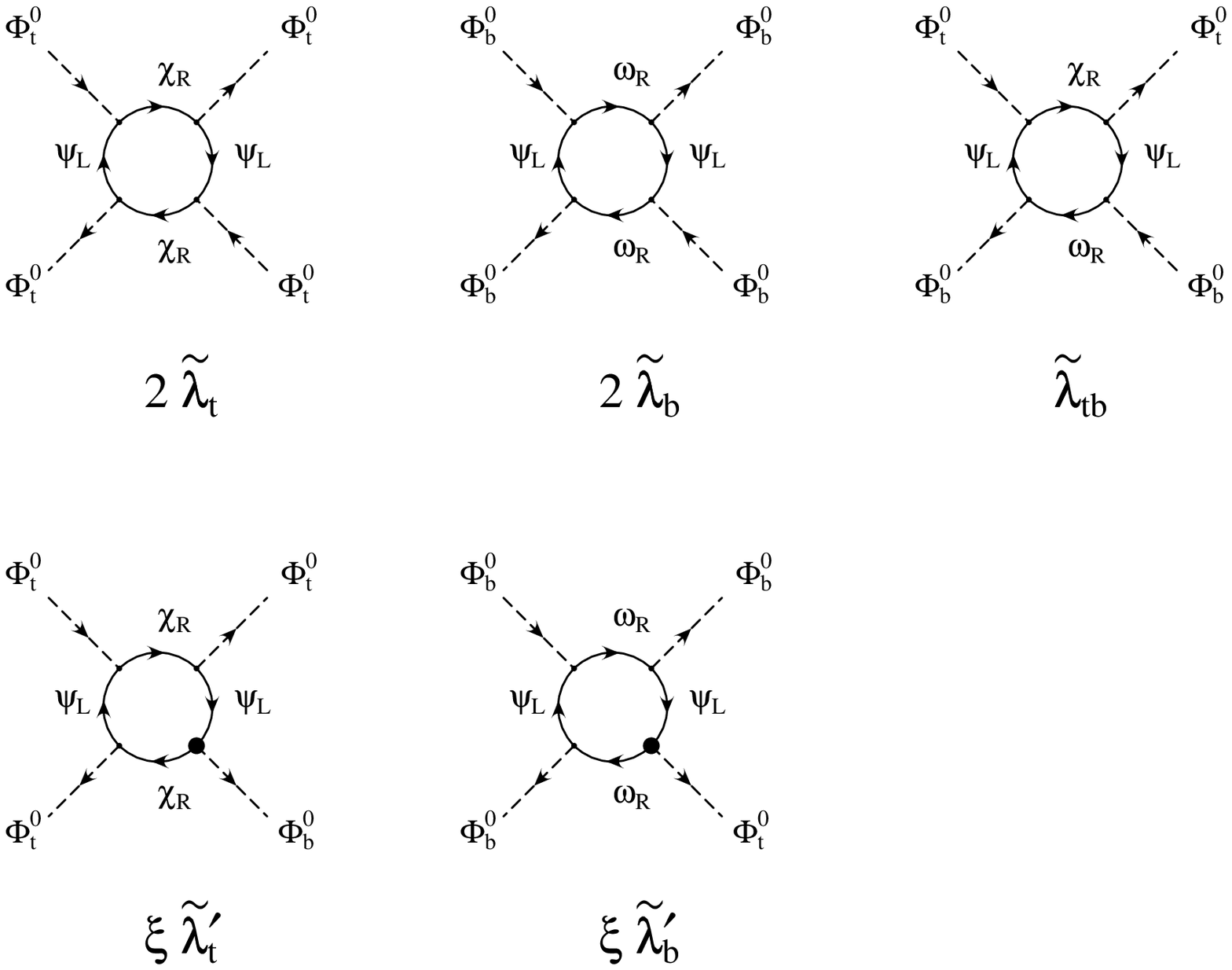} 
\end{center}
%\vspace*{5mm}
\caption{
Effective mass terms and quartic self-couplings in the Higgs potential.
}
\label{fig:mass-coup}
\end{figure}

In this Appendix, we present an independent derivation of the
composite Higgs spectrum by analyzing the Higgs potential 
in the extended model with a bottom seesaw.
The potential analysis confirms the results derived in
Sec.\,3 where the Higgs masses are explicitly computed 
in the broken phase including the exact
seesaw mass-diagonalizations.
We start from the gauge-invariant Lagrangian 
$\,{\cal L}_{\rm mass}+{\cal L}_{\rm int}+{\cal L}_{\rm PQB}$
in Sec.\,3 
[cf. Eqs.\,(\ref{eq:xwmass}), (\ref{eq:Lxw0}) and (\ref{eq:LPQ})],  
and evolve it down to the scale $\mu \(< M_{\x,\w} \leq \cut\)$.
We can thus derive the gauge-invariant effective Lagrangian
with two dynamical Higgs-doublet fields and their interaction terms,
up to $\O(\xi)$,
\beq
\label{eq:LLmu}
\ba{ll}
{\cal L}_{\mu < M_{\x,\w}} \,=&
-h_1\[s_t\;\ov{\psi}t_R\Phi_{t0}+ s_b\;\ov{\psi}b_R\Phi_{t0}+{\rm h.c.}\] +
 Z_{t} | D_\mu \Phi_{t0} |^2
+Z_{b} | D_\mu \Phi_{b0} |^2   \\[3mm]
& +\xi\(Z_{t}+Z_{b}\)\ep^{\alpha\beta}
  \[( D^\mu \Phi_{t0} )^\alpha ( D_\mu \Phi_{b0} )^\beta + {\rm h.c.}\] 
  -V_H\,, %[\Phi_{t0},\Phi_{b0}]
\ea  
\eeq 
where $(s_t,\,s_b)\equiv (\sin\theta_t,\,\sin\theta_b)$ and
$(\tan\theta_t,\,\tan\theta_b) \equiv (\muxt/\muxx,\,\muwb/\muww)$,
with $(\theta_t,\theta_b)$ the partial rotation angles
for $(t_R,\,\x_R)$ and $(b_R,\,\w_R)$  
[cf. Eqs.\,(\ref{eq:partialrotation})-(\ref{eq:thetaRapp}) 
in Sec.\,2.2].
The Higgs potential $V_H$ can be written as,
\beq
\label{eq:V0}
\ba{ll}
V_H \,= &
\widetilde{M}_t^2 | \Phi_{t0} |^2 
+\widetilde{M}_b^2 | \Phi_{b0} |^2 
+\xi\widetilde{M}_{tb}^2 \left[ \epsilon^{\alpha \beta} 
     \Phi_{t0}^{\alpha} \Phi_{b0}^{\beta} + {\rm h.c.} \right]
+ \widetilde{\lambda}_t ( \Phi_{t0}^{\dagger} \Phi_{t0} )^2 
+ \widetilde{\lambda}_b ( \Phi_{b0}^{\dagger} \Phi_{b0} )^2
  \nonumber 
\\ [3mm]
& +\widetilde{\lambda}_{tb} ( \Phi_{t0}^{\dagger} \Phi_{b0} )
                            ( \Phi_{b0}^{\dagger} \Phi_{t0} )
  +\xi \left[ \widetilde{\lambda}_t^\prime | \Phi_{t0} |^2 +
               \widetilde{\lambda}_b^\prime | \Phi_{b0} |^2 \right]
    \left[ \ep^{\alpha \beta} \Phi_{t0}^{\alpha} 
                              \Phi_{b0}^{\beta} + {\rm h.c.} \right] ,
\ea
\eeq
where the loop-induced Higgs mass terms and couplings are graphically
defined in Fig.\,\ref{fig:mass-coup}. For simplicity, the fermion
lines of $\x_R$ and $\w_R$ represent the fields before the partial
rotations mentioned above, but we keep in mind that such rotations just 
split each graph into two; this will not affect our 
current general derivation as is easy to check. 
From Eq.\,(\ref{eq:V0}) and Fig.\,\ref{fig:mass-coup},
we can derive three general relations, up to $\O(\xi)$,
\beq
\label{eq:relation}
\xi\widetilde{M}^2_{tb} =
\xi \[  \widetilde{M}^2_t + \widetilde{M}^2_b -\cut^2
\] ,~~~~
\widetilde{\lambda}_t^\prime =2\widetilde{\lambda}_t\,,~~~~
\widetilde{\lambda}_b^\prime =2\widetilde{\lambda}_b\,.
\eeq

The next step is to write down the renormalized Higgs potential,
analyze the physical vacuum and derive the Higgs mass spectrum.
So, we first express $V_H$ in terms of renormalized quantities, 
\beq
\label{eq:VR}
\ba{ll}
V_H \,= &
{M}_t^2 | \Phi_{t} |^2 
+{M}_b^2 | \Phi_{b} |^2 
+\xi {M}_{tb}^2 \left[ \epsilon^{\alpha \beta} 
     \Phi_{t}^{\alpha} \Phi_{b}^{\beta} + {\rm h.c.} \right]
+ {\lambda}_t ( \Phi_{t}^{\dagger} \Phi_{t} )^2 
+ {\lambda}_b ( \Phi_{b}^{\dagger} \Phi_{b} )^2
  \nonumber 
\\ [1mm]
& +{\lambda}_{tb} ( \Phi_{t}^{\dagger} \Phi_{b} )
                            ( \Phi_{b}^{\dagger} \Phi_{t} )
  +\xi \left[ {\lambda}_t^\prime | \Phi_{t} |^2 +
              {\lambda}_b^\prime | \Phi_{b} |^2 \right]
    \left[ \ep^{\alpha \beta} \Phi_{t}^{\alpha} 
                              \Phi_{b}^{\beta} + {\rm h.c.} \right] ,
\ea
\eeq
with the following renormalization relations,
\bea
\label{eq:RR}
%\ba{l}
&
M_t^2  = \widetilde{M}_t^2 Z_t^{-1},~~~
M_b^2  = \widetilde{M}_b^2 Z_b^{-1},~~~
M_{tb}^2  = \widetilde{M}_{tb}^2 \(Z_t Z_b\)^{-\half} ,
&
\\[3mm]
&
\lambda_t = \widetilde{\lambda}_t Z_t^{-2},~~~
\lambda_t^\prime = \widetilde{\lambda}_t^\prime {Z_t^{-\f{3}{2}} Z_b^{-\half}},~~~
\lambda_b = \widetilde{\lambda}_b Z_b^{-2},~~~
\lambda_b^\prime = \widetilde{\lambda}_b^\prime {Z_t^{-\half} Z_b^{-\f{3}{2}}} ,~~~
\lambda_{tb} = \widetilde{\lambda}_{tb} \(Z_t Z_b\)^{-1}.
&
\nonumber
\eea
%\eeq
where $(Z_t,\,Z_b)$ are the wavefunction renormalization constants
of $(\Phi_{t0},\,\Phi_{b0})$, defined as,
$\Phi_{t0}=Z_t^{-1/2}\Phi_t$ and 
$\Phi_{b0}=Z_b^{-1/2}\Phi_b$.
Then, we can shift the VEVs of the two renormalized
Higgs-doublets, $(\Phi_t,\,\Phi_b)$,
similar Eq.\,(\ref{eq:hthb}) in Sec.\,3.2.
For the analysis of physical vacuum and the Higgs mass spectrum, it is
convenient to choose the unitary gauge, in which
the three physical combinations (orthogonal to the would-be Goldstone
bosons) are defined as,
$A^0 = \sin \!\beta ' \pi_b^0 + \cos \beta ' \pi_t^0$ and
$H^\pm = \sin \!\beta ' \pi^\pm_b + \cos \!\beta ' \pi^\pm_t$,
with $\tanb ' = (v_t + \xi v_b) / (v_b + \xi v_t) = \tanb 
+ \xi (1 - \tanb)+\O(\xi^2)$.

Minimizing the effective Higgs potential $V_H$ in (\ref{eq:VR}), we 
derive two extremum conditions,
\bea
\label{eq:gapV}
\[M_t^2 + \lambda_t v_t^2\] 
+ \xi \cot \beta \left[ 
M_{tb}^2 + \frac{3}{2} \lambda_t^\prime v_t^2 
+ \frac{1}{2} \lambda_b^\prime v_b^2 \right] &=& 0 \,,\nonumber \\
\[M_b^2 + \lambda_b v_b^2 \]
+ \xi \tan \beta \left[ 
M_{tb}^2 + \frac{1}{2} \lambda_t^\prime v_t^2 
+ \frac{3}{2} \lambda_b^\prime v_b^2 \right] &=& 0 \, ,
\eea
which determine the physical vacuum and is formally equivalent to the
gap equations (tadpole conditions) derived 
in Eq.\,(\ref{eq:gaptb}) of Sec.\,3.2.
These conditions are needed in our derivation and can be used to simplify
the mass formulae for the Higgs bosons. 
We start by extracting the $A^0$ mass term from Higgs potential (\ref{eq:VR})
and obtain, up to $\O(\xi)$,
\bea
\label{eq:V-MA}
M^2_A &=& \frac{\xi}{2 \sin \beta \, \cos \beta} \left[
-2 M_{tb}^2 - \lambda_t^\prime v_t^2 - \lambda_b^\prime v_b^2 \right]
= \frac{2 \xi \Lambda^2}{\dis\sin\!2\beta \sqrt{Z_{t}Z_{b}}}  \,,
\eea
where we have used the 
minimal conditions in Eq.\,(\ref{eq:gapV}), and the relations in
Eqs.\,(\ref{eq:relation}) and (\ref{eq:RR}).
This result
confirms our explicit one-loop calculation of $M_A$ in Eq.\,(\ref{eq:MA}) of
Sec.\,3.3. We proceed by deriving the mass formulae for 
the neutral and charged Higgs bosons $(h_t^0,\,h_b^0,\,H^\pm)$, 
which can be summarized up to $\O(\xi)$,
\bea
M_{11}^2 &\!\!=\!\!& 2 \lambda_b v_b^2
+ \dis\xi \left[ 3 \lambda_b^\prime v_t v_b - \tanb \left(
M^2_{tb} + \frac{1}{2} \lambda_t^\prime v_t^2 
+ \frac{3}{2} \lambda_b^\prime v_b^2 \right) \right] 
\nonumber \\
&\simeq&  4 \mbw^2 + \sin^2 \!\beta \, M^2_A \,,
\\[3mm]
M_{22}^2 &\!\!=\!\!&  2 \lambda_t v_t^2 
+ \dis\xi \left[ 3 \lambda_t^\prime v_t v_b - \cot \!\beta \left(
M^2_{tb} + \frac{3}{2} \lambda_t^\prime v_t^2 + \frac{1}{2}
\lambda_b^\prime v_b^2 \right) \right] 
\nonumber \\
&\simeq&  4 \mtx^2 + \cos^2 \!\beta \, M^2_A \,, 
\\[3mm]
\xi M_{12}^2 &\!\!=\!\!&  \dis\xi \left[ 
M_{tb}^2 + \f{3}{2} \lambda_t^\prime v_t^2 
         + \f{3}{2} \lambda_b^\prime v_b^2 \right] 
\nonumber \\
&\simeq&  - \sin \!\beta \, \cos \!\beta \; M^2_A 
+ 4 \xi \left( \mtx^2 + \mbw^2 \right) ,
\eea
and
\bea
M^2_{H^\pm} &=& \frac{\lambda_{tb}}{2} 
                \left[ v_t\sin\!\beta^\prime  + v_b\cos\!\beta^\prime 
               \right]^2
-\dis\f{\xi}{\sin \!2\beta}\[2M_{tb}^2+\la_t'v_t^2+\la_b'v_b^2\]
\nonumber \\[3mm] 
& \simeq &
2 \(\mtx^2 + \mbw^2\) + \[M^2_A 
- 4 \xi \( \mtx^2\cot\!\beta + \mbw^2\tan\!\beta \)\] ,
\eea
where  mass notations of $M_{11,22,12}^2$ are the same as 
Eq.\,(\ref{eq:Mdef}) in Sec.\,3.3, and
for simplification we have used the relation,
$M_\chi \simeq M_\omega$,
which results in $Z_t \simeq Z_b \simeq Z_{tb} / 2$
and 
$\lambda_t \simeq \lambda_b \simeq \lambda_{tb} / 2$.
These are good approximations since the heavy masses
$\MX$ and $\MW$ only affect them via weak logarithmic dependences
(due to the decoupling theorem)
and $M_\chi \simeq M_\omega$ is also justified
from our physical seesaw solutions in Fig.\,\ref{fig:tbss_soall}.
In summary, the above analysis agrees with our calculations 
in Sec.\,3.3, and is particularly simple in extracting leading
corrections at $\O(\xi)$.  It is also remarkable that in this 
analysis we derive all relations in a rather general and formal
manner in which no explicit one-loop calculation is needed
for the quantities such as  $Z_{t,b}$ and $\lambda_{t,b,tb}$\,.\\

%\pagebreak
%
\vspace*{-5mm}

\pagebreak

\end{document}